\documentclass[twocolumn,showpacs,amsmath,amssymb,prd]{revtex4}
\usepackage{graphicx}
\usepackage{graphicx}
\usepackage{epsf}

\def\agt{\mathrel{\raise.3ex\hbox{$>$}\mkern-14mu\lower0.6ex\hbox{$\sim$}}}
\def\alt{\mathrel{\raise.3ex\hbox{$<$}\mkern-14mu\lower0.6ex\hbox{$\sim$}}}
\newcommand{\beq}{\begin{equation}}
\newcommand{\eeq}{\end{equation}}
\newcommand{\beqn}{\begin{eqnarray}}
\newcommand{\eeqn}{\end{eqnarray}}

\newcommand{\cB}{{\cal{B}}}

\begin{document}

\title{Collapse of magnetized hypermassive neutron stars in general 
relativity: \\  Disk evolution and outflows}

\author{Branson C. Stephens}
\altaffiliation{Present address:  Princeton Center for Theoretical 
Physics, Princeton, NJ 08544}

\author{Stuart L. Shapiro}
\altaffiliation{Also at the Department of Astronomy and NCSA, University
of Illinois at Urbana-Champaign, Urbana, IL 61801}

\author{Yuk Tung Liu}

\affiliation{Department of Physics, University of Illinois at 
Urbana-Champaign, Urbana, IL 61801, USA}

\begin{abstract}
We study the evolution in axisymmetry of accretion disks formed 
self-consistently through collapse 
of magnetized hypermassive neutron stars to black holes.  %
Such stars can arise
following the merger of binary neutron stars.  They are differentially 
rotating, dynamically stable, and 
have rest masses exceeding the mass limit for uniform rotation.  
However, hypermassive neutron stars  are secularly unstable to collapse due 
to MHD-driven angular momentum  transport.  The rotating black hole 
which forms in this process is surrounded by a hot, massive, magnetized torus 
and a magnetic field collimated along the spin axis.  This system is a 
candidate for the central engine of a short-hard gamma-ray burst (GRB).   %
Our code integrates the coupled 
Einstein-Maxwell-MHD equations and is used to follow the collapse 
of magnetized hypermassive neutron star models in full general relativity
until the spacetime settles down to a quasi-stationary state. 
We then employ the Cowling approximation, in which the spacetime is frozen, 
to track the subsequent evolution of the disk.  This approximation allows us to 
greatly extend the disk evolutions and study the resulting outflows, which may 
be relevant to the generation of a GRB.  %
We find
that outflows are suppressed when a stiff equation of state is assumed
for low density disk material and are sensitive to the initial magnetic 
field configuration.  
 
\end{abstract}


\maketitle

\section{Introduction}

Binary neutron star coalescence has been proposed
for many years as an explanation of short-hard GRBs~\cite{GRB,GRB-BNS}. 
Possible associations between short GRBs and elliptical galaxies
reported recently~\cite{short} make it unlikely that short GRBs are
related to collapse of massive stars and supernovae.  The merger of 
compact-object binaries (neutron star-neutron star or black hole-neutron star) 
is now the favored hypothesis for explaining short GRBs.  According to this 
scenario, the merger results in the formation of a stellar-mass black hole
surrounded by a hot accretion torus which contains $\sim 1$--$10\%$ of 
the total mass of the system.  Energy extracted 
from this system, either by MHD processes or neutrino-radiation, powers 
the GRB fireball.  The viability of this model depends in part on the presence 
of a sufficiently massive accretion disk following collapse and
on whether the accretion flow produces sufficiently energetic outflows.

Instead of directly collapsing to a black hole, some binary
neutron star (NS) mergers could form hypermassive neutron stars (HMNSs)
as an intermediate state.  Such stars are supported 
against collapse by strong differential rotation~\cite{BSS,SBS,MBS04}, 
which naturally arises from the merger~\cite{RS99,SU00,FR00}.
The latest binary NS merger
simulations in full general relativity~\cite{STUa,STUb,STUc} have
confirmed that HMNS formation is indeed a possible outcome. 
We note, however, that HMNSs could also result from core collapse 
of massive stars, since collapse generates strong differential 
rotation~\cite{zm97,diffrot} (see also~\cite{liu01}).

Differentially rotating stars tend to approach rigid rotation when 
acted upon by processes which transport angular momentum. HMNSs,
however, cannot settle down to rigidly rotating NSs since 
their masses exceed the maximum allowed by rigid rotation. These objects
thus undergo `delayed' collapse to a black hole (BH).
Several processes can act to transport angular momentum and drive the 
HMNS to collapse.  For example, previous calculations in full general 
relativity have modeled HMNS evolution driven by viscous angular momentum 
transport~\cite{viscosity} and by angular momentum loss due to gravitational 
radiation~\cite{STUb}.  

HMNS collapse due to magnetic field effects has recently been studied
numerically in~\cite{DLSSS,grbletter,DLSSS2} using codes for evolving 
magnetized fluids in full general relativity~\cite{DLSS,SS05} 
(see also~\cite{valencia:fn,byucode,aeicode}).  The secular angular momentum 
transport in this case 
is provided by two primary MHD effects.  The magnetic winding 
effect~\cite{spruit} refers to the twisting of ``frozen-in'' magnetic field 
lines by a shear flow.  The resulting magnetic stresses transport 
angular momentum so as to drive the fluid toward uniform rotation.  
This magnetic braking occurs on the Alfv\'en timescale~\cite{Shapiro2000}, 
which is typically much longer than the dynamical time.  In contrast 
to this smooth winding process, the magnetorotational instability 
(MRI)~\cite{MRI0,MRIrev} leads to exponential growth of field line distortions 
on a timescale comparable to the rotation period.  The nonlinear outcome 
of this instability is MHD turbulence, which enhances angular 
momentum transport.  Thus, magnetic braking and the MRI ultimately lead 
to collapse of the HMNS and the 
formation of a hot, magnetized accretion disk surrounding the central BH.  
This hot disk produces copious $\nu\bar{\nu}$ 
pairs, and the resulting annihilation energy is roughly commensurate 
with the requirements for a short-hard GRB (as long as the emission is 
somewhat beamed)~\cite{grbletter}.

In addition to $\nu\bar{\nu}$ pairs, 
MHD processes have also been suggested as mechanisms of powering 
GRBs.  In this paper, we explore the possibility of MHD-induced jet 
formation by simulating the self-consistent formation of disks
through HMNS collapse in full general relativity, continuing the 
disk evolution for up to $\sim 2000M$ after the collapse, and 
examining the physical processes at work. 
The long duration of these simulations is achieved by using the Cowling 
approximation (in which the spacetime metric is fixed) following the 
phase of black hole excision and live evolution of the spacetime metric.
(We note that by imposing the Cowling approximation, we cannot take into account 
changes in the metric due to the rearrangement of mass in the disk
and/or accretion.   However, we find that the change in disk mass 
during the Cowling phase is $\alt 3$\% of the total mass in all of the 
cases we study.   Hence, the Cowling approximation is expected to 
be fairly accurate.)  
We also demonstrate 
our code's ability to handle magnetized accretion flows in stationary 
spacetimes by reproducing the results of~\cite{mg04} for accretion of a 
magnetized Fishbone-Moncrief torus~\cite{fishmon} onto a fixed Kerr BH.  
The good agreement of our results with the published results provides 
confidence in our code's ability to handle complex MHD accretion scenarios.

While the HMNS calculations presented here employ the Cowling
approximation for the late phases of the evolution, we emphasize
that these runs are performed using a code capable of handling
{\it dynamical} spacetimes.  Using the Cowling approximation
allows us to probe the quasistationary behavior at much later
times than presented in our earlier study [19].  We
also note that the final spacetime is more general than the Kerr
spacetime (our code allows for the presence of a
massive accretion disk).  Accretion disk dynamics in such a
spacetime have not been explored by previous fixed-background
{\it stationary} spacetime GRMHD simulations.

The evolution of accretion flows around BHs in stationary spacetimes and 
the consequent jet formation  has been studied numerically
by several groups~\cite{mg04,dvhk03,dvhkh05,dvso05,mcnum,mnkhf06,kskm02}.  The initial
data for the studies of McKinney and Gammie~\cite{mg04}
and of De Villiers {\it et al.}~\cite{dvhk03} consist of thick tori 
with weak poloidal field loops surrounding Kerr BHs of varying spins.  
The magnetic field is subject to the MRI, and the resulting MHD turbulence 
drives accretion onto the central BH.  Magnetic field lines 
carried into the BH open outward and take on a stationary, split-monopole 
like structure.  A relativistic, Poynting dominated outflow develops 
in this funnel region, while a mildly relativistic, matter dominated 
outflow moves along the outer edge of the funnel.  McKinney~\cite{mcnum}
considered the evolution of these outflows as they propagate to 
large radius and found that the terminal Lorentz factors for the 
inner, fast jet range up to $\sim 10^3$, easily accommodating observational
constraints on GRB jets.   McKinney also showed that the opening 
angle of the outflow is controlled at small radii by the corona pressure, 
at intermediate radii by the funnel wall outflow, and at large distances
by internal magnetic stresses.  In contrast, Mizuno 
{\it et al.}~\cite{mnkhf06} consider a thin, Keplerian disk threaded by a 
uniform, vertical magnetic field.  Accretion onto the BH again 
leads to a faster inner jet and a slower outer jet, though both 
are mildly relativistic. 

In this paper, we find that HMNS collapse leads to a magnetically 
dominated funnel region surrounded (in some cases) by a mildly relativistic, 
unbound outflow.  This flow has a similar morphology to those of the 
magnetized accretion disk simulations in stationary spacetimes 
described above~\cite{mg04,dvhk03}.   Though we do not find 
a Poynting-dominated jet in the funnel, the evolution in this region
is sensitive to the numerical handling of matter in highly 
magnetically dominated regimes.  As discussed in~\cite{DLSS}, accurate 
evolution is such regions is a significant challenge.

Below, we first give a brief description of our formulation and numerical methods.  
In Section~\ref{disk}, we present results for the evolution of a magnetized 
torus surrounding a Kerr 
BH, following~\cite{mg04}.  We describe our results for disk evolution 
following HMNS collapse in Section~\ref{sec:results} and summarize in 
Section~\ref{sec:conclusions}.  
  
\section{Formulation}
\label{sec:formulation}

\subsection{Basic equations and numerical methods}
\label{sec:basic_eqs}

The formulation and numerical scheme for our general relativistic,
magnetohydrodynamic (GRMHD) simulations are
the same as those reported in~\cite{DLSS}, to which the reader 
may refer for details.
Here we briefly summarize the method and introduce our notation.
We assume geometrized units ($G = c = 1$) except where stated explicitly.

We adopt the Baumgarte-Shapiro-Shibata-Nakamura (BSSN)
formalism~\cite{BSSN} to evolve the spacetime metric. In this
formalism, the evolution variables are the conformal exponent $\phi
\equiv \ln \gamma/12$, the conformal 3-metric $\tilde
\gamma_{ij}=e^{-4\phi}\gamma_{ij}$, three auxiliary functions
$\tilde{\Gamma}^i \equiv -\tilde \gamma^{ij}{}_{,j}$, the trace of
the extrinsic curvature $K$, and the tracefree part of the conformal extrinsic
curvature $\tilde A_{ij} \equiv e^{-4\phi}(K_{ij}-\gamma_{ij} K/3)$.
Here, $\gamma={\rm det}(\gamma_{ij})$, and $\gamma_{ij}$ is the spatial
3-metric. The full spacetime metric $g_{\mu \nu}$
is related to the three-metric $\gamma_{\mu \nu}$ by $\gamma_{\mu \nu}
= g_{\mu \nu} + n_{\mu} n_{\nu}$, where the future-directed, timelike
unit vector $n^{\mu}$ normal to the time slice can be written in terms
of the lapse $\alpha$ and shift $\beta^i$ as $n^{\mu} = \alpha^{-1}
(1,-\beta^i)$.  

In this paper, we assume both equatorial and axisymmetry, 
and so we only evolve the region with $x>0$ (where $x$ represents
the cylindrical radius $\varpi$) and $z>0$. We adopt 
the Cartoon method~\cite{cartoon} to impose axisymmetry in the metric 
evolution, and use a cylindrical grid to evolve the MHD and Maxwell 
equations. For the gauge conditions, 
we adopt hyperbolic driver conditions as in~\cite{excision_hydro} to 
evolve the lapse $\alpha$ and shift $\beta^i$. 

The fundamental variables in ideal MHD are the rest-mass density 
$\rho$, specific internal energy $\varepsilon$, pressure $P$, four-velocity 
$u^{\mu}$, and magnetic field $B^{\mu}$ measured by a normal
observer moving with a 4-velocity $n^{\mu}$ (note that $B^{\mu} n_{\mu}=0$). 
During the evolution, we also need the three-velocity $v^i = u^i/u^t$.
The ideal MHD condition is written as $u_{\mu} F^{\mu\nu}=0$, 
where $F^{\mu\nu}$ is the electromagnetic tensor. The tensor 
$F^{\mu\nu}$ and its dual in the ideal MHD approximation are 
given by 
\beqn
&&F^{\mu\nu}=\epsilon^{\mu\nu\alpha\beta}u_{\alpha}b_{\beta}, \label{eqFF}\\
&&F^*_{\mu\nu} \equiv {1 \over 2}\epsilon_{\mu\nu\alpha\beta} F^{\alpha\beta}
=b_{\mu} u_{\nu}- b_{\nu} u_{\mu}, 
\eeqn
where $\epsilon_{\mu\nu\alpha\beta}$ is the Levi-Civita tensor.  
Here we have introduced an auxiliary magnetic 4-vector 
$b^{\mu}=B^{\mu}_{(u)}/\sqrt{4\pi}$, where $B^{\mu}_{(u)}$ is the 
magnetic field measured by an observer comoving with the fluid and 
is related to $B^{\mu}$ by 
\beq
  B^{\mu}_{(u)} = -\frac{(\delta^{\mu}{}_{\nu} + u^{\mu} u_{\nu}) B^{\nu}} 
  {n_{\lambda}u^{\lambda}} \ .
\eeq 

The energy-momentum tensor is written as
\beqn
T_{\mu\nu}=T_{\mu\nu}^{\rm Fluid} + T_{\mu\nu}^{\rm EM}, 
\eeqn
where $T_{\mu\nu}^{\rm Fluid}$ and $T_{\mu\nu}^{\rm EM}$ denote the
fluid and electromagnetic pieces of the stress-energy 
tensor. They are given by 
\beqn
&&T_{\mu\nu}^{\rm Fluid}=
\rho h u_{\mu} u_{\nu} + P g_{\mu\nu}, \\
&&T_{\mu\nu}^{\rm EM}= \frac{1}{4\pi} \left(
F_{\mu\sigma} F^{~\sigma}_{\nu}-{1 \over 4}g_{\mu\nu}
F_{\alpha\beta} F^{\alpha\beta} \right) \nonumber \\
&&~~~~~~=\biggl({1 \over 2}g_{\mu\nu}+u_{\mu}u_{\nu}\biggr)b^2
-b_{\mu}b_{\nu}, 
\eeqn
where $h\equiv 1+\varepsilon+P/\rho$ is the specific enthalpy, and 
$b^2\equiv b^{\mu}b_{\mu}$. Hence, the total stress-energy tensor becomes
\beq
  T_{\mu\nu}= (\rho h + b^2) u_{\mu} u_{\nu} + \left( P + \frac{b^2}{2}
\right) g_{\mu\nu} - b_{\mu} b_{\nu} \ .
\label{eq:mhdTab}
\eeq
The magnetic pressure is defined as suggested by the second 
term in the above equation:  $P_{\rm mag} \equiv b^2/2$.  

In our numerical implementation of the GRMHD and magnetic 
induction equations, 
we evolve the following conserved variables: 
\beqn
&&\rho_* \equiv - \sqrt{\gamma}\, \rho n_{\mu} u^{\mu}, 
\label{eq:rhos} \\
&& \tilde{S}_i \equiv -  \sqrt{\gamma}\, T_{\mu \nu}n^{\mu} \gamma^{\nu}_{~i}, \\
&& \tilde{\tau} \equiv  \sqrt{\gamma}\, T_{\mu \nu}n^{\mu} n^{\nu} - \rho_*, 
\label{eq:S0} \\
&& \cB^i \equiv  \sqrt{\gamma}\, B^i. 
\eeqn 
The evolution equations are integrated in conservative form 
using a high-resolution shock-capturing (HRSC) scheme. Specifically, we 
use the monotonized central (MC) scheme~\cite{vL77} for data reconstruction 
and the HLL (Harten, Lax and van-Leer) scheme~\cite{HLL} to compute 
the flux. The magnetic field $\cB^i$ must satisfy the no monopole 
constraint $\partial_i \cB^i=0$. Thus, we adopt the flux constrained 
transport (flux-CT) scheme~\cite{t00}.  In this scheme, the induction 
equation is differenced in such a way that a second order, corner-centered 
representation of the divergence is preserved as a numerical 
identity.  As in~\cite{DLSSS2},
we apply outer boundary conditions on the primitive variables $\rho,P,v^i,$
and $B^i$.  Outflow boundary conditions are imposed on the hydrodynamic 
variables (i.e., the variables are copied along the grid directions with the
condition that the velocities be positive or zero in the outer grid zones).  
The magnetic field is linearly interpolated onto the boundaries.  Finally, 
the conserved variables are recomputed on the boundary.   

At each timestep, the
primitive variables $(\rho,P,v^i)$ must be computed from the evolution
variables $(\rho_*,\tilde{\tau},\tilde{S}_i)$. This is done by 
numerically solving the
algebraic equations~(\ref{eq:rhos})--(\ref{eq:S0}) together with an 
equation of state (EOS), $P=P(\rho,\varepsilon)$.
We perform evolutions with two types of EOS.  For the Fishbone-Moncrief
disk~\cite{fishmon} in Section~\ref{disk} as well as one of the HMNS models 
(star~A, see Section~\ref{hmnsinitialdata}), we use the $\Gamma$-law EOS:
\beq
P = (\Gamma-1)\rho\varepsilon \ .
\eeq
The corresponding cold EOS is a simple polytrope, $P_{\rm cold} = K\rho^{\Gamma}$,
where $K$ is a constant.
For the second HMNS model (star~C in Section~\ref{hmnsinitialdata}), we assume 
a more realistic hybrid EOS~\cite{zm97,SS05}, in which the total pressure is written as a 
sum of cold and thermal parts
\beq
P = P_{\rm cold} + P_{\rm th}
\eeq
The cold contribution to the pressure depends only on the 
density, and is defined as follows:
\beq
P_{\rm cold} = \left\{ \begin{array}{ll}
                           K_1\rho^{\Gamma_1} & \mbox{for $\rho \leq \rho_{\rm nuc}$} \\    
                           K_2\rho^{\Gamma_2} & \mbox{for $\rho \geq \rho_{\rm nuc}$} 
			   \end{array} \right. \ .
\label{hybrid1}
\eeq
We set $\Gamma_1=1.3$, $\Gamma_2=2.75$, $K_1=5.16 \times 10^{14}$~cgs,
$K_2=K_1\rho_{\rm nuc}^{\Gamma_1-\Gamma_2}$, and $\rho_{\rm nuc}=1.8
\times 10^{14}~{\rm g/cm^3}$.  This EOS has the 
desirable property that the dependence of pressure on density
stiffens above nuclear density and the resulting maximum 
Tolman-Oppenheimer-Volkov 
mass ($2.01 M_{\odot}$) is in line with predictions of realistic EOSs~\cite{EOS}.  
Shock heating will increase the pressure above its cold 
value at a given density given by Eq.~(\ref{hybrid1}).  This 
is reflected by the thermal contribution to the pressure:
\beq
P_{\rm th} = (\Gamma_{\rm th} - 1)\rho\varepsilon_{\rm th} , 
\label{hybrid2}
\eeq
where $\varepsilon_{\rm th} = \varepsilon - \varepsilon_{\rm cold}$, and
$\varepsilon_{\rm cold}$ is the specific internal energy consistent
with the cold pressure:
\beq
\varepsilon_{\rm cold} (\rho) = 
-\int P_{\rm cold}(\rho)\, d \left( \frac{1}{\rho}\right) \ .
\eeq
In this paper, we take $\Gamma_{\rm th} = 1.3$.

In low-density and/or highly magnetically dominated regions, we 
find that the inversion procedure for obtaining the primitive 
variables $(\rho,P,v^i)$ from Eqs.~(\ref{eq:rhos})--(\ref{eq:S0}) 
sometimes returns an unphysical solution (i.e., a solution with negative 
pressure).  In such grid zones, we apply the fix suggested by Font 
{\it et al.}~\cite{fmst00}, which consists of replacing the energy 
equation~(\ref{eq:S0}) by the cold EOS, $P = P_{\rm cold}(\rho)$ when solving 
the system of equations.
This substitution guarantees a positive pressure.  
In rare cases, this revised system also fails to give a solution 
and we repair the 
zone by averaging from nearby zones.  (Averaging is not applied
to the magnetic field, since this would introduce monopoles.)
For a typical run with $500^2$ resolution, we find that 
$<10$ zones require this secondary fix on a given timestep.

The code used here has been tested in multiple relativistic MHD 
simulations, including MHD shocks, nonlinear MHD wave propagation, 
magnetized Bondi accretion, and MHD waves induced by linear 
gravitational waves~\cite{DLSS}.  We have also compared this code with 
the GRMHD code of Shibata and Sekiguchi~\cite{SS05} by performing 
simulations of the evolution of magnetized HMNSs~\cite{DLSSS,DLSSS2}, 
and of magnetorotational collapse of stellar cores~\cite{SLSS}. We 
obtain good agreement between these two independent codes.

\subsection{Diagnostics}

We keep track of the rest mass $M_0$ and angular momentum $J$ on our grid
by computing the following volume integrals: 
\beqn
 M_0 &=& \int_V \rho_* d^3x \ , \label{eq:m0} \\
 J &=& \int_V \tilde{S}_{\varphi} d^3x \ . \label{eq:j_vol}
\eeqn
Note that this formula for $J$ is only valid in an axisymmetric 
spacetime~\cite{Wald}.  The total rest mass $M_0$ is 
conserved (baryon number conservation), and angular momentum $J$ is 
conserved in axisymmetry since gravitational radiation carries away 
no angular momentum.  Our finite differencing scheme is designed
to conserve $M_0$ and $J$ as a numerical identity.  This is possible 
since the continuity equation $\nabla_{\mu}(\rho u^{\mu})=0$ 
and momentum equation $\nabla_{\mu} T^{\mu}{}_i=0$ 
can be written as 
\beqn
 \partial_t \rho_* + \partial_j (\rho_* v^j) & = & 0  \ , \label{eq:conteq} \\
 \partial_t \tilde{S}_i + \partial_j (\alpha \sqrt{\gamma} T^j{}_i) &=& 
 \frac{1}{2} \alpha \sqrt{\gamma} T^{\mu \nu} g_{\mu \nu,i} \ .
\eeqn
Taking the $\varphi$-component of the second equation, we obtain the equation 
for $\tilde{S}_{\varphi}$ in axisymmetry:
\beq
 \partial_t \tilde{S}_{\varphi} + 
\partial_j (\alpha \sqrt{\gamma} T^j{}_{\varphi}) = 0 \ . \label{eq:Sphieq}
\eeq
Note that both Eqs.~(\ref{eq:conteq}) and (\ref{eq:Sphieq}) are written 
in conservative form with no source terms. This allows us to design a 
finite differencing scheme to conserve $M_0$ and $J$ as a numerical 
identity.

In practice, the conservation of $M_0$ and $J$ will not be exact for 
three reasons.  Most significantly, outflows 
from the computational grid remove both rest mass and angular momentum.   
Secondly, as described in Section~\ref{sec:basic_eqs}, the inversion from 
conserved to primitive variables requires on rare occasion a fix in which primitive 
variables are averaged from nearby cells.  This averaging can affect the 
total rest mass and angular momentum.  (In contrast, the fix suggested by 
Font {\it et al.}, which is much more commonly employed in our code, does 
not affect the rest mass and angular momentum since it does not change
$\rho_*$ and $\tilde{S}_{\varphi}$.)  The third factor preventing strict conservation
is the imposition of floor values for the rest mass density (see below).
In particular, applying a density floor increases the total rest mass.
We also find that the floor tends to increase the angular momentum 
as well.  The rate of increase for these quantities can be judged from the
early part of the simulation (before any outflow from the computational
grid).  Based on these rates, we find that the fractional increases
in $M_0$ and $J$ due to the floor is at most $\sim {\rm few} \times 10^{-4}$
for the entire duration of the runs.  

Soon after an apparent horizon forms, we excise the BH interior
to continue the evolution~\cite{excision_hydro,DLSSS}.
During the post-excision evolution, we compute the rest mass 
$M_{\rm out}$ and angular momentum $J_{\rm out}$ of the 
material outside the BH by computing integrals~(\ref{eq:m0}) 
and (\ref{eq:j_vol}) over the volume outside the apparent horizon. This
material includes the disk and corona, as well as any outbound material
which may be on the grid.  The irreducible mass 
$M_{\rm irr}$ of the BH is given by $M_{\rm irr} = \sqrt{A/16\pi}$, 
where $A$ is the surface area of the apparent horizon. Since $J$ is 
conserved, we can compute the BH's angular momentum $J_h$ by 
\beq
  J_h = J - J_{\rm out} .
\label{jhole}
\eeq
This would be exact only if the total angular momentum were strictly
conserved.  However, as mentioned above, strict conservation
is broken by our atmosphere treatment, by our treatment of zones
in which the primitive variable inversion fails, and because of
outflows from the grid.  We calculate that the angular momentum 
loss through the outer boundary is at most a few percent. Nevertheless, 
we assume $J$ to be perfectly conserved after excision and 
hence use the value of $J$ just before excision when 
computing $J_h$.  The BH's mass $M_h$ is then computed from the formula 
\beq
  M_h = \sqrt{M_{\rm irr}^2+(J_h/2M_{\rm irr})^2} \ ,
\label{mhole}
\eeq
which is exact for a Kerr spacetime, and is in accord with the 
formula derived using the isolated and dynamical 
horizon formalism~\cite{iso-dyn-hor}. 

Shortly after BH excision 
(hundreds of $M$), the spacetime settles down to an approximately 
stationary state, and it is possible to define an (approximately) 
conserved energy:
\beq
E =-\int \alpha \sqrt{\gamma}\, T^t_{~t} d^3x \ .
\eeq
We can then define the fluxes of rest mass, energy, and angular momentum 
across any closed two-dimensional surface $S$ in a time slice:
\beqn
&&\dot{M}_0 = \oint_S \alpha \rho v^i d^2 \Sigma_i \ ,\\
&&\dot{E}   =-\oint_S \alpha T^i_{~t} d^2 \Sigma_i \ ,\\
&&\dot{J}   = \oint_S \alpha T^i_{~\varphi} d^2 \Sigma_i \ , 
\eeqn
where 
\beq
  d^2 \Sigma_i = \frac{1}{2} \epsilon_{ijk} dx^j \wedge dx^k \ , 
\eeq
and $\epsilon_{ijk} = n_{\mu} \epsilon^{\mu}{}_{ijk}$ is the Levi-Civita 
tensor associated with the three-metric $\gamma_{ij}$. We use the above
formulae for calculating fluxes through the apparent horizon when the 
apparent horizon has a general shape.  
However, the disk simulations in Section~\ref{disk} are performed 
with a fixed Kerr background metric for which the horizon surface
is spherical in our adopted coordinates.      
In such cases, the surface $S$ is a sphere with radius $r$, and the
flux expressions reduce to
\beqn
&&\dot{M}_0(r)= \oint_{r={\rm const}} dA \rho_* v^r r^2 \label{mdotsph} \\
&&\dot{E}(r)=-\oint_{r={\rm const}} dA \alpha \sqrt{\gamma}\, T^r_{~t}, \label{edotsph} \\
&&\dot{J}(r)= \oint_{r={\rm const}} dA \alpha \sqrt{\gamma}\, T^r_{~\varphi}, \label{jdotsph}
\eeqn
where $dA=r^2 \sin\theta d\theta d\phi$. 

To determine whether a fluid particle is unbound, we 
compute $u_t$. In a stationary spacetime, the value of $u_t$ for a particle 
moving on a geodesic is conserved. If the particle is unbound, the radial 
velocity $v^r>0$ and $-u_t = 1/\sqrt{1-v^2} > 1$ at infinity. Hence $v^r$ 
and $u_t$ provide an approximate criterion to determine whether a fluid 
element is unbound, provided that the fluid motion is predominantly 
ballistic and pressure and electromagnetic forces can be neglected.
(We note that, in general, this condition is necessary
but not sufficient even in the absence of external forces~\cite{ansorg}.)

\section{Magnetized Disk Evolutions}
\label{disk}

\begin{figure*}
\begin{center}
\epsfxsize=2.2in
\leavevmode
\hspace{-0.7cm}\epsffile{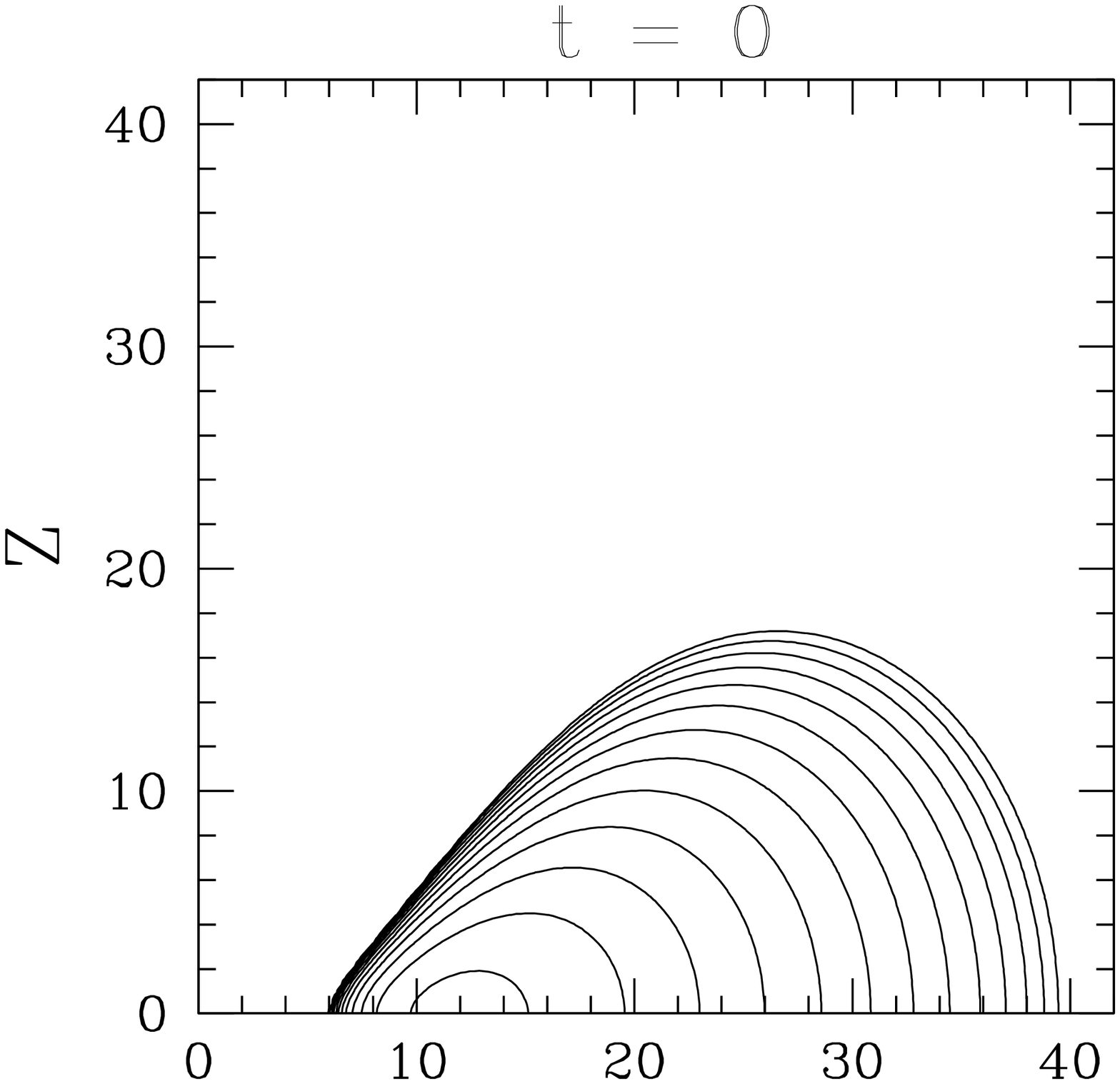}
\epsfxsize=2.2in
\leavevmode
\hspace{-0.5cm}\epsffile{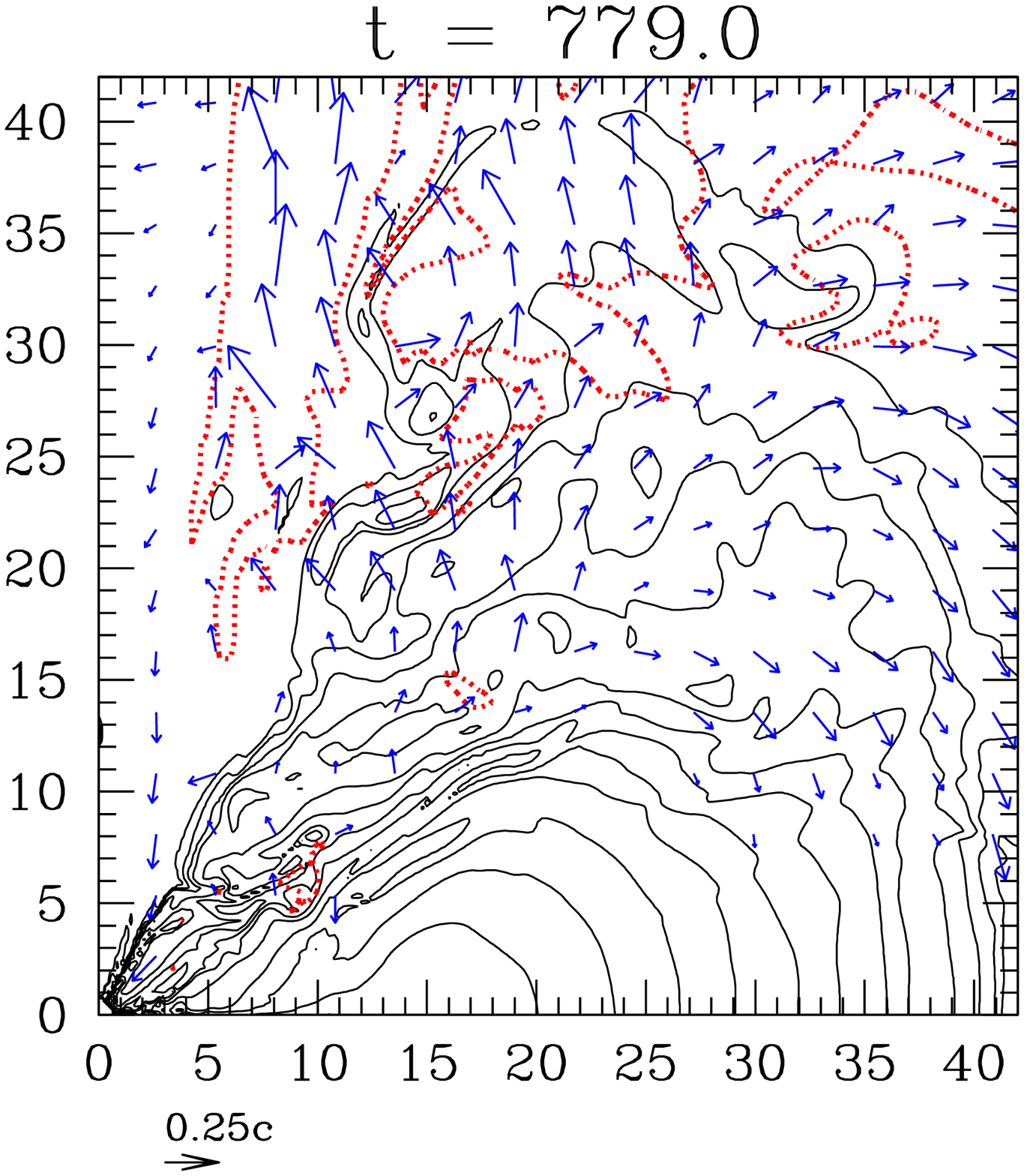}
\epsfxsize=2.2in
\leavevmode
\hspace{-0.5cm}\epsffile{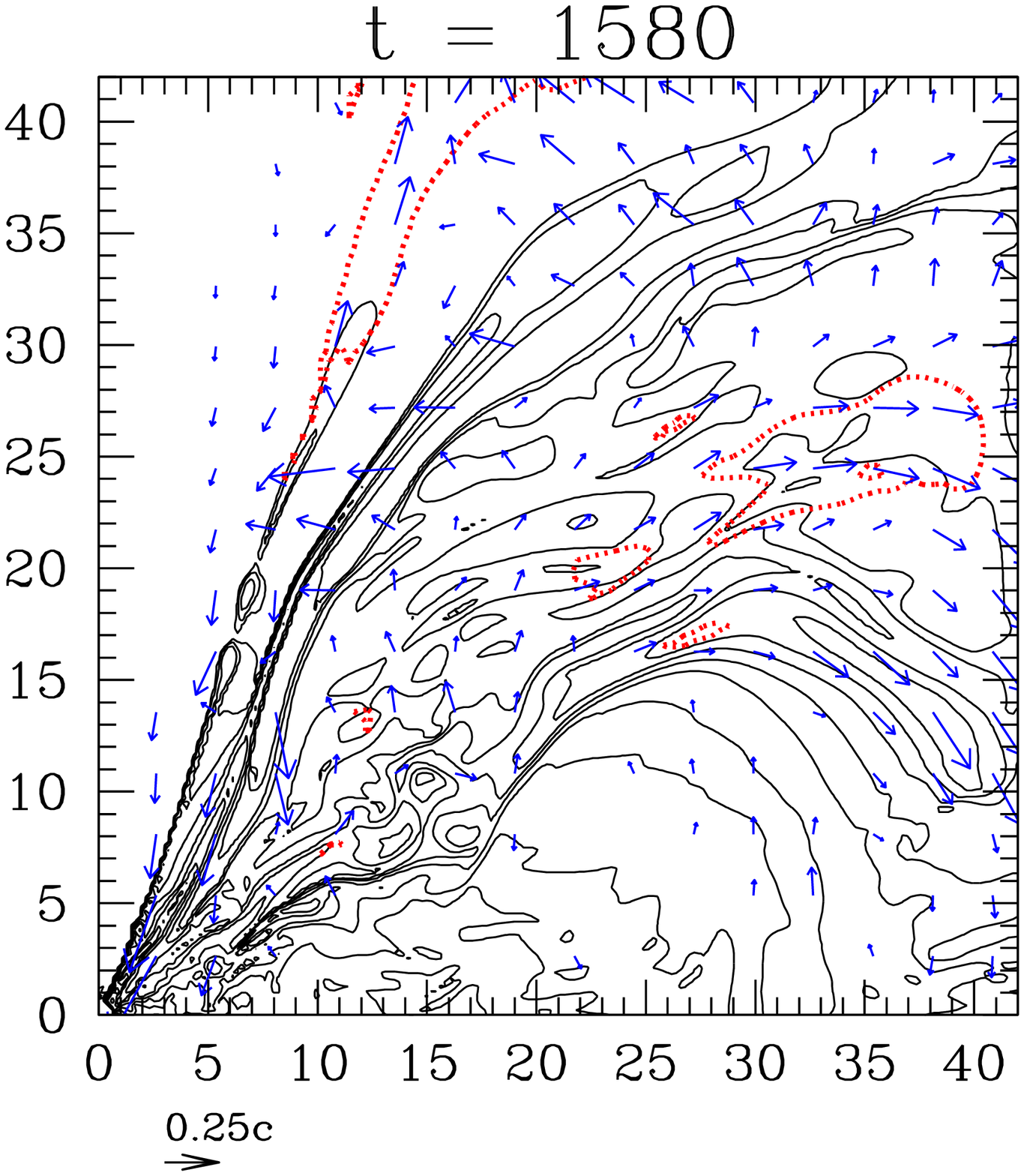} \\
\vspace{-0.5cm}
\epsfxsize=2.2in
\leavevmode
\hspace{-0.7cm}\epsffile{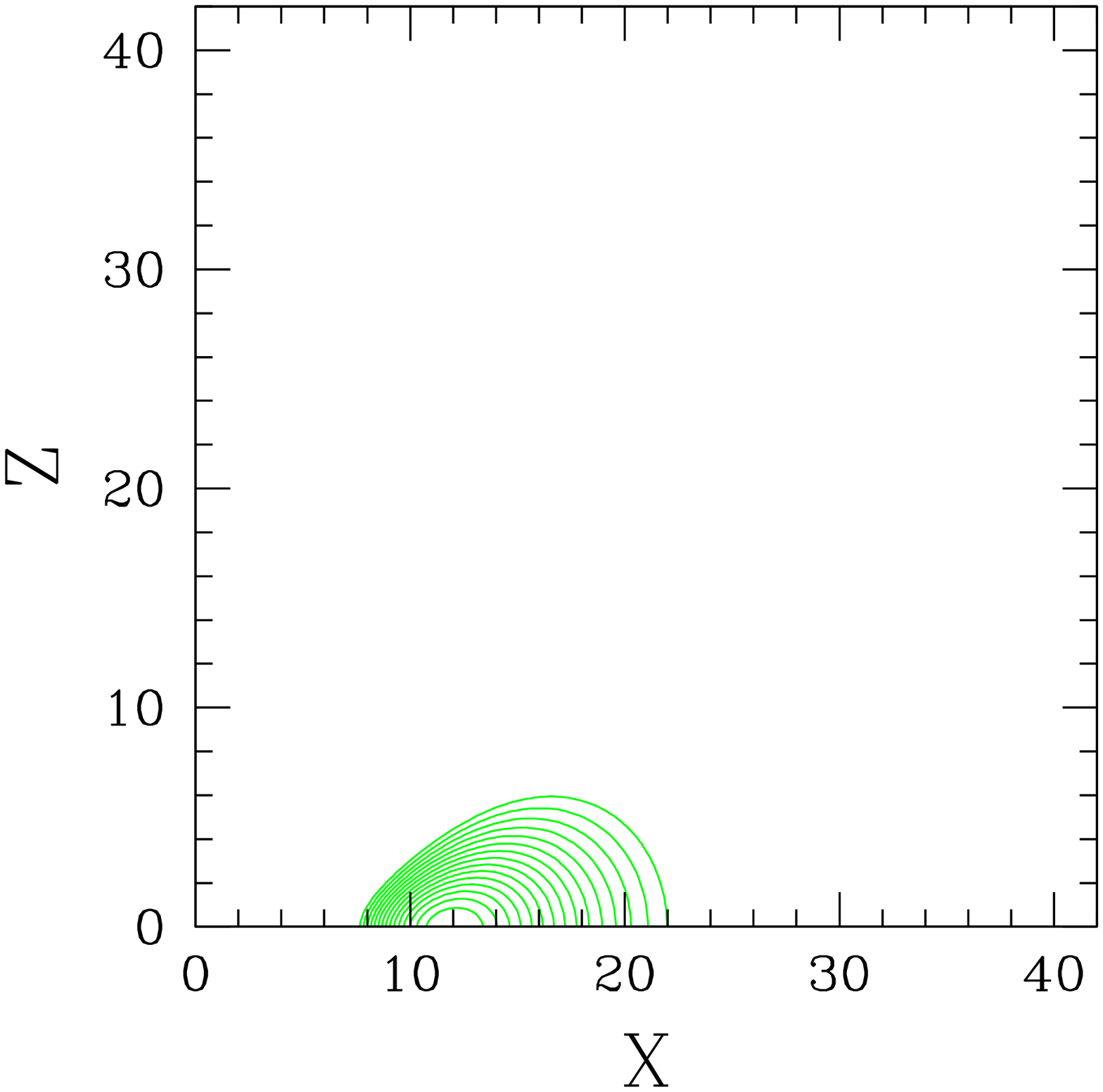}
\epsfxsize=2.2in
\leavevmode
\hspace{-0.5cm}\epsffile{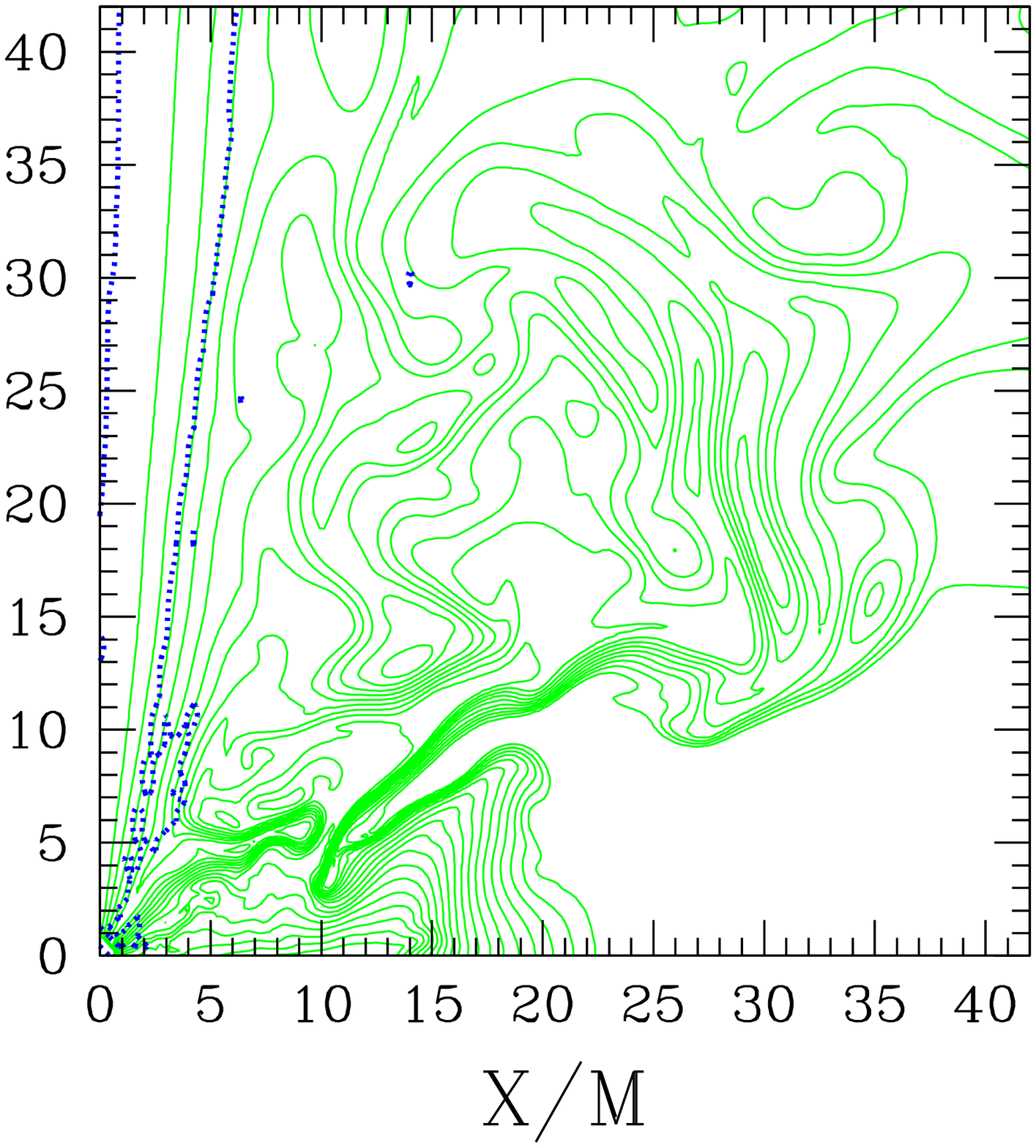}
\epsfxsize=2.2in
\leavevmode
\hspace{-0.5cm}\epsffile{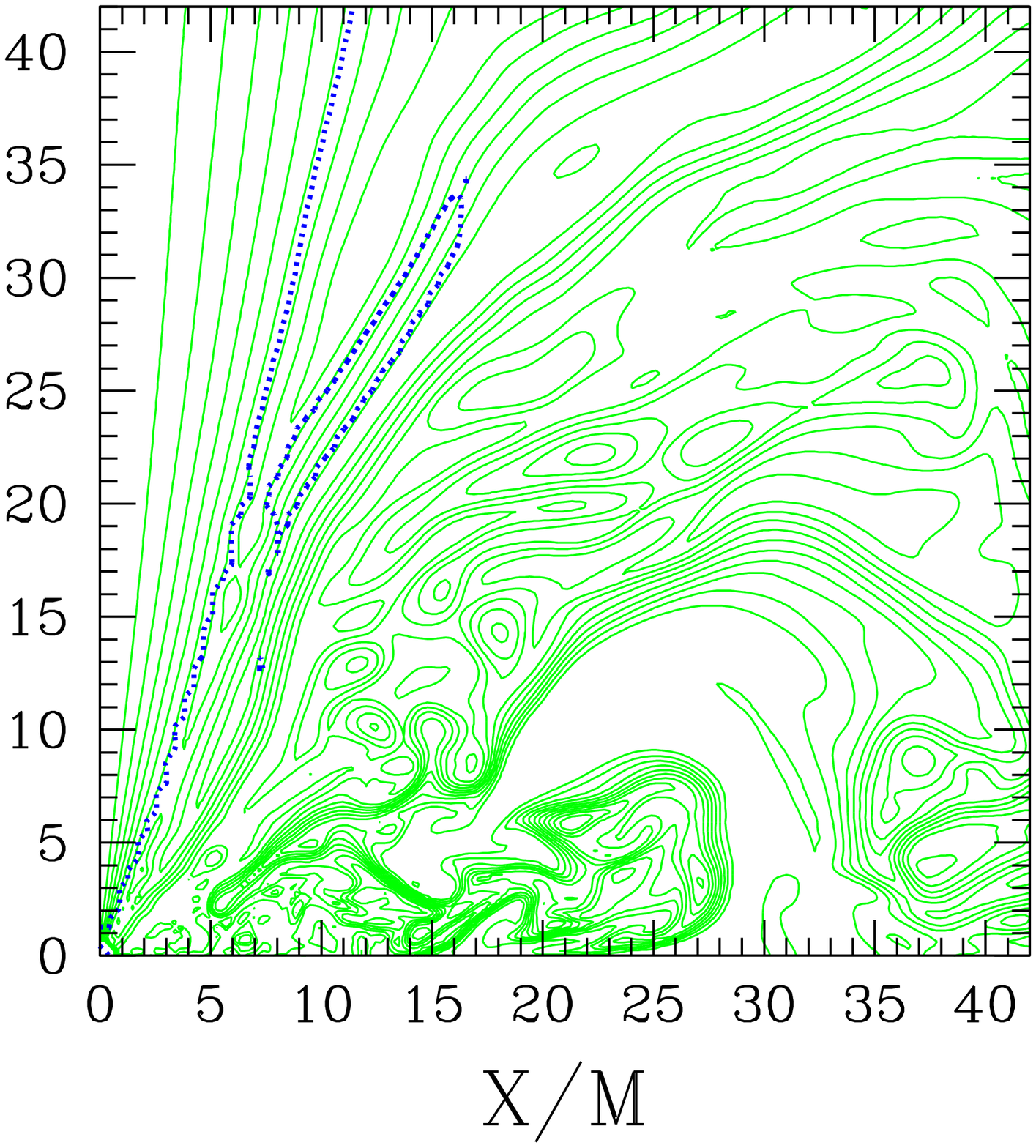}
\caption{Panels in the first row show density contours (solid black lines) 
and velocity vectors at selected times during the evolution of the magnetized
FM torus.  The dotted (red) contour lines enclose regions with 
unbound outflows (having $-u_t > 1$ and positive radial velocity).  The second 
row shows the poloidal magnetic field structure at the corresponding times.  The
dotted blue contours correspond to $\beta = P/P_{\rm mag} = 0.01$ and
thus enclose magnetically dominated regions (such as the funnel region 
near the polar axis in the last two snapshots).  
In this and all subsequent density contour plots, levels are defined by 
$\rho/\rho_{\rm max}= 10^{-0.3 i - 0.09}~(i=0$--12).  
Likewise, for all plots of magnetic field lines, the field
lines are drawn as contours of $A_{\varphi}$ according to $A_{\varphi} = A_{\varphi,\rm min}
+ (A_{\varphi,\rm max} - A_{\varphi,\rm min}) i/20~(i=1$--19),
where $A_{\varphi,\rm max}$ and $A_{\varphi,\rm min}$ are the maximum
and minimum values of $A_{\varphi}$.
\label{disk_contours}}
\end{center}
\end{figure*}

In order to verify the ability of our code to handle magnetized accretion
flows in stationary spacetimes, we perform runs to compare with published 
results of McKinney and 
Gammie~\cite{mg04} (see also~\cite{harm}) for the evolution of a magnetized 
Fishbone-Moncrief (FM) torus~\cite{fishmon}. In particular, we consider the case 
referred to in~\cite{mg04} as the fiducial run.  The spacetime 
metric corresponds to a fixed Kerr BH with $M=1$ in Kerr-Schild coordinates~\cite{mg04}:
\beqn
ds^2 & = & -\left(1-\frac{2r}{\Sigma}\right)dt^2 + \left(\frac{4r}{\Sigma}\right)drdt 
+ \left(1+\frac{2r}{\Sigma}\right)dr^2  \nonumber \\
& & + \Sigma d\theta^2 + 
 \sin^2\theta\left[\Sigma + a^2\left(1+\frac{2r}{\Sigma}\right)\sin^2\theta \right]d\phi^2 
\nonumber \\
& & -\left(\frac{4ar^2 \sin^2\theta}{\Sigma}\right)dtd\phi  \nonumber \\ 
& & - 2a\left(1+\frac{2r}{\Sigma}\right)\sin^2\theta drd\phi \ ,
\eeqn
where $\Sigma = r^2 + a^2 \cos^2\theta$ and the spin parameter is chosen as 
$a = 0.938 $.  (In this section, units with $M=1$ will be assumed.)  
The BH is surrounded by a FM torus specified by 
$u^t u_{\phi} = 4.281$ and inner disk radius $r_{\rm in} = 6$.  The torus has an outer 
radius of $42$ in the equatorial plane and the pressure maximum is located at 
$r = 12$.  The FM solution provides the specific enthalpy 
distribution, from which the rest-mass density and pressure are derived 
assuming a cold polytropic EOS, $P = K\rho^\Gamma$, with $\Gamma = 4/3$.  
The constant $K$ is chosen so that the maximum rest-mass density at $t=0$ is unity.  
The torus is evolved assuming the equation of state 
$P = (\Gamma-1)\rho\varepsilon$ (again with $\Gamma = 4/3$), in order to account for 
entropy generation in shocks.

\begin{figure}
\vspace{-4mm}
\begin{center}
\epsfxsize=3.in
\leavevmode
\epsffile{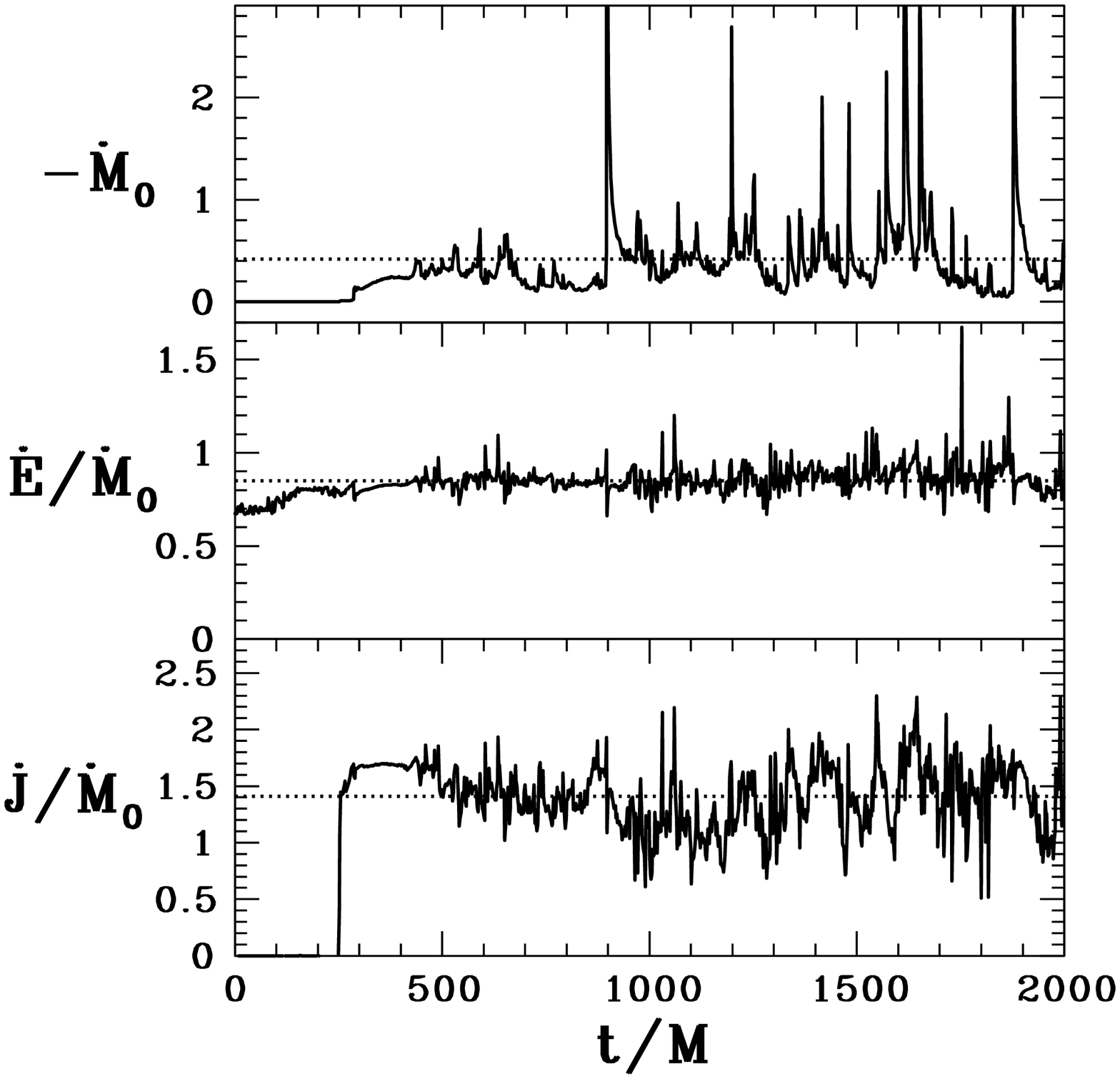}
\caption{Upper panel shows the rest-mass accretion rate (solid line), along with its
time-averaged value 0.42 (dotted line).  (Here, time averages are performed 
over the period $500  < t < 2000 $.)  The middle (lower) panel shows the ratio of 
the accretion rate of total energy (angular momentum) to the rest mass 
accretion rate.  The time averaged values (dotted lines) are $\dot{E}/\dot{M}_0 = 0.85$ 
and $\dot{J}/\dot{M}_0 = 1.41$.  (Here, we assume units with BH mass $M=1$.) 
\label{disk_fluxes}}
\end{center}
\end{figure}

\begin{table*}
\caption{Resolution Study}
\begin{center}
\begin{tabular}{c c c c c}
\hline 
\hline
\vspace{0.03in}
Resolution & $-\dot{M}_0$${}^{\rm a}$ & $\dot{E}/\dot{M}_0$ & $\dot{J}/\dot{M}_0$ 
 & $\lambda_{\rm max}/\Delta$${}^{\rm b}$ \\
\hline\hline
$200^2$ & 0.29  & 0.87  & 1.66  & 6 \\
$320^2$ & 0.36  & 0.82  & 1.45  & 10 \\
$400^2$ & 0.37  & 0.82  & 1.35  & 12  \\
\hline \hline
\end{tabular}
\end{center}
\vskip 12pt
\begin{minipage}{6cm}
\raggedright
${}^{\rm a}$ {The averaging period for the values of $-\dot{M}_0$, 
$\dot{E}/\dot{M}_0$, and $\dot{J}/\dot{M}_0$ in this table is 
$500  < t < 1000.$ (Here, we assume units with BH mass $M=1$.)} \\
${}^{\rm b}$ {The ratio of the typical MRI wavelength at $t=0$ 
to the grid resolution in the neighborhood of the (gas) pressure 
maximum.} \\
\end{minipage}
\label{rescomp}
\end{table*}

In the absence of magnetic fields and viscosity, the torus is in equilibrium.  
However, following~\cite{mg04}, we add a small poloidal magnetic field by 
specifying the azimuthal component of the magnetic vector potential:
\beq
A_{\varphi} \propto \max[(\rho-\rho_{\rm cut}),0] \ ,    \label{aphi_disk}
\eeq
where the cutoff density $\rho_{\rm cut}$ is chosen as $0.2\rho_{\rm max} = 0.2$.  
This form of $A_{\varphi}$ results in magnetic field loops confined within the 
torus.  The proportionality constant determines the strength of the magnetic field 
and is chosen so that $\max(P_{\rm mag})/\max(P) = 0.01$.  Thus, the 
dynamical equilibrium of the torus is only slightly perturbed by the addition 
of the magnetic field.

We evolve these initial data assuming axial and equatorial symmetry and using 
cylindrical coordinates $(\varpi,z)$.  In addition, we introduce a logarithmic 
radial coordinate transformation in order to concentrate zones near the BH 
event horizon.  In particular, we take 
\beq
r = e^{(\bar{r} - r_0)}, \ \ \ \ r_0 = 0.2 \ , 
\eeq
and then $\bar{\varpi} = (\bar{r}/r)\varpi$ and $\bar{z} = (\bar{r}/r)z$.  
We perform three runs having $200^2$, $320^2$, and $400^2$ zones, with
uniform resolution in $(\bar{\varpi},\bar{z})$ and with the outer boundaries 
held fixed at $\bar{\varpi}_{\rm max} = \bar{z}_{\rm max} = 4.0$.  
The resulting grid is non-uniform in $\varpi$ and $z$, and the outer boundary
in the $(\varpi,z)$-plane is not square.  At its point of closest
approach to the origin  in the $(\varpi,z)$-plane, the outer boundary is 
located at a radius $r = 44.7$. In contrast to our distorted rectangular 
grid, McKinney and Gammie~\cite{mg04} employ spherical polar coordinates with a 
logarithmic radius and $\theta$-zones concentrated toward the disk 
plane. As in many hydrodynamic simulations in astrophysics, we add a tenuous 
``atmosphere'' that covers the computational grid outside the star. 
Following~\cite{mg04}, this is done by imposing floors on both the rest-mass 
density and pressure as follows: 
$\rho_{\rm atm} = 10^{-4} r^{-3/2}$ and $P_{\rm atm} = 3.3 \times 10^{-7} r^{-5/2}$.

We first discuss the general features of the evolution.  Figure~\ref{disk_contours} 
shows snapshots of isodensity contours and velocity vectors 
(in the top row), and of the poloidal magnetic field lines (in the 
bottom row, drawn as contours of constant $A_{\varphi}$).  The snapshots 
at $t=0$ show the initial density distribution of the torus and the loops of 
magnetic field confined within it.  The MRI and magnetic braking cause 
angular momentum transport in the disk, leading to accretion onto the central 
BH and ejection of some material.  Magnetic field lines 
carried into this central region are stretched upward by outgoing material, leading 
to the collimation of field lines in the region of the spin axis.  This collimated 
structure (first seen in the snapshots at $t = 779 $ in Fig.~\ref{disk_contours}) 
persists through the rest of the simulation.  The panels 
in Fig.~\ref{disk_contours} showing the magnetic field lines also have dotted 
contours corresponding to $\beta = P/P_{\rm mag} = 10^{-2}$.  These 
contours enclose the collimated magnetic field lines, showing that this funnel 
region is strongly magnetically dominated.

The MRI-driven turbulence causes the magnetic field remaining in the disk to become 
highly tangled.  The violent motions of the disk 
lead to the ejection of material, especially near the outer edge 
of the collimated magnetic field region (i.e., the funnel wall, or funnel-corona 
interface as discussed in~\cite{mg04}).  In the top row of panels, the 
dotted contours surround regions which have $-u_t > 1$ and positive radial velocity, 
roughly corresponding to 
regions of outflow.  Such regions can be seen just outside the magnetically dominated 
funnel in the last two sets of snapshots in Fig.~\ref{disk_contours}.

Figure~\ref{disk_fluxes} shows the mass accretion rate $\dot{M}_0$ as a function 
of time for the $400^2$ case, along with the ratios $\dot{E}/\dot{M}_0$, and 
$\dot{J}/\dot{M}_0$ (where $\dot{E}$ and $\dot{J}$ are the total energy and angular momentum
fluxes through the horizon).  These quantities are calculated
as integrals over the horizon surface using Eqs.~(\ref{mdotsph})-(\ref{jdotsph}).  
(We exclude from these integrals those zones which may have 
been affected by a failed primitive variable inversion.  These failures 
tend to occur in the small region near the polar axis which is highly magnetically
dominated.  In particular, by examining these failures at several times, we 
find that $>99\%$ of the failures occur in regions where 
$\beta = P/P_{\rm mag} < 10^{-4}$, while $>70\%$ of the failures occur for 
$\beta <  10^{-5}$.)  The time averaged values (for the period $500 < t < 2000$) are 
$-\dot{M}_0 = 0.42$, $\dot{E}/\dot{M}_0 = 0.85$,
and $\dot{J}/\dot{M}_0 = 1.41$ and are shown as dotted lines in 
Fig.~\ref{disk_fluxes}.  These are in good agreement with the values given
by McKinney and Gammie~\cite{mg04} for the same averaging period:   
$-\dot{M}_0 = 0.35$, $\dot{E}/\dot{M}_0 = 0.87$, and $\dot{J}/\dot{M}_0 = 1.46$. 
Finally, we show in Table~\ref{rescomp} that these characteristics of the 
accretion flow are consistent for all three resolutions considered.  
In addition, the table gives the ratio of the typical unstable wavelength of the
MRI to the spatial resolution.
As an estimate of the typical MRI wavelength at $t=0$, we take a typical value of 
\beq
\lambda_{\rm max} \simeq \frac{8\pi v_A^z}{\sqrt{15}\Omega}   
\eeq
in the equatorial plane, where ${\bf v}_A = {\bf B}/\sqrt{4\pi\rho}$ is 
the (Newtonian) Alfv\'en velocity 
and $\Omega = v^{\varphi}$ is the angular velocity~\cite{MRIrev,SLSS}.  
(This expression is exact for Newtonian flows with a Keplerian angular velocity profile.) 
Our highest resolution ($400^2$) thus gives $\sim 12$ points across the MRI wavelength.  
We expect that the MRI is fairly well resolved for the two higher resolution 
runs ($320^2$ and $400^2$) and perhaps marginally resolved in the $200^2$ run.
The rough quantitative agreement between our results and those
of~\cite{mg04} is quite reasonable given the very different grid structures 
employed by the two codes and gives us confidence in the ability of our code
to handle complex accretion flows.

\section{Results for HMNS evolution}
\label{sec:results}
We now turn to magnetized HMNS collapse and the remnant accretion
disk evolution.  We describe the initial data models and the three 
evolution cases in Section~\ref{hmnsinitialdata}.  We then describe 
the results and implications for jet formation from HMNS collapse.  

\subsection{Initial Data}
\label{hmnsinitialdata}

\begin{figure*}
\begin{center}
\epsfxsize=2.2in
\leavevmode
\hspace{-0.7cm}\epsffile{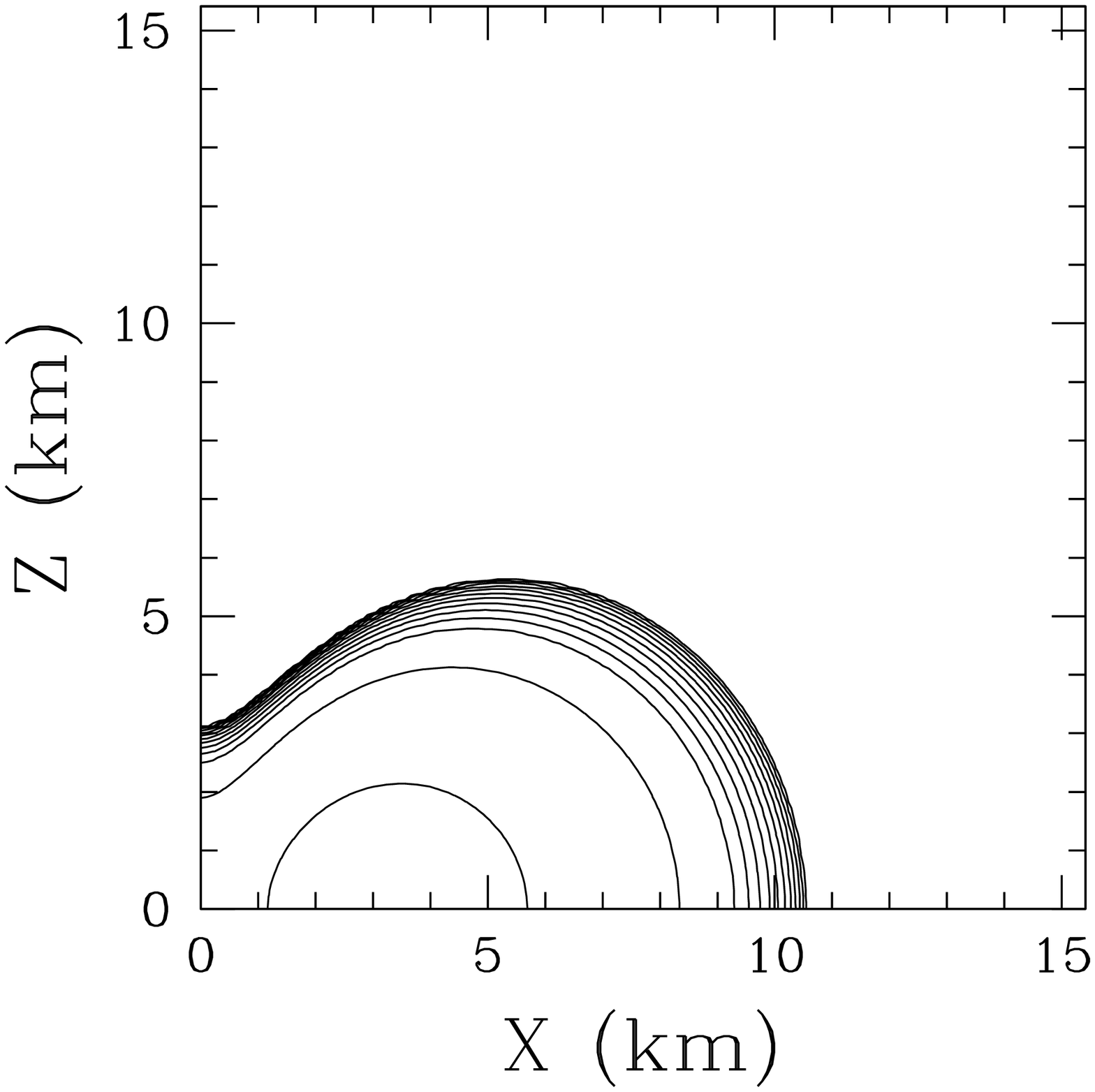}
\epsfxsize=2.2in
\leavevmode
\hspace{-0.5cm}\epsffile{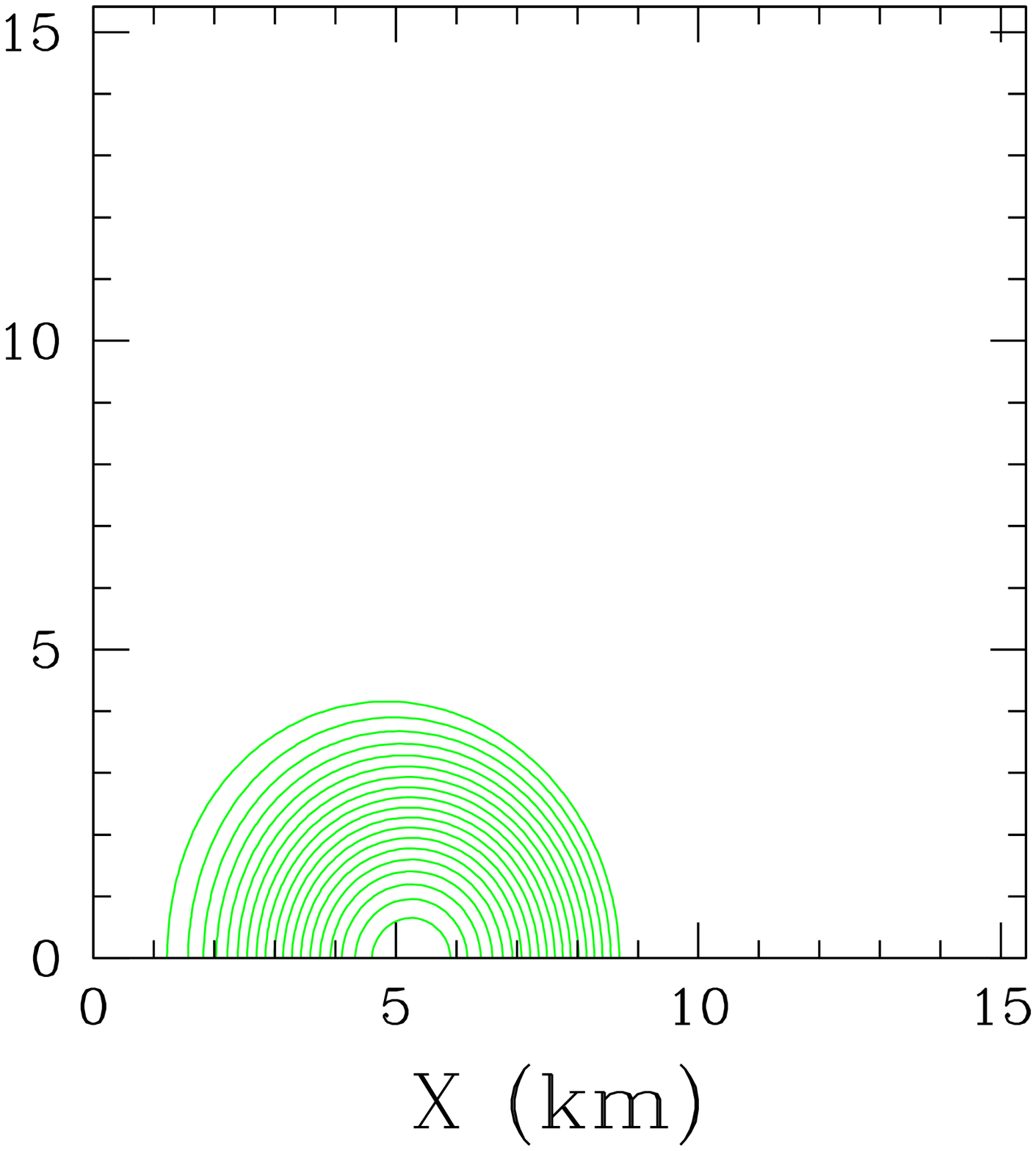}
\epsfxsize=2.2in
\leavevmode
\hspace{-0.5cm}\epsffile{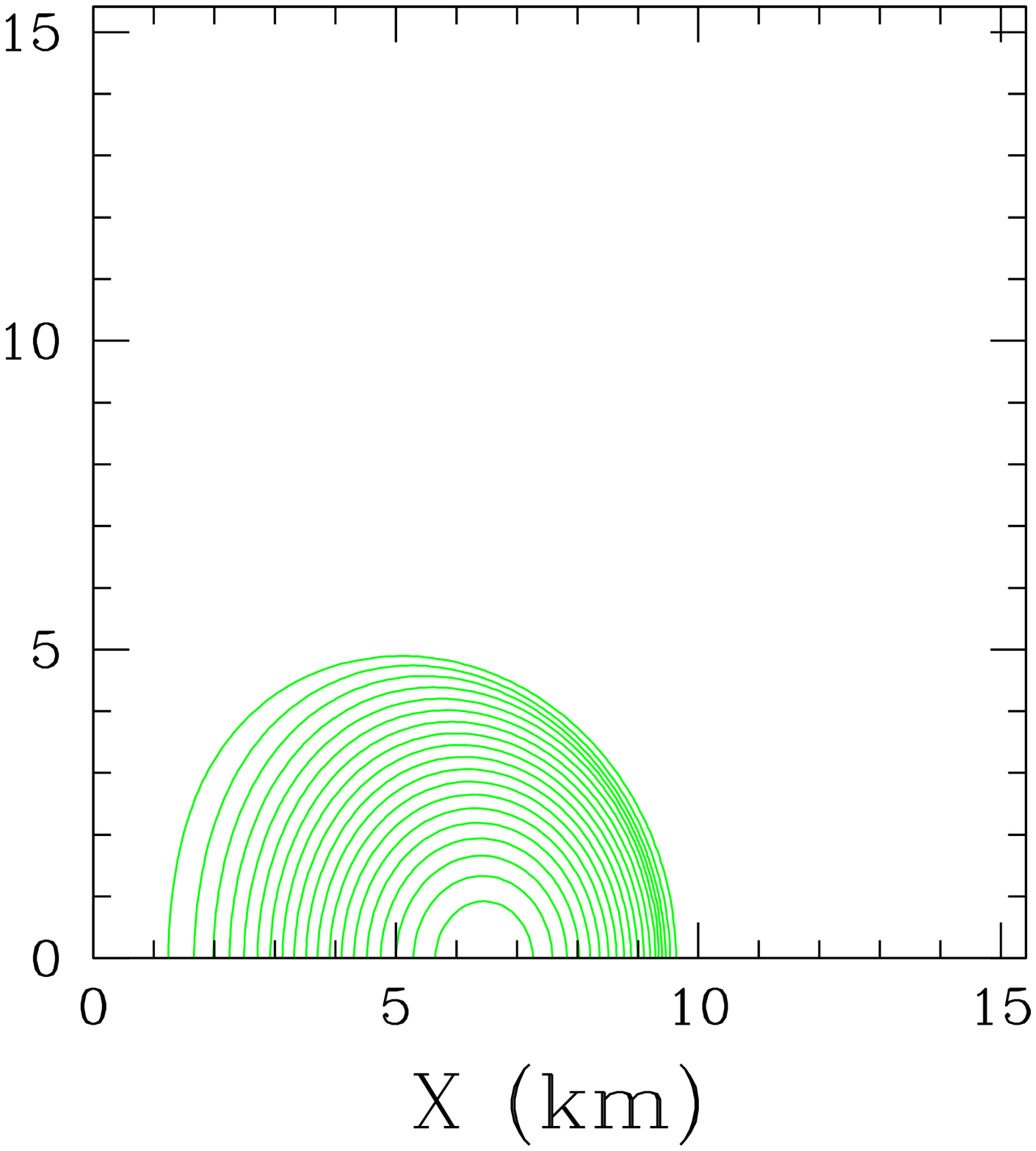} 
\caption{The left panel shows density contour lines for star~C at $t=0$.  
The middle and right panels show initial magnetic field lines for cases C1 
and C2, respectively.  Contour lines have the same meaning as in 
Fig.~\ref{disk_contours}.
\label{starC_init}}
\end{center}
\end{figure*}

\begin{table*}
\caption{Initial Models}
\begin{center}
\begin{tabular}{c c c c c c c c }
\hline 
\hline
\vspace{0.03in}
Model &  \small{$M(M_{\odot})$} & 
\small{$M_0(M_{\odot})$}  & $R_{\rm eq}/M$${}^{\rm a}$ & $J/M^2$ 
& $T_{\rm rot}/|W|$${}^{\rm b}$ & $\Omega_{\rm eq}/\Omega_c$${}^{\rm c}$ &
$P_c/M{}^{\rm d} $ \\
\hline\hline
  A  & 2.71 & 2.95 & 4.48  &  1.0  & 0.249 & 0.33  & 38.4 \\
\hline
  C  & 2.64 & 2.95 & 2.75  &  0.82 & 0.241 & 0.185 & 15.5 \\
\hline \hline
\end{tabular}
\end{center}
\vskip 12pt
\begin{minipage}{12cm}
\raggedright
${}^{\rm a}$ {The equatorial coordinate radius $R_{\rm eq}$ normalized by 
the Arnowitt-Deser-Misner (ADM) mass.}
\\
${}^{\rm b}$ {The ratio of the rotational kinetic energy to the gravitational 
binding energy.}
\\
${}^{\rm c}$ {The ratio of the angular velocity at the equator to the 
central angular velocity.}
\\
${}^{\rm d}$ {The initial central rotation period $P_c$.}
\end{minipage}
\label{startable}
\end{table*}

The evolution cases described below employ the two HMNS models 
referred to in~\cite{DLSSS2} as stars~A and C. (For the sake of 
consistency, those labels will be retained in the present paper.)  

Since star~A is constructed using a $\Gamma=2$ polytropic EOS, 
$P=K\rho^{\Gamma}$, this model may be scaled to any desired 
physical mass by adjusting the value of $K$~\cite{CST}.  
In contrast, star~C is constructed with the more realistic cold hybrid
EOS given in Eq.~(\ref{hybrid1}).  This hybrid EOS introduces a 
physical density scale, and star~C thus does not have the scale freedom 
enjoyed by star~A.
For definiteness and ease of comparison between the evolution cases, we 
fix the value of the polytropic constant for star~A in the discussion 
below to $K= 1.37 \times 10^{5} {\rm g}^{-1} {\rm cm}^5 {\rm s}^{-2}$.  
This value was chosen so that stars~A and C have the same rest mass 
($M_0 = 2.95 M_{\odot}$).  With this choice, Table~\ref{startable} lists 
some properties of these stars.  We note that the rest masses of stars~A 
and C exceed the 
supramassive limits (i.e., the mass limits for rigid rotation) by 46\%  
and 14\%, respectively.  These stars rotate very rapidly and are highly 
flattened due to centrifugal force.  We also note that the
parameters of star~C are chosen in order to mimic the HMNSs formed through 
binary NS mergers with realistic EOSs in~\cite{STUb}.

As in previous papers (e.g, \cite{CST,BSS,SBS,DLSS,DLSSS2}), we choose the 
initial rotation law $u^t u_{\varphi}=A^2(\Omega_c-\Omega)$, where $u^{\mu}$ 
is the four-velocity, $\Omega_c$ is the angular velocity along the rotational 
axis, and $\Omega \equiv u^{\varphi}/u^t$ is the angular velocity. In the 
Newtonian limit, this differential rotation law becomes
\beq
  \Omega = \frac{\Omega_c}{1 + {\frac{\varpi^2}{A^2}}}\ .
\eeq
The constant $A$ has units of length and determines the steepness of the
differential rotation. In this paper, $A$ is set equal to the
coordinate equatorial radius $R_{\rm eq}$ for star~A and to
$0.8R_{\rm eq}$ for star~C. The corresponding values of 
$\Omega_{\rm eq}/\Omega_c$ are shown in Table~\ref{startable} 
(where $\Omega_{\rm eq}$ is the angular velocity at the equatorial 
surface).  

We must also specify initial conditions for the magnetic field.  We choose
to add a weak poloidal magnetic field to the equilibrium model 
by introducing a vector potential taking one of the following forms: 
\beq
A_{\varphi} = A_b \varpi^2 {\rm max}(P-P_{\rm cut}, 0) \ ,
\label{Aphi1}
\eeq
or 
\beq
A_{\varphi} = A_b \varpi^2 {\rm max}(\rho^{3/2}-\rho^{3/2}_{\rm cut}, 0) \ ,
\label{Aphi2}
\eeq
with cutoffs $\rho_{\rm cut} = 0.04 \rho_{\rm max}$ and 
 $P_{\rm cut} = 0.04 P_{\rm max}$. [Note that our
previous study~\cite{DLSSS2} uses the form in Eq.~(\ref{Aphi1}).]  
As with Eq.~(\ref{aphi_disk}), these prescriptions result in poloidal
loops of magnetic field confined within the stars.  Since the
physically realistic magnetic field configuration is unknown,
we adopt this simple prescription as a first step.  This is 
numerically convenient since
confining the initial magnetic field to the high density regions
allows us to avoid strongly magnetically dominated regions in the
initial data.
The proportionality constant $A_b$ determines the initial strength of 
the magnetic field. We characterize the strength of the initial magnetic 
field by $C\equiv {\rm max}(b^2/P)$.  We choose
$A_b$ such that $C\sim 10^{-3}$--$10^{-2}$. We have verified 
that such small initial magnetic fields introduce negligible violations
of the Hamiltonian and momentum constraints in the initial data.
For example, without the magnetic field, the normalized Hamiltonian 
constraint violation due to discretization error is $0.57\%$ for 
Star~C.  For the strongest magnetic field case, C2 (see below), 
adding the magnetic field increases the constraint violation by a 
fraction $1.8\times 10^{-5}$.  

We discuss results of three evolutions, one with star~A, and two 
with star~C (labeled C1 and C2).  Case~A is a continuation of the star~A run
in~\cite{DLSSS2} starting from the point of excision 
($t = 2570 M = 66.9 P_c$).  In the present paper, we evolve the system 
through the post-excision phase and then, for an extended period, in 
the Cowling approximation.  Thus, we consider the longer-timescale behavior of 
the same model presented in~\cite{DLSSS2}.  For case~A, the magnetic field at
$t=0$ takes the form given in Eq.~(\ref{Aphi1}), with $C = 2.5 \times 10^{-3}$,
giving $B^x_{\rm max} = 9.63\times 10^{15}~{\rm G}$.  This
run uses a constant atmosphere floor density 
$\rho_{\rm atm} = 10^{-7}\rho_{\rm max}(t=0) = 4.5 \times 10^7~{\rm g}/{\rm cm}^2$.  
We also impose a pressure floor given by $P_{\rm atm} = 5 \times 10^{-15} P_{\rm max}$.  
This pressure floor is quite small because it is determined by 
taking half of the cold polytropic pressure at the atmosphere density, and the 
EOS has $\Gamma=2$.  In practice, this pressure floor is rarely invoked.  These
values for the pressure and density floors are chosen as in~\cite{DLSSS2} 
since this run is a continuation of the earlier run.

\begin{figure}
\vspace{-4mm}
\begin{center}
\epsfxsize=3.in
\leavevmode
\epsffile{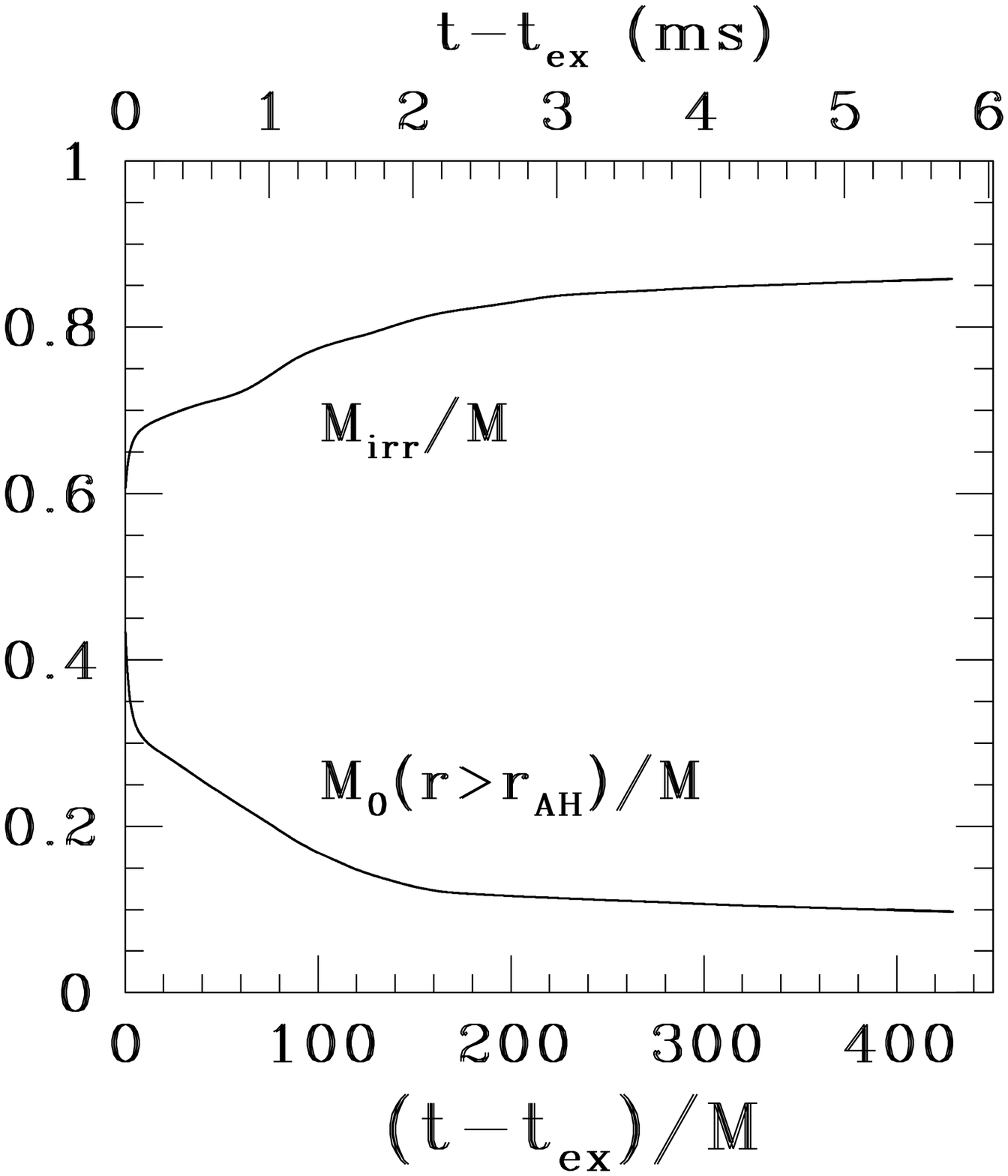}
\caption{Irreducible mass ($M_{\rm irr}$)  and
the rest mass remaining outside the apparent horizon for case~A, 
normalized by the ADM mass ($M$) at $t=0$.  Time is given in units of $M$ 
(bottom axis) and in units of ms (top axis).  By the end of the 
post-excision run, which lasts $427 M$, the BH has settled to a 
quasi-stationary state. \label{excisionA}}
\end{center}
\end{figure}

The two runs involving star~C differ only in their magnetic field configurations.
Case~C1 has an initial magnetic field of the form given by Eq.~(\ref{Aphi1}), and
$C = 8.3 \times 10^{-3}$ ($B^x_{\rm max} = 1.81\times 10^{16}~{\rm G}$).  Case~C2, on the other hand, uses the form in
Eq.~(\ref{Aphi2}), which has the effect of shifting the initial distribution of  
magnetic field energy toward the outer layers of the star.  
Figure~\ref{starC_init} shows the initial density contours and the
magnetic field lines obtained from the two prescriptions for $A_{\varphi}$.
We choose the 
coefficient $A_b$ for case~C2 such that $C = 9.1 \times 10^{-2}$ ($B^x_{\rm max} = 1.84\times 10^{16}~{\rm G}$).  Though this value of $C$ is larger
than our previous choices, we note that the {\it average} value of $b^2/P$ 
is $5 \times 10^{-3}$.  Thus, the overall perturbation to the star is still 
small, and we do not observe any artifacts from it during the evolution.
Both cases~C1 and C2 use an atmosphere prescription inspired by the disk evolutions
in~\cite{mg04}, in which floors on pressure and density fall off with radius 
according to:  $\rho_{\rm atm} = 10^{-7}\rho_{\rm max}(t=0) r^{-3/2}$ and 
$P_{\rm atm} = 1.37 \times 10^{-4} P_{\rm max}(t=0) r^{-5/2}$.  
This prescription for the pressure floor gives an atmosphere which is significantly
hotter than the cold pressure corresponding to $\rho_{\rm atm}$ determined by the
hybrid EOS.  We find that using a hot atmosphere with this EOS leads
to fewer failures of the primitive variable inversion.  The radial dependencies
are {\it ad hoc}, but the purpose of the atmosphere to stabilize
the evolution without significantly affecting the physical outcome.  Allowing 
the density floor to decrease with radius may further reduce the impact of
the atmosphere on the physical behavior~\cite{mcnumalt}. 
Whereas case~A begins 
with the post-excision phase after the HMNS has collapsed, cases~C1 and C2 are evolved 
from $t=0$.    

All three cases are evolved with $500^2$ uniform spatial resolution in $(\varpi,z)$.  
In case~A, the outer boundaries are located at $4.5 R_{\rm eq} = 20.1 M = 80.3~{\rm km}$.  
For the runs with star~C, the outer boundaries are placed at 
$5 R_{\rm eq} = 13.7 M = 53.6~{\rm km}$.

\begin{figure*}
\begin{center}
\epsfxsize=2.2in
\leavevmode
\hspace{-0.7cm}\epsffile{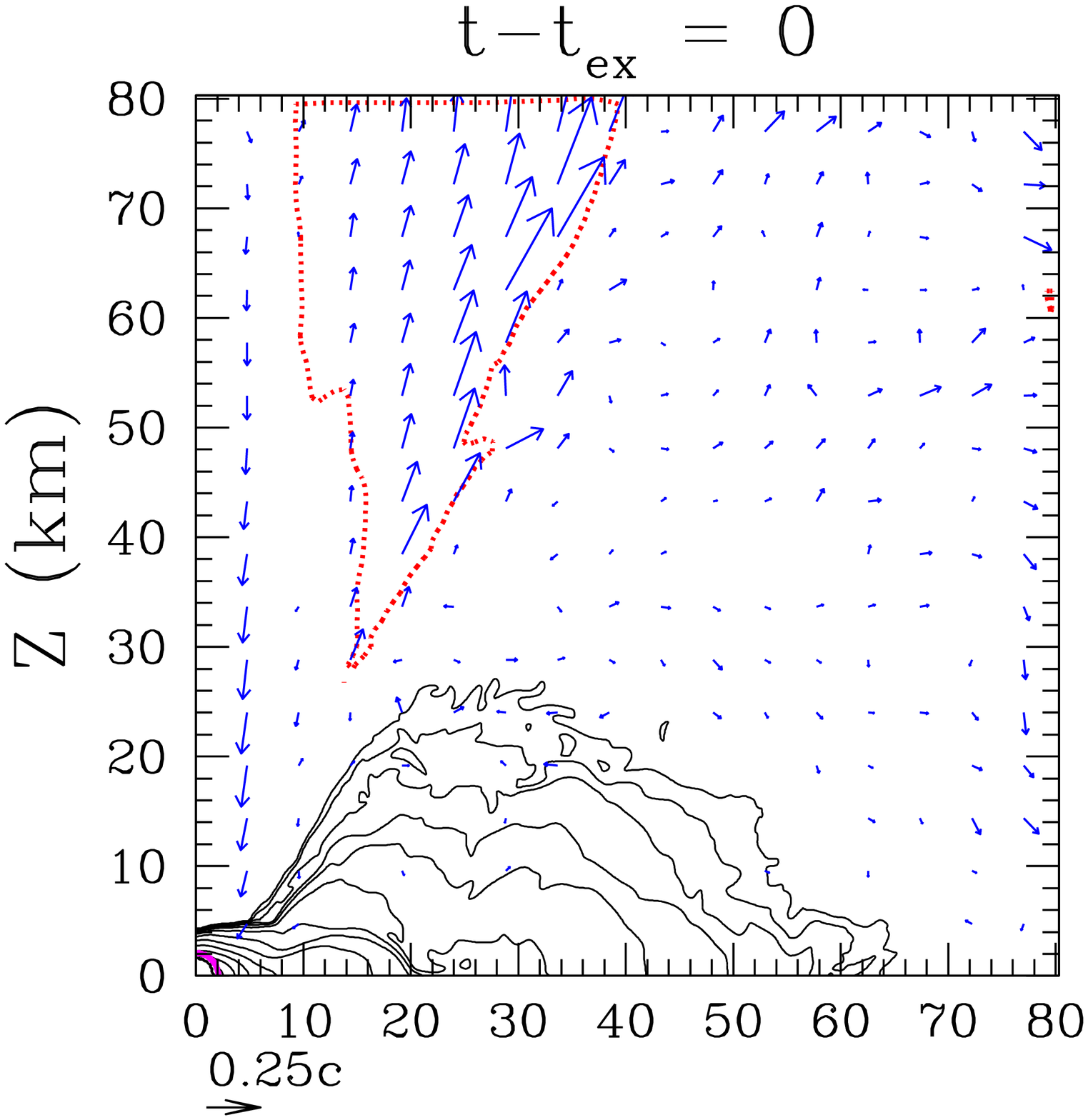}
\epsfxsize=2.2in
\leavevmode
\hspace{-0.5cm}\epsffile{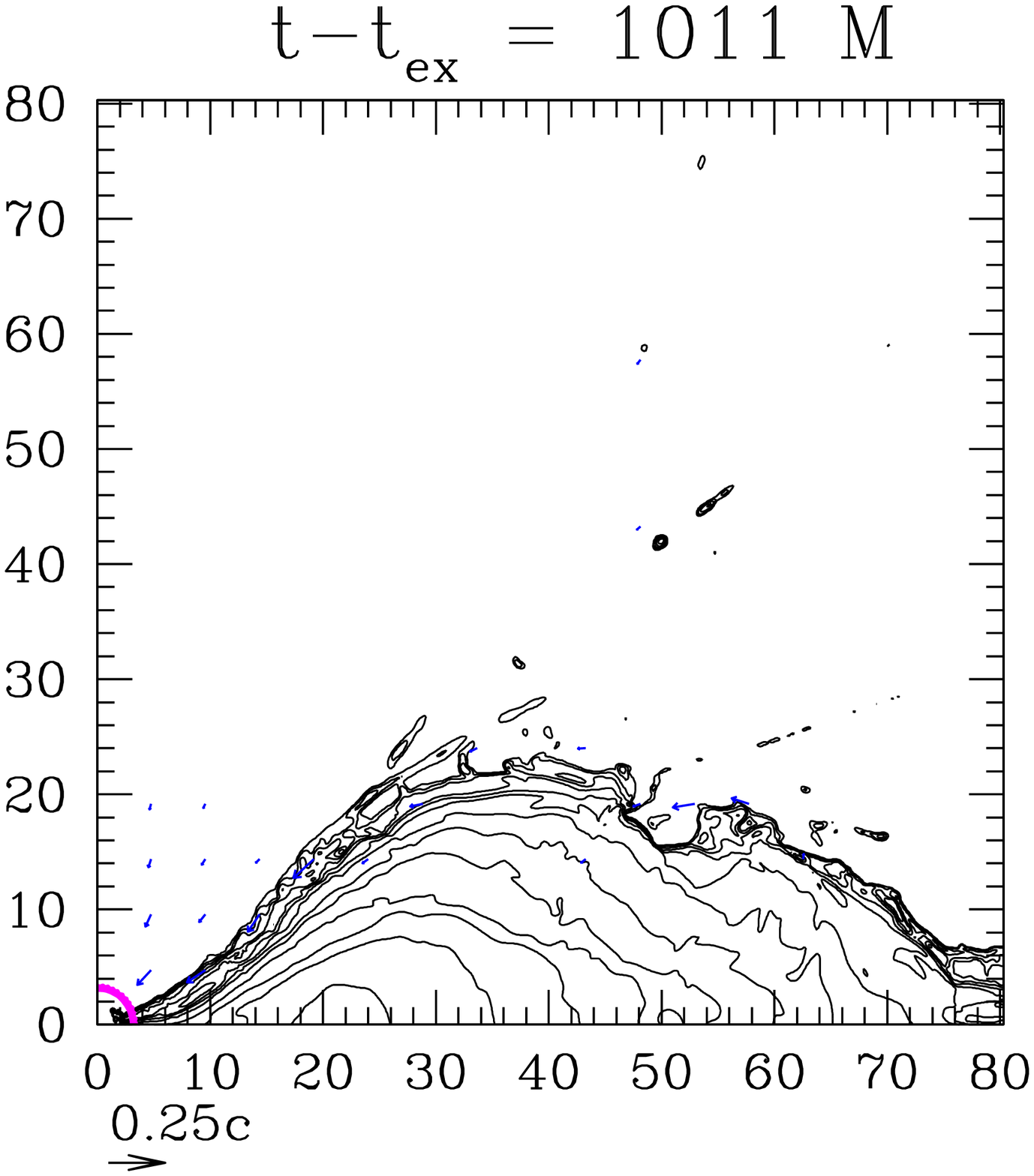}
\epsfxsize=2.2in
\leavevmode
\hspace{-0.5cm}\epsffile{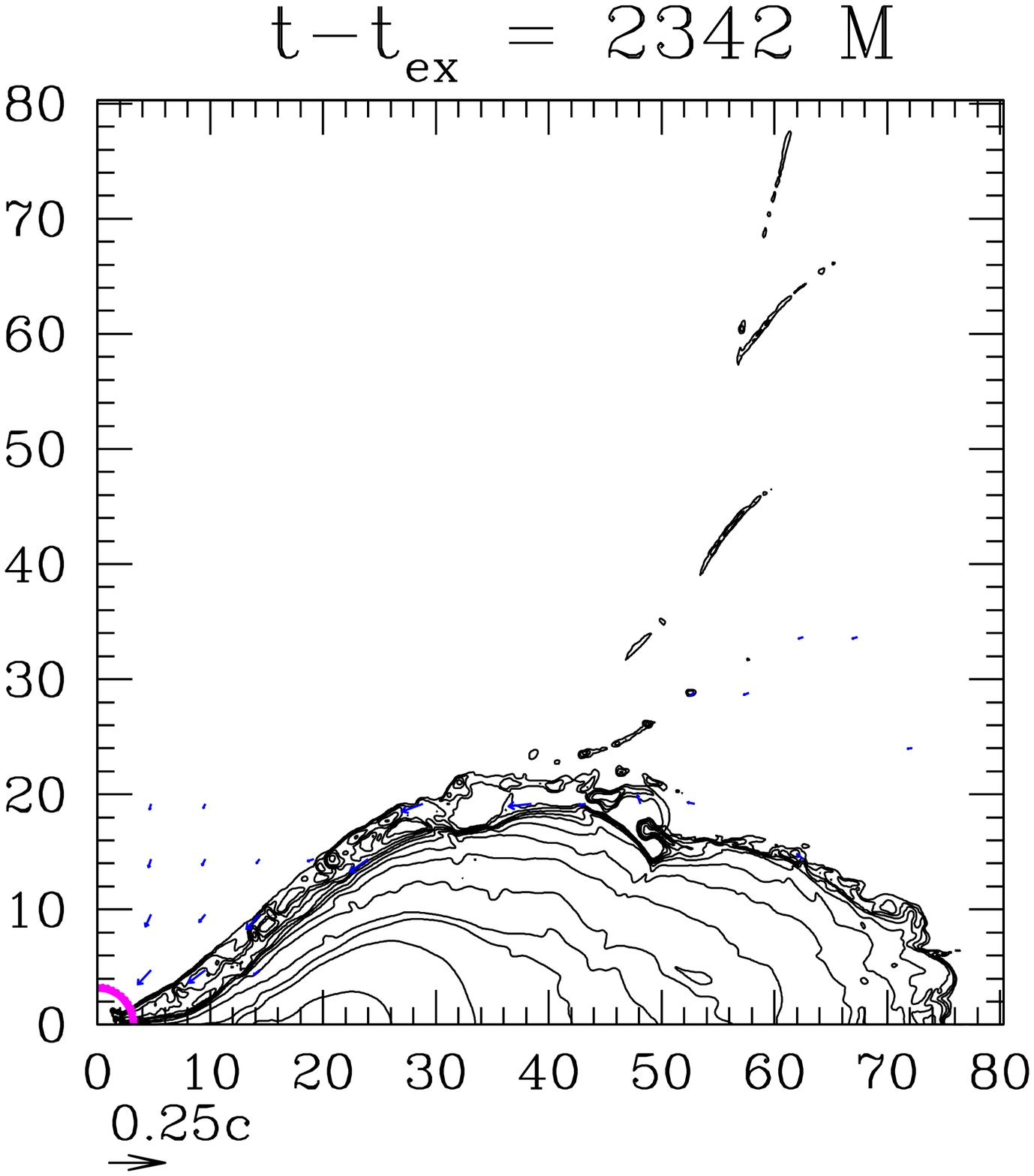} \\
\vspace{-0.5cm}
\epsfxsize=2.2in
\leavevmode
\hspace{-0.7cm}\epsffile{
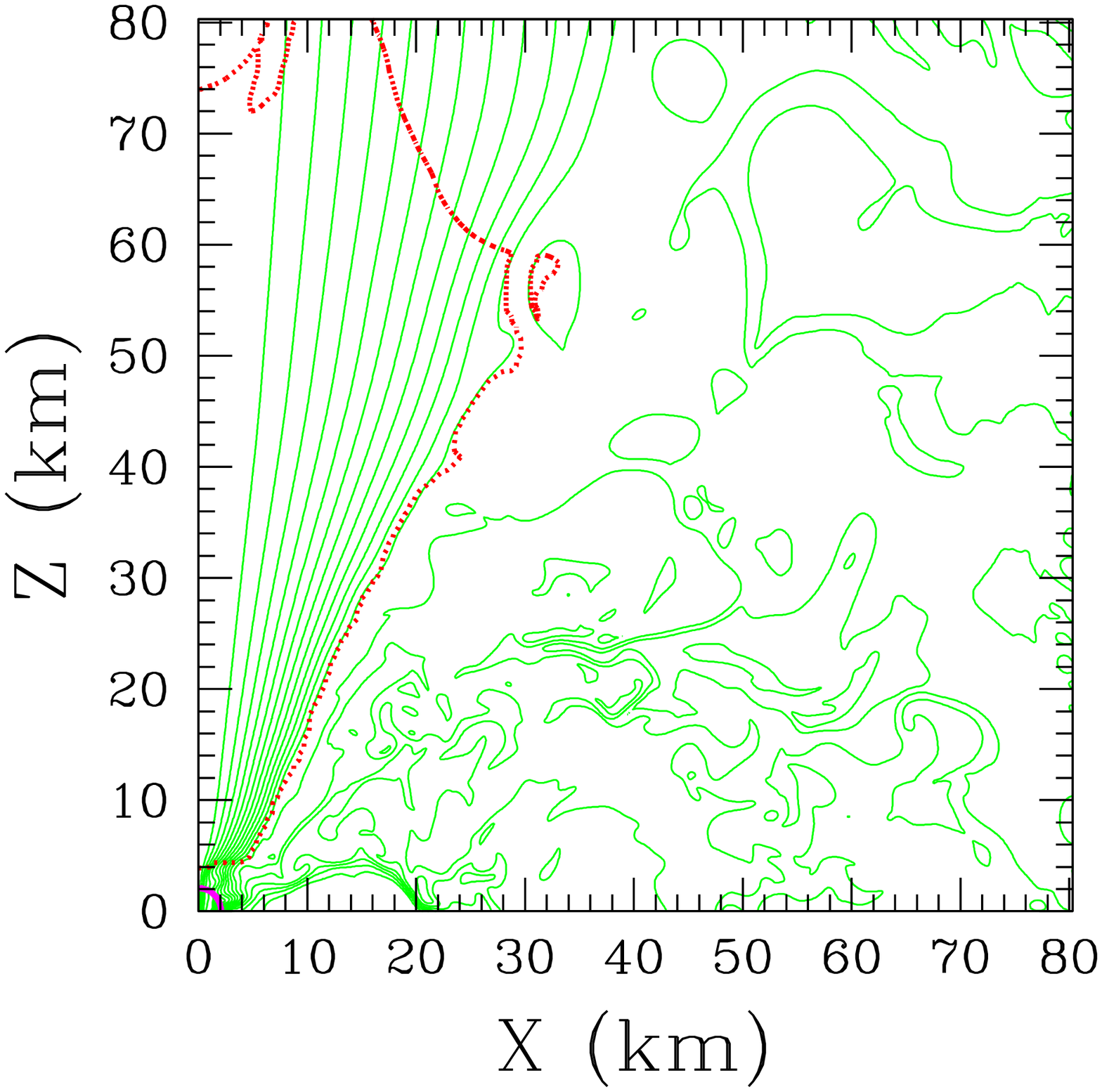}
\epsfxsize=2.2in
\leavevmode
\hspace{-0.5cm}\epsffile{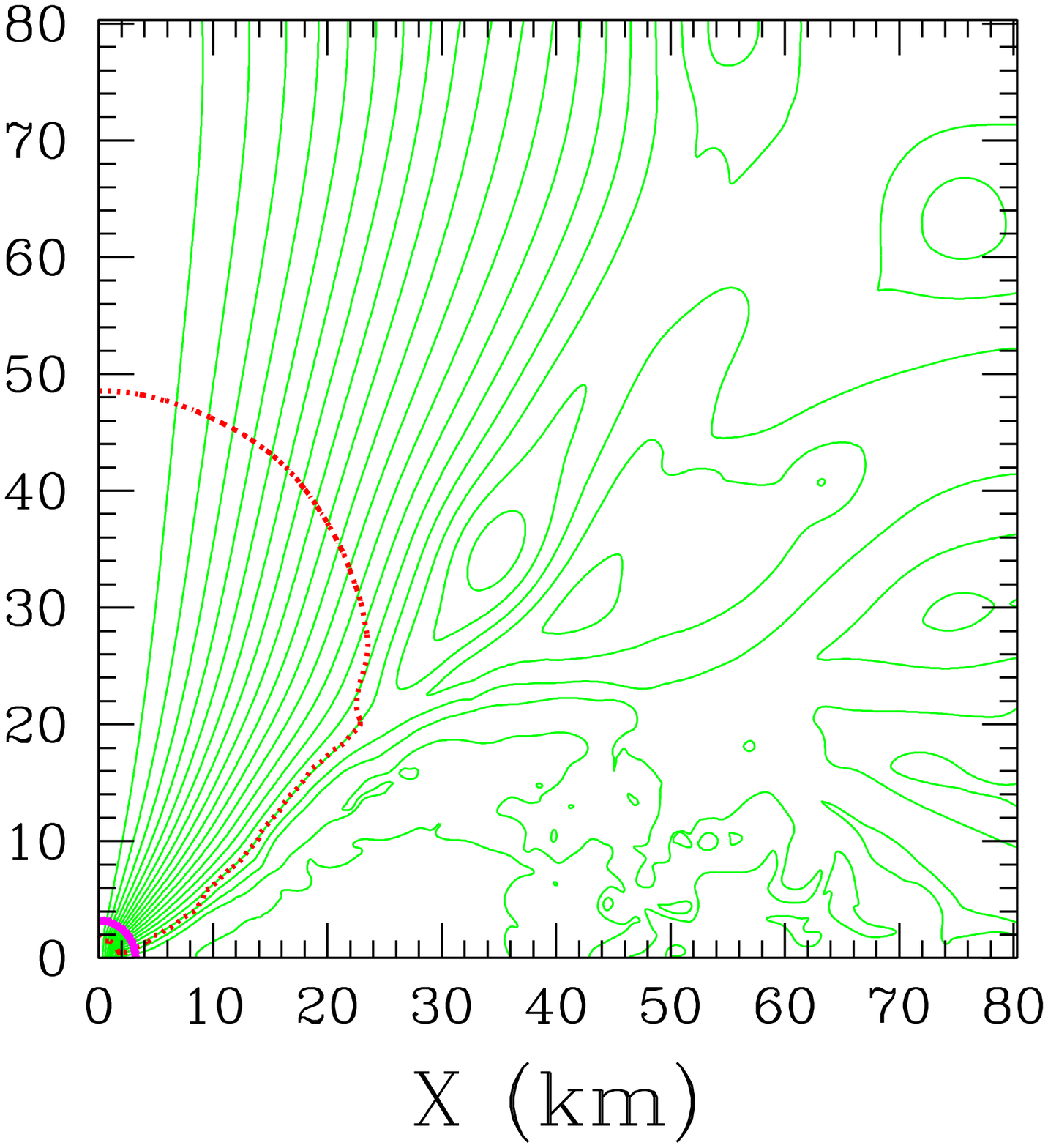}
\epsfxsize=2.2in
\leavevmode
\hspace{-0.5cm}\epsffile{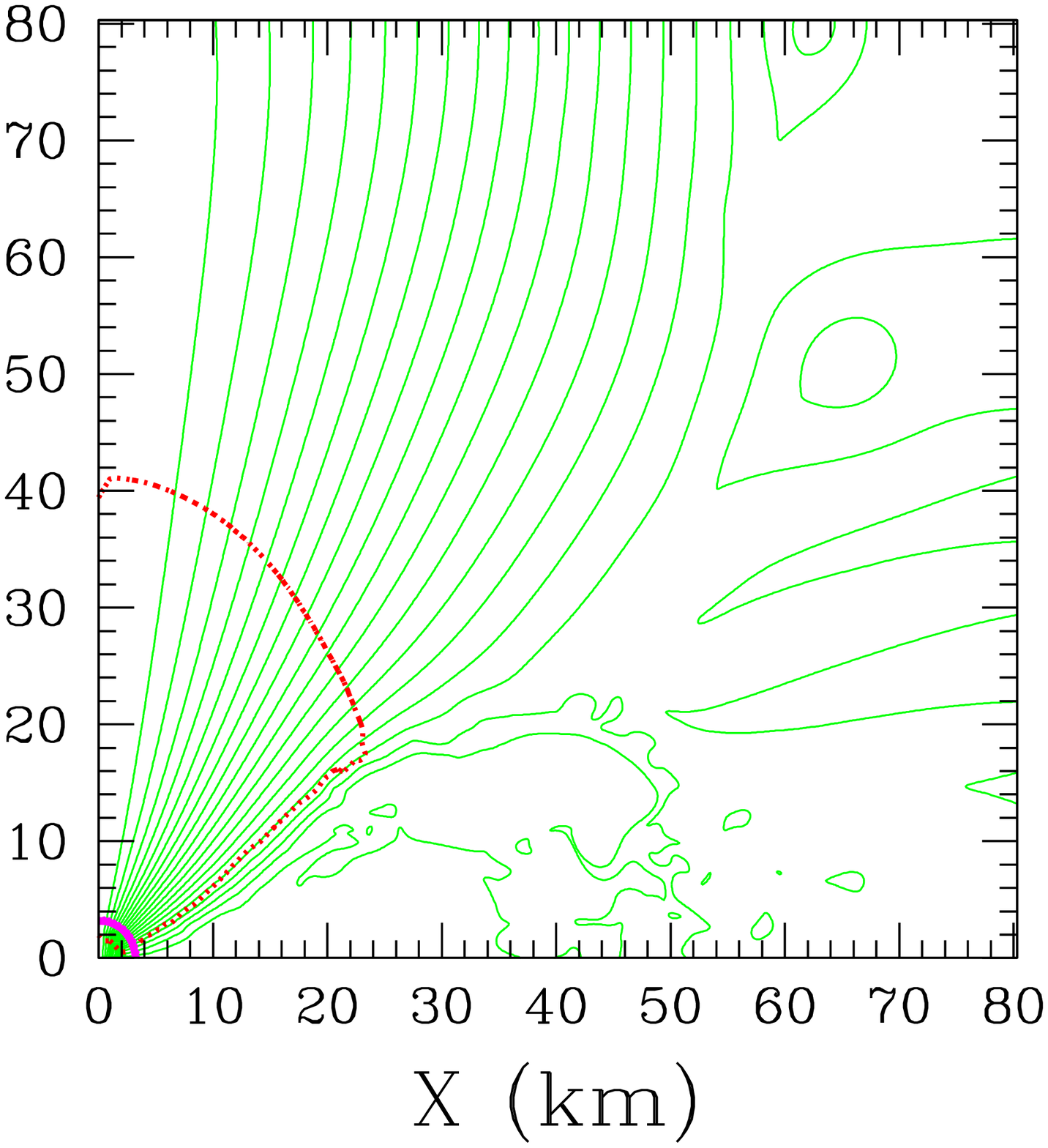}

\caption{Post-excision and Cowling evolution of star~A at selected times.  
The upper panels show density contours and velocity vectors and the lower 
panels show the poloidal magnetic field lines.  The dotted (red) contour 
lines enclose regions with unbound outflows (upper panels) and regions with 
$b^2/\rho \geq 1$ (lower panels).  The heavy magenta circle centered on the 
origin marks the location of the apparent horizon.
\label{star_A_contours}}
\end{center}
\end{figure*}

\begin{figure}
\vspace{-4mm}
\begin{center}
\epsfxsize=3.in
\leavevmode
\epsffile{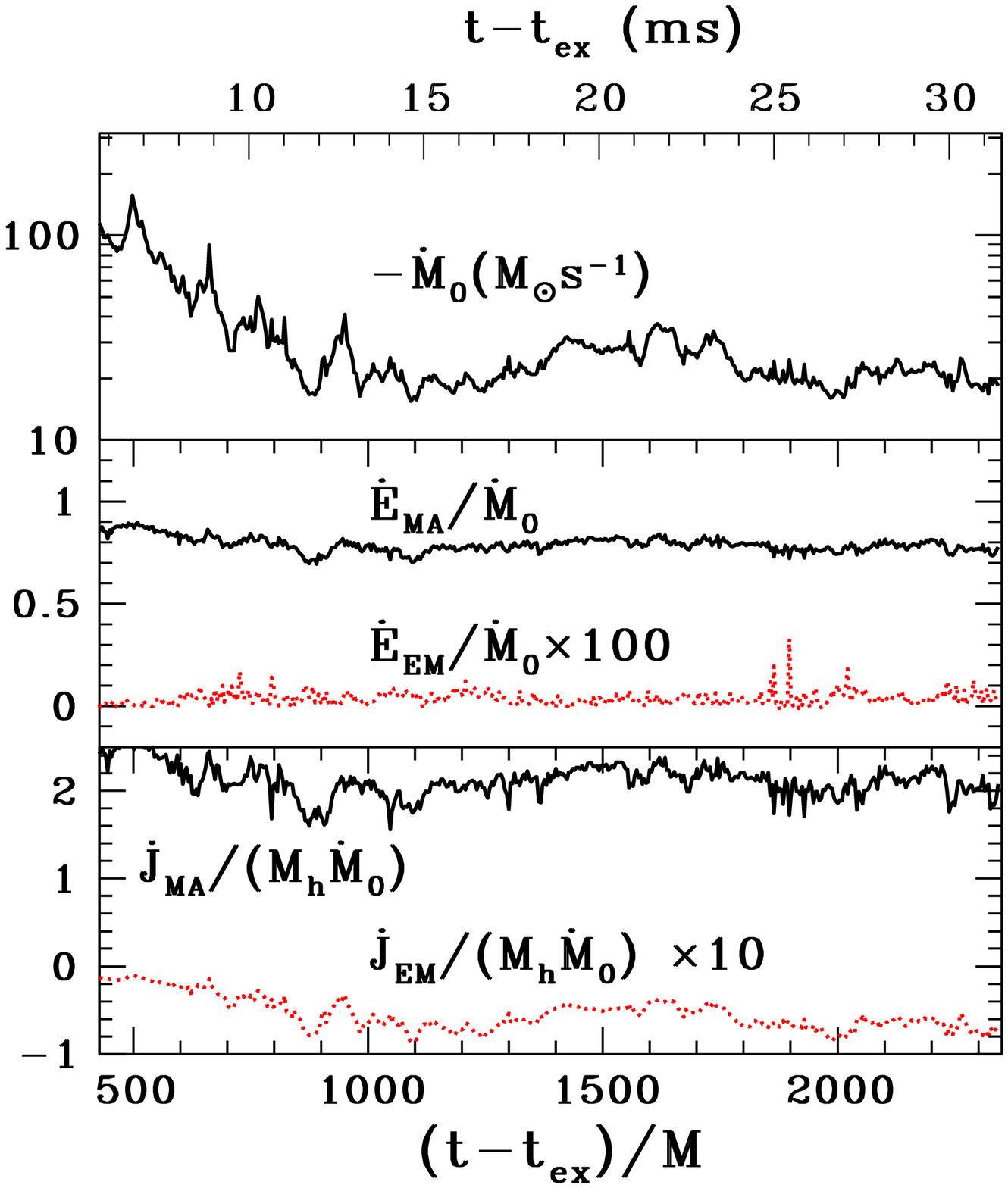}
\caption{Fluxes through the apparent horizon during the Cowling evolution 
of star~A.  The upper panel shows the negative of the rest mass flux in 
units of solar masses per second (where $\dot{M}_0 < 0$ indicates flow 
{\it into} the horizon).  The middle panel shows separately the matter 
and electromagnetic contributions to the total energy flux 
($\dot{E}_{\rm MA}$ and $\dot{E}_{\rm EM}$ respectively), normalized by 
$\dot{M}_0$.  Note that the electromagnetic contribution has been multiplied 
by a factor of 100.  The lower panel shows the electromagnetic and matter 
contributions to the angular momentum flux through the horizon 
($\dot{J}_{\rm MA}$ and $\dot{J}_{\rm EM}$ respectively), normalized by 
$M_h\dot{M}_0$ to yield a non-dimensional ratio.  The electromagnetic 
contribution has been multiplied by a factor of 10.
\label{fluxesA}}
\end{center}
\end{figure}

\subsection{Results for Star A}
The presence of the small seed magnetic field in star~A renders the star
secularly unstable.  Magnetic winding and the MRI transport angular momentum
outward, leading to contraction of the core and expansion of the outer layers.  
The core eventually becomes radially unstable to collapse, which occurs at 
$t \approx 2535 M = 66 P_c$.  This leads to the formation of a BH surrounded 
by a magnetized accretion disk.  (For a detailed discussion of the pre-collapse
and early disk evolution of this model, see~\cite{DLSSS2}.)  

Our code quickly loses accuracy after the formation of the BH due to grid 
stretching on the BH throat.  In order to prevent this, we excise the interior
region surrounding the singularity from the grid at $t = t_{\rm ex} = 2570 M$.  
Excision may not be necessary.  Alternative methods have been suggested 
to handle the black hole spacetime in the presence of matter~\cite{br06,fbest07}.  Whether this 
technique would be effective for the simulations described here 
deserves further study.  For the present, we will employ the 
standard excision technique.
We note that, during the pre-excision evolution, the L2 norms 
of the Hamiltonian and momentum constraints (as defined in~\cite{DLSSS2}) are 
satisfied to better than 1\%.  During the excision evolution, the maximum
values of the constraints are given by 
$(\mathcal{H},\mathcal{M}^x,\mathcal{M}^y,\mathcal{M}^z) = 
(0.5\%,0.5\%,3\%,0.75\%)$.  

As discussed in~\cite{DLSSS2}, we are only able to perform the post-excision evolution 
accurately for $\sim 400M$.  In order to consider the disk behavior on longer 
timescales, we freeze the spacetime metric after the BH has settled down to 
a quasi-stationary state.  We then evolve the MHD
equations in this fixed spacetime (i.e., the Cowling approximation).
We start the Cowling phase at $t = 2997 M$.  The 
duration of the post-excision phase (with live BSSN evolution) is thus $427 M$.  
(We note that the accretion flow, which has little effect on the spacetime metric,
is strongly time dependent when the Cowling approximation is first imposed.  The
evolution of the disk itself is thus not yet quasi-stationary.)  
The Cowling phase of the evolution lasts for a further $1912 M$.
At the beginning of the Cowling phase, the rotation period near the middle 
of the disk is $\sim 164M$.
Thus the post-excision and Cowling phases encompass $\sim 14$ rotation periods of 
the disk.  We note that the change in the disk mass during the Cowling phase is about 3\% of 
the total rest mass.   The Cowling approximation does not take this change into 
account, but the approximation should be fairly accurate since 
the self-gravity of the disk is estimated to affect the 
dynamics by $\sim (M_{\rm disk}/M)$, and the error is thus
$\sim \Delta M_{\rm disk}/M \sim 3$\% (where $\Delta M_{\rm disk}$ is the change in the disk 
mass during the Cowling phase).

Figure~\ref{excisionA} shows the evolution of the irreducible mass and the 
rest mass remaining outside the horizon during the post-excision phase.  Following a 
period of rapid accretion (representing the final stages of collapse), the 
accretion slows down to a quasi-stationary rate.  This transition
occurs at $t-t_{\rm ex} \sim 170 M$.  We can estimate the BH mass $M_h$ and 
angular momentum $J_h$ at this 
time [see Eqs.~(\ref{jhole}) and (\ref{mhole})], and we find that $M_h = 0.91M$ 
and $J_h/M_h^2 = 0.79$.  Figure~\ref{star_A_contours} shows 
snapshots during the post-excision and Cowling phases.  We find
no evidence for significant unbound outflows in this case (aside from a 
transient which can be seen in the first set of panels).  Accretion from the 
disk takes place primarily near the equatorial plane, and very little material
is churned up from the disk into a corona.  The disk thus reaches an 
essentially quasi-stationary state.  To check that the lack
of outflows in this case is not caused by our constant density floor, we 
also performed this evolution with floors which fall off with radius.
In particular, we set $\rho_{\rm atm} = 10^{-7}\rho_{\rm max}(t=0) r^{-3/2}$ and 
$P_{\rm atm} = 10^{-14} P_{\rm max}(t=0) r^{-5/2}$.  However, this did not
make any qualitative difference in the outcome, and no significant outflows
were observed.

Finally, we plot accretion rates through the apparent horizon during the 
Cowling phase in Figure~\ref{fluxesA}.  Averaging $\dot{E}$, $\dot{J}$, 
and $\dot{M}_0$ individually over the duration of the Cowling run, we find
the ratios $\dot{E}/\dot{M}_0 = 0.81$ and $\dot{J}/(M_h \dot{M}_0) = 2.2$.  These
are in rough agreement with Table~2 of~\cite{mg04}, 
which gives results for disk evolutions with varying BH spin.  Though their
table does not give results for a BH with the specific spin parameter 
$J_h/M_h^2 = 0.8$, (which is the estimate for the BH in case~A), the nearby 
table entries suggest $\dot{E}/\dot{M}_0 \sim 0.9$ and 
$\dot{J}/(M_h \dot{M}_0) \sim 2$.  (Note that $M_h = 1$ 
in~\cite{mg04}.)  Roughly speaking, the $\dot{J}/(M_h \dot{M}_0)$ ratio 
increases as the BH spin decreases because the innermost stable circular 
orbit (from which material plunges into the BH) moves outward in radius as 
the spin decreases.  This explains why accretion onto star~A gives a higher
value for $\dot{J}/(M_h \dot{M}_0)$ than the disk simulation in 
Section~\ref{disk}, for which the BH has higher spin $J_h/M_h^2 = 0.938$.

\begin{figure}
\vspace{-4mm}
\begin{center}
\epsfxsize=3.in
\leavevmode
\epsffile{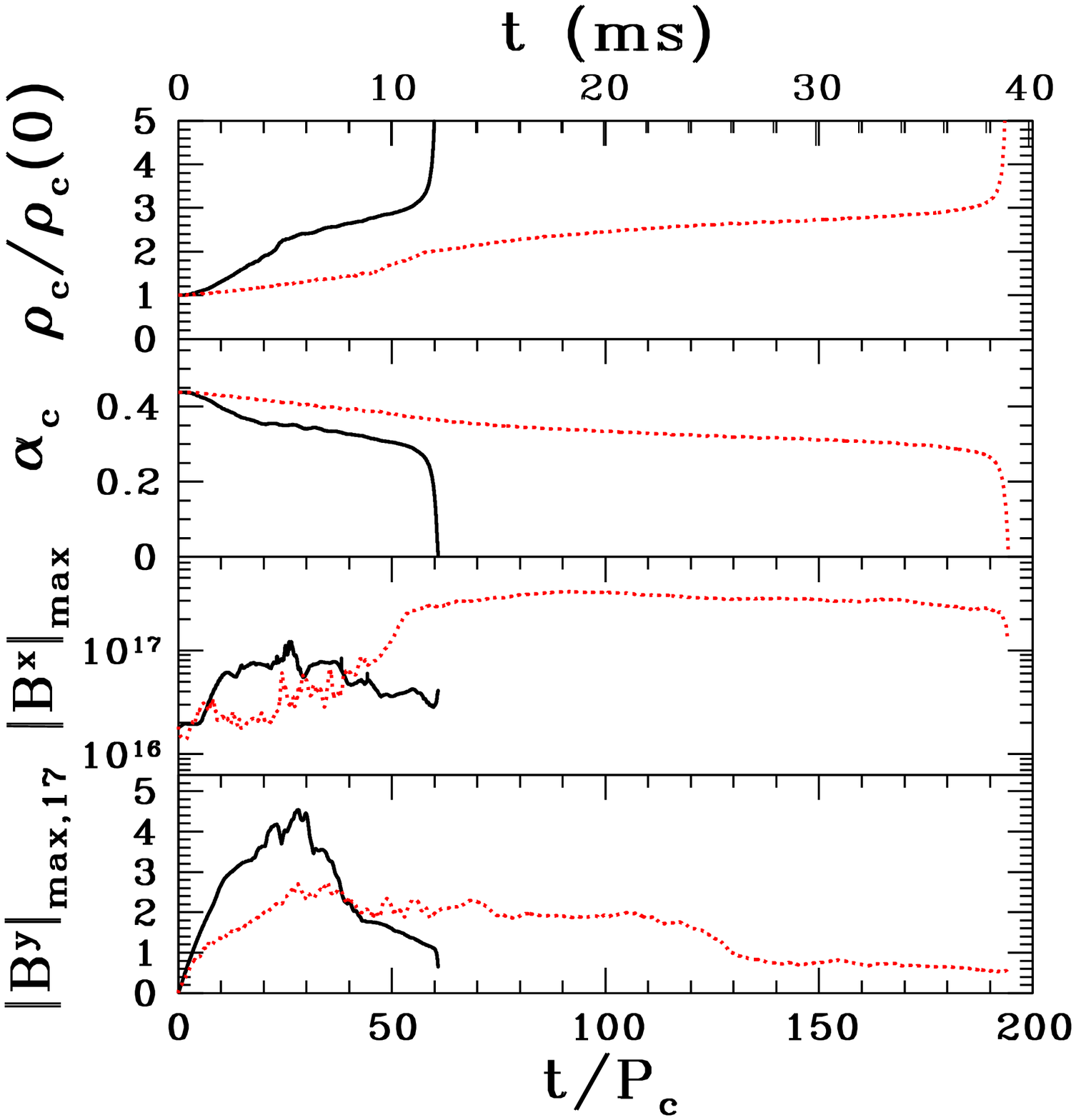}
\caption{Pre-excision evolution for cases~C1 (solid black line) and 
C2 (red dashed line).  From top to bottom, the quantities plotted are: 
(i) the central rest-mass density normalized by its initial value, 
(ii) the central lapse, (iii) the maximum value of $|B^x|$ in units 
of gauss (G), and (iv) the maximum value of $|B^y|$ in units of 
$10^{17}~{\rm G}$.  Collapse occurs in case~C1 much earlier than in case~C2.
\label{preex}}
\end{center}
\end{figure}

\subsection{Results for Star C}
\begin{figure*}
\begin{center}
\epsfxsize=2.2in
\leavevmode
\hspace{-0.7cm}\epsffile{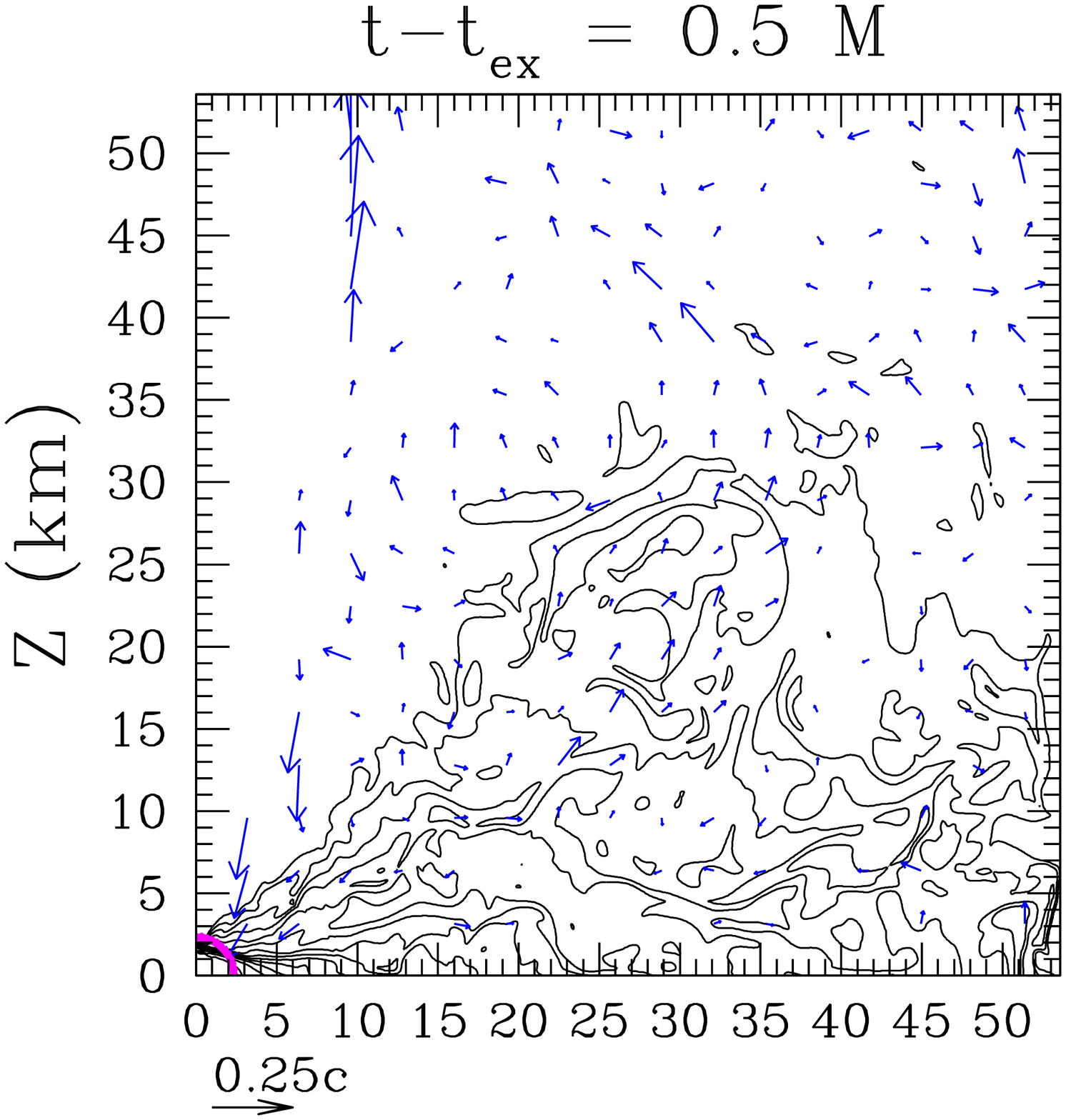}
\epsfxsize=2.2in
\leavevmode
\hspace{-0.5cm}\epsffile{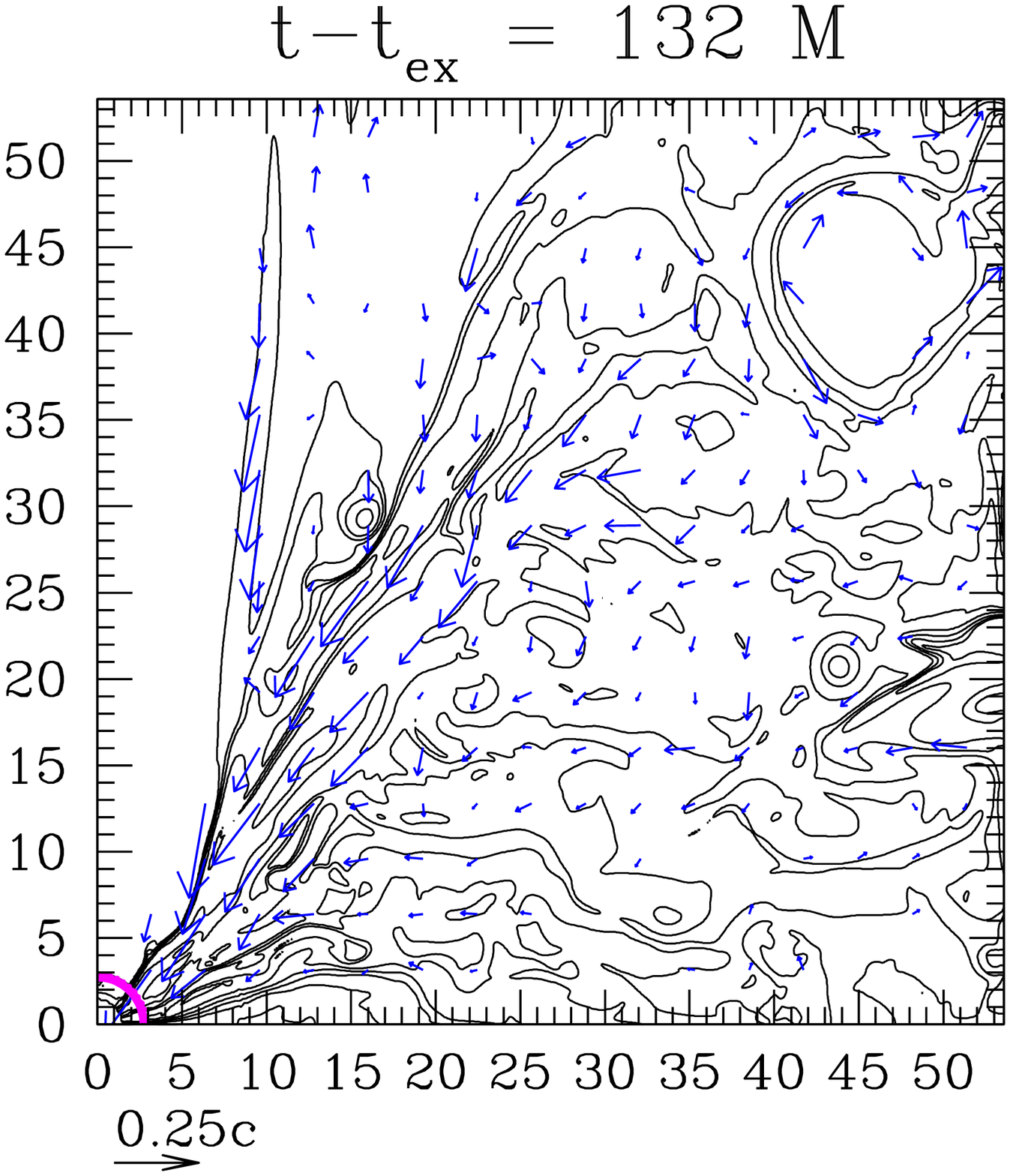}
\epsfxsize=2.2in
\leavevmode
\hspace{-0.5cm}\epsffile{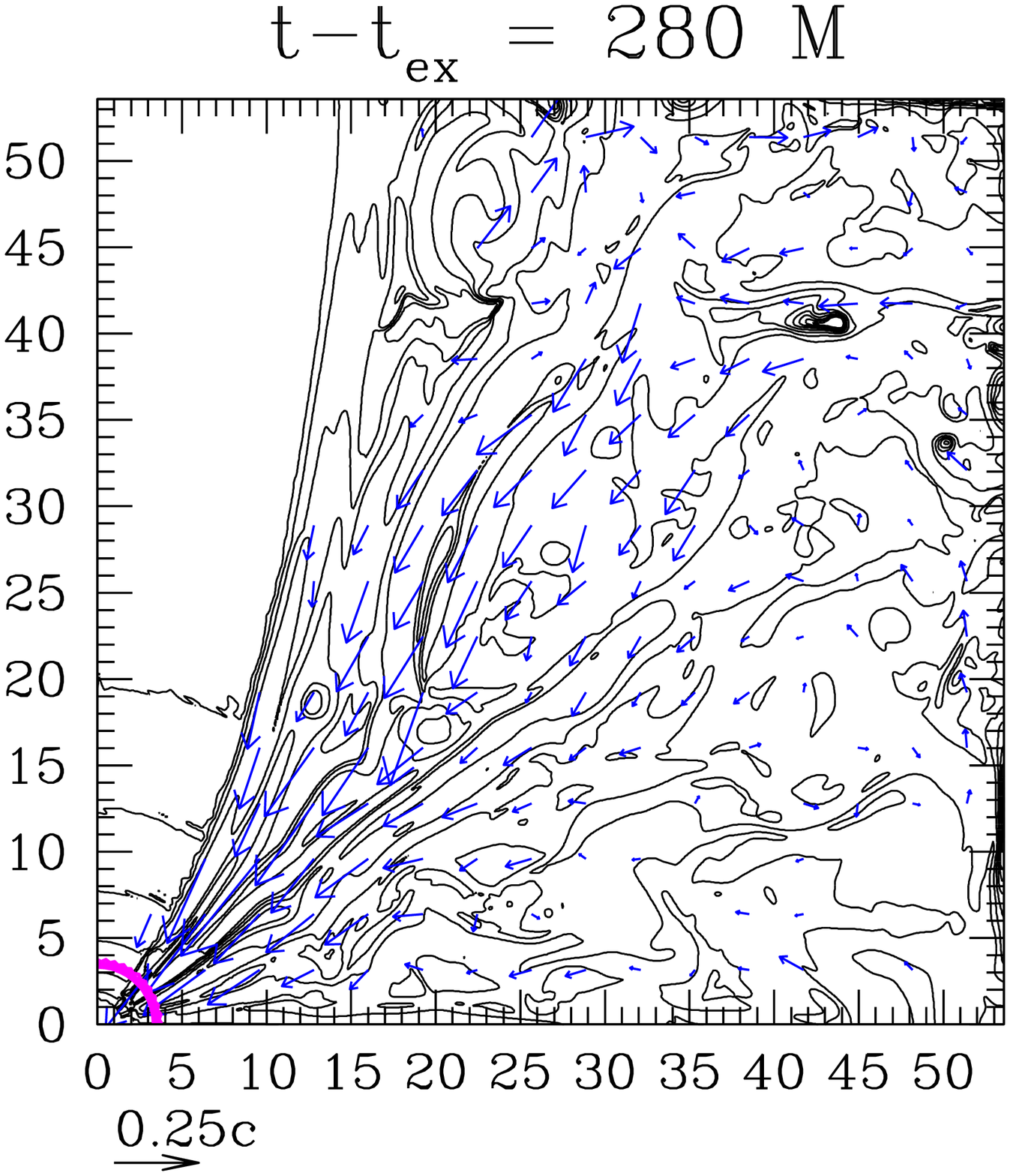} \\
\vspace{-0.5cm}
\epsfxsize=2.2in
\leavevmode
\hspace{-0.7cm}\epsffile{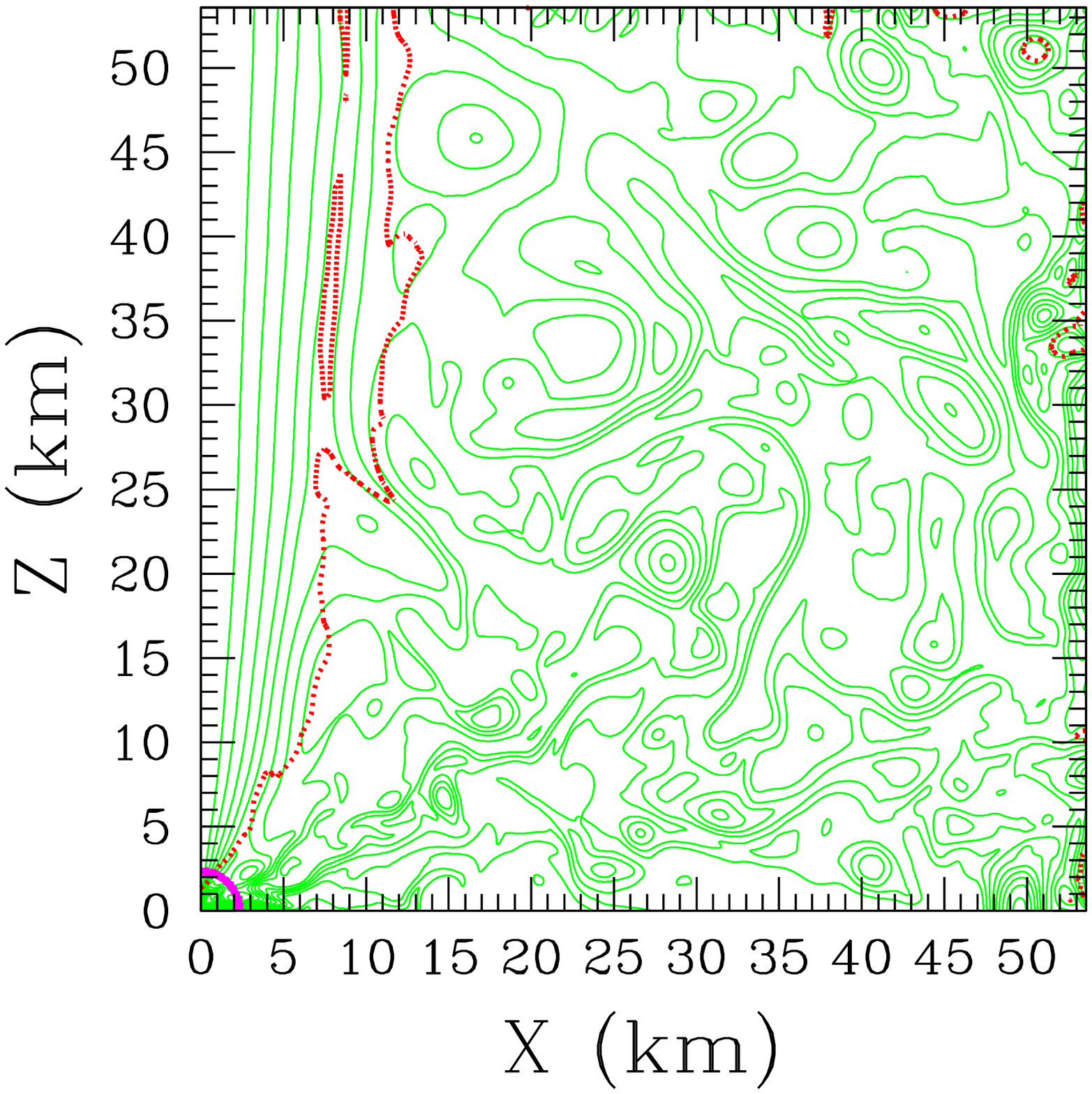}
\epsfxsize=2.2in
\leavevmode
\hspace{-0.5cm}\epsffile{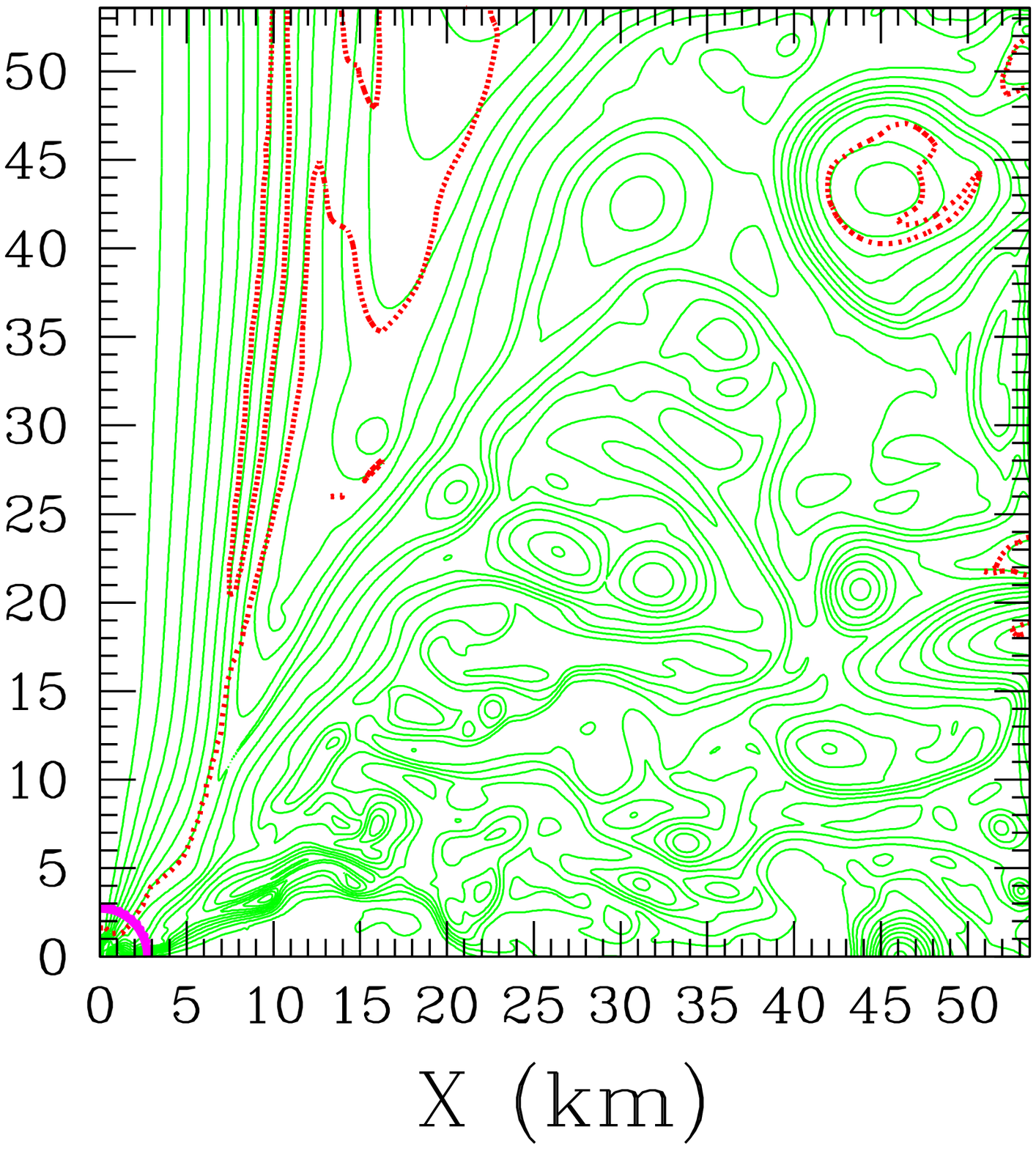}
\epsfxsize=2.2in
\leavevmode
\hspace{-0.5cm}\epsffile{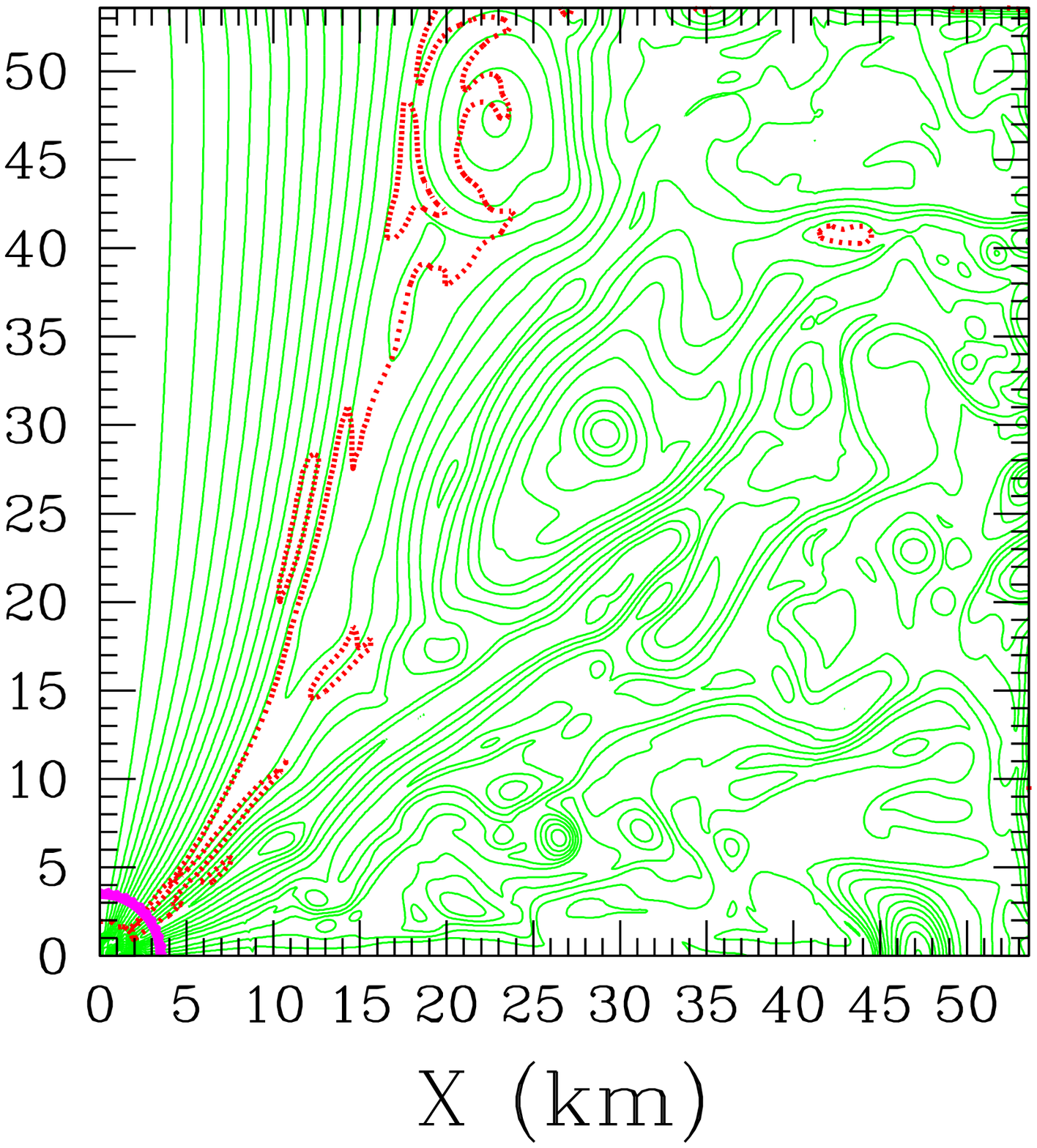}
\end{center}

\begin{center}
\epsfxsize=2.2in
\leavevmode
\hspace{-0.7cm}\epsffile{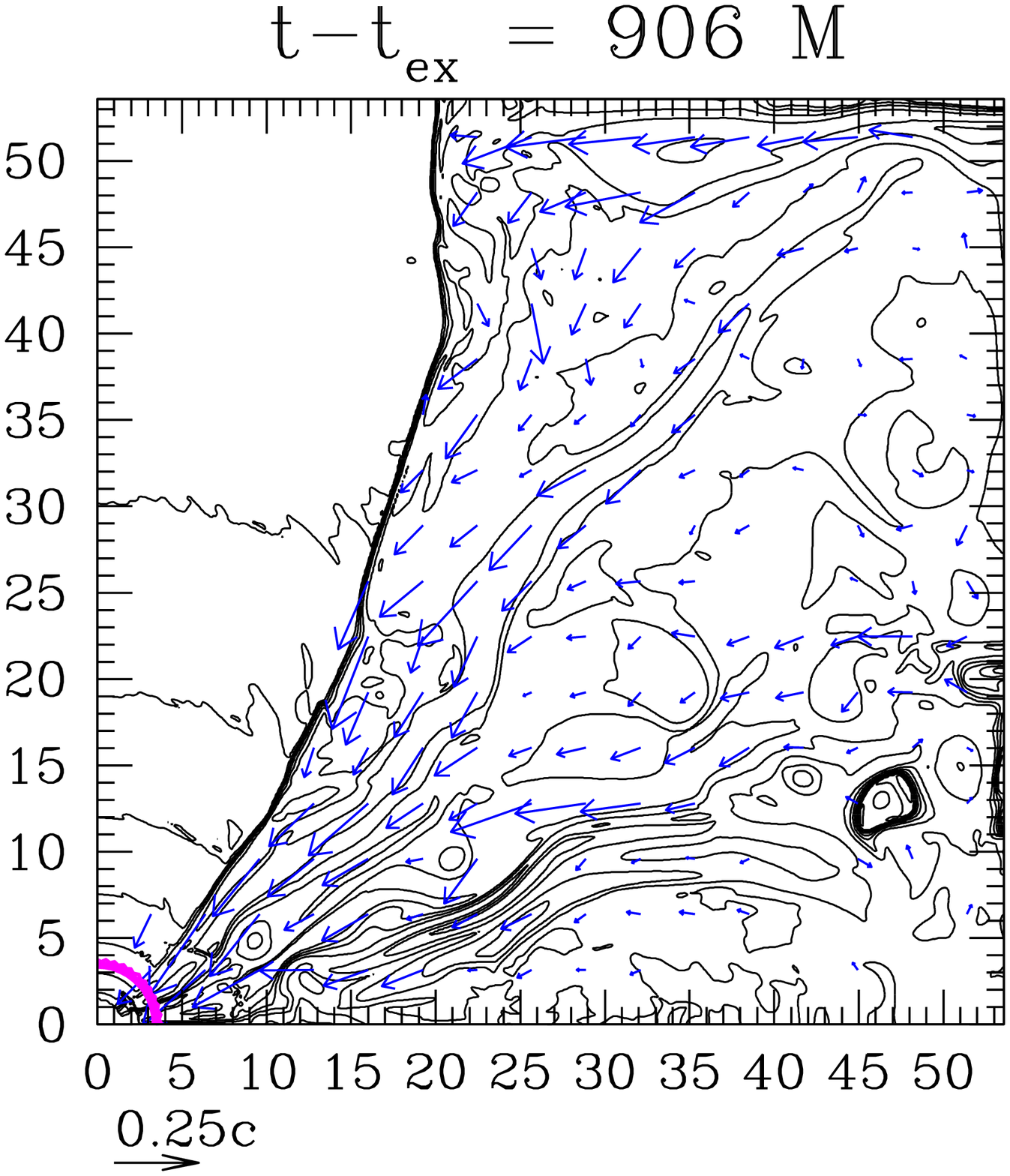}
\epsfxsize=2.2in
\leavevmode
\hspace{-0.5cm}\epsffile{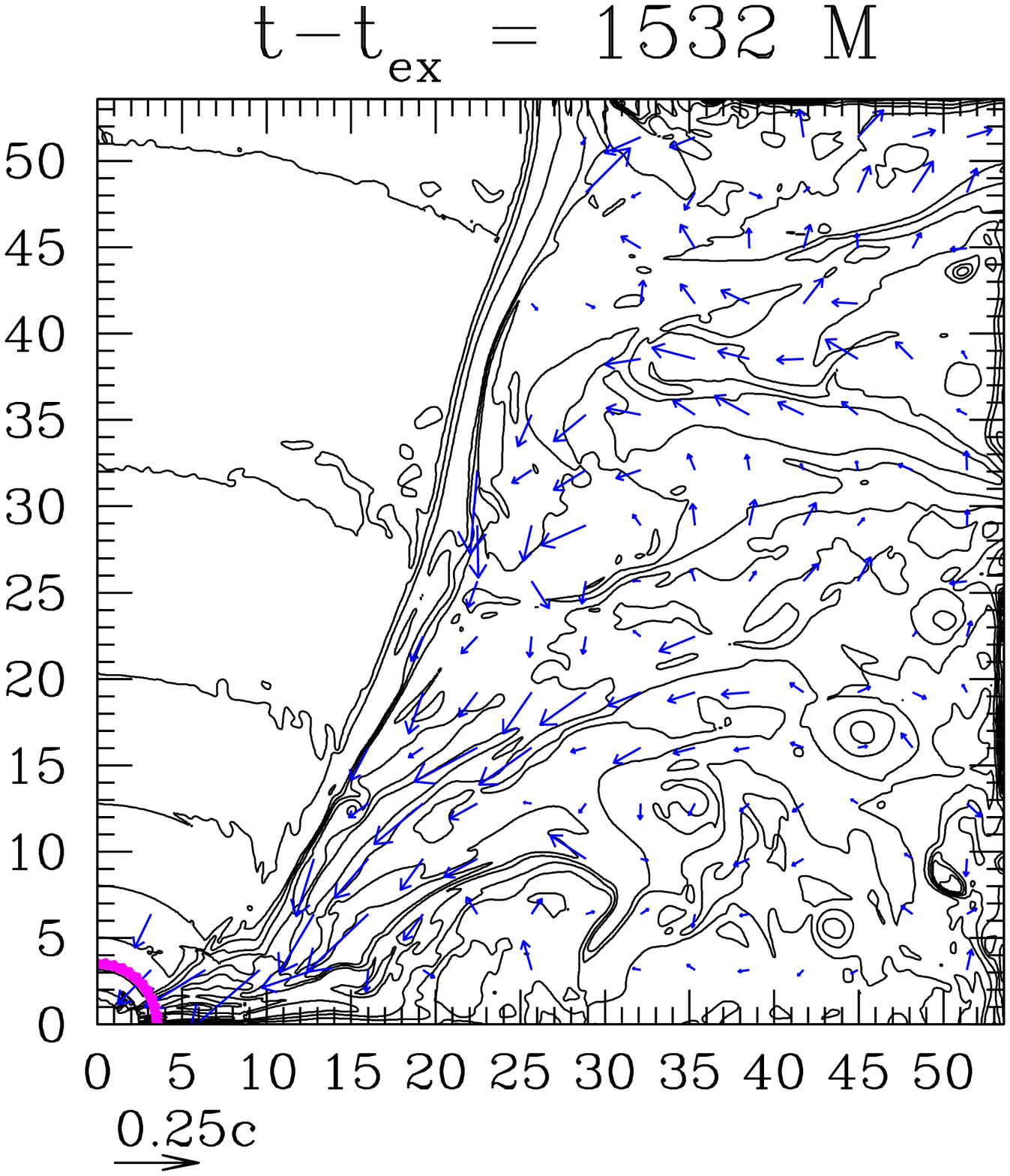}
\epsfxsize=2.2in
\leavevmode
\hspace{-0.5cm}\epsffile{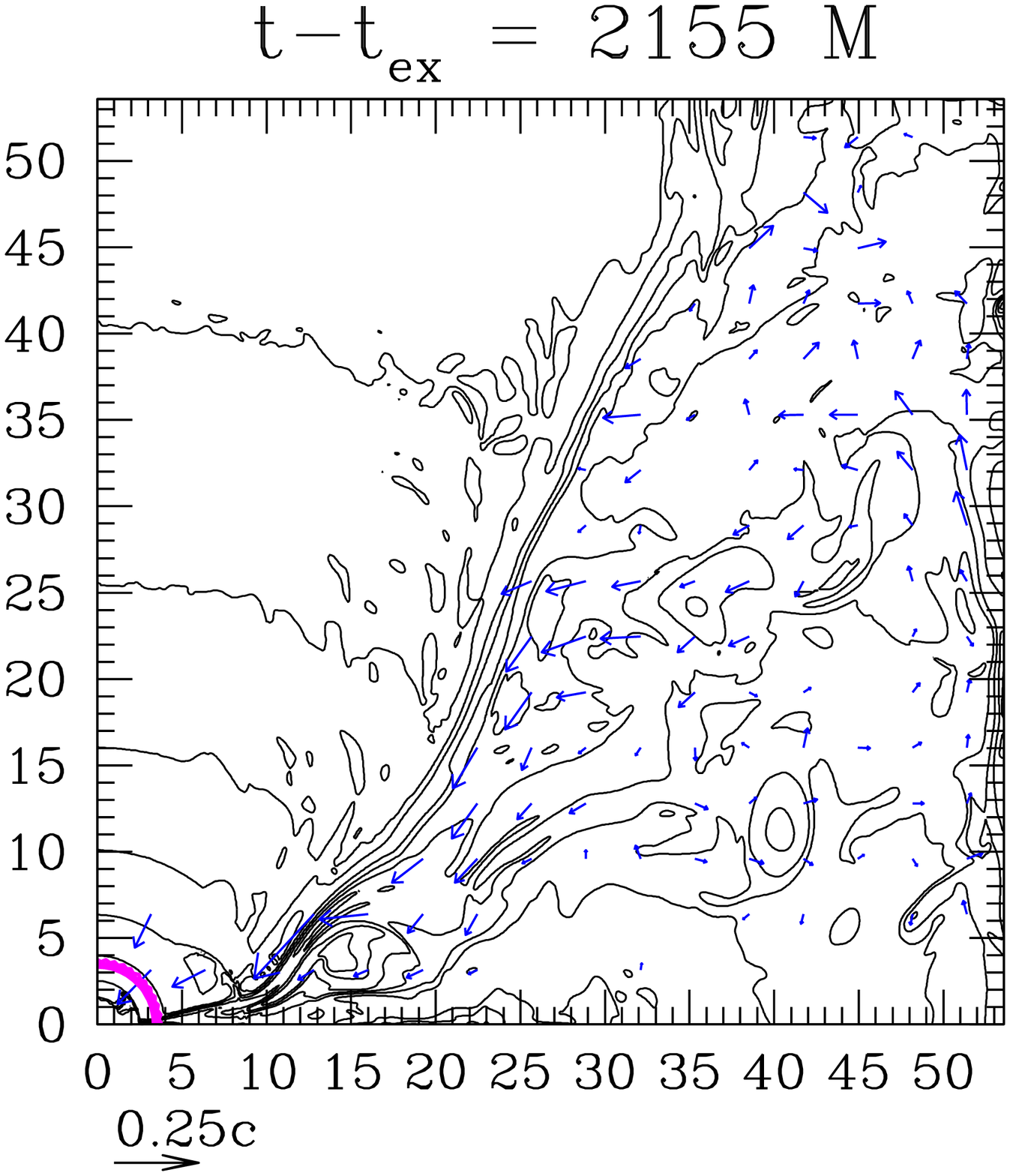} \\
\vspace{-0.5cm}
\epsfxsize=2.2in
\leavevmode
\hspace{-0.7cm}\epsffile{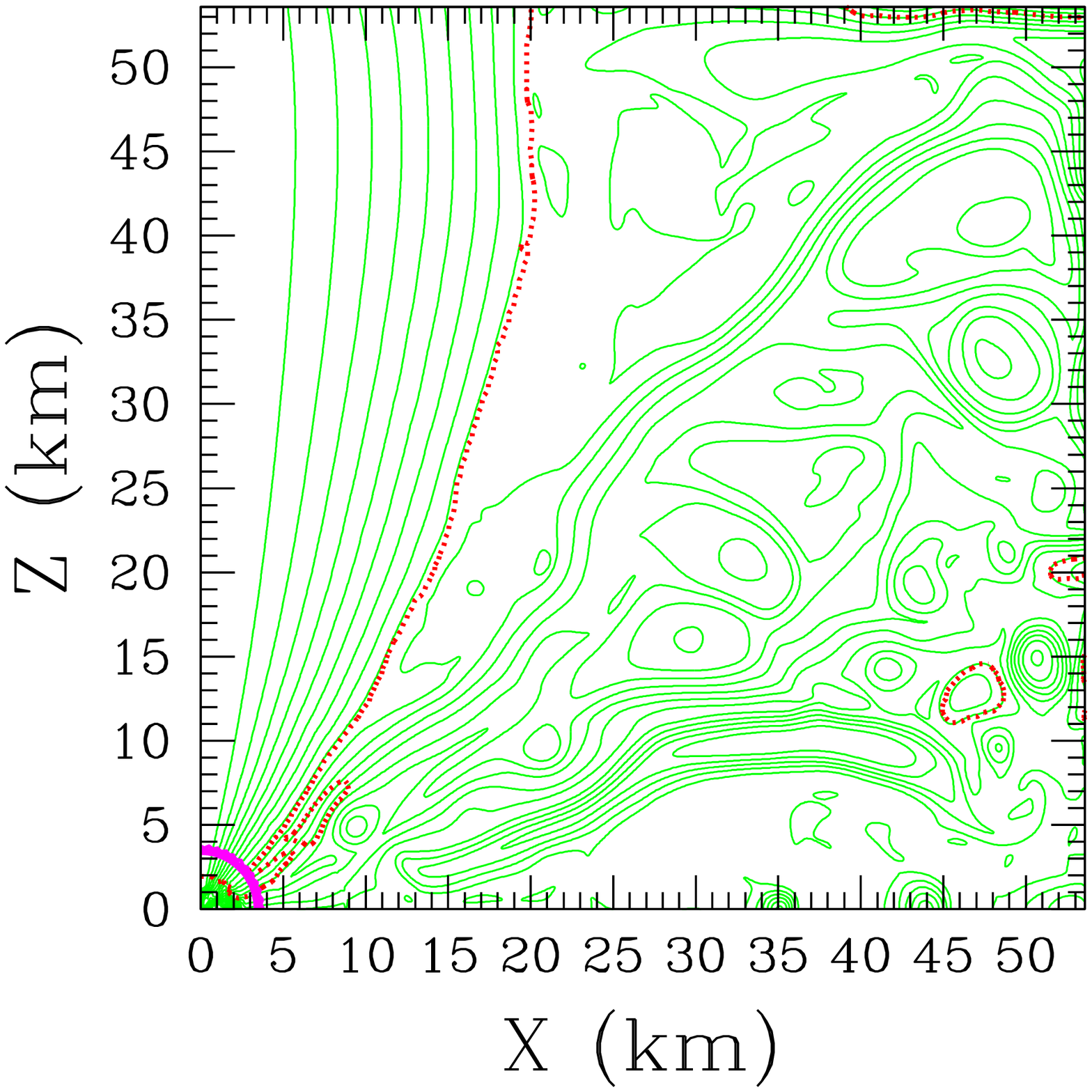}
\epsfxsize=2.2in
\leavevmode
\hspace{-0.5cm}\epsffile{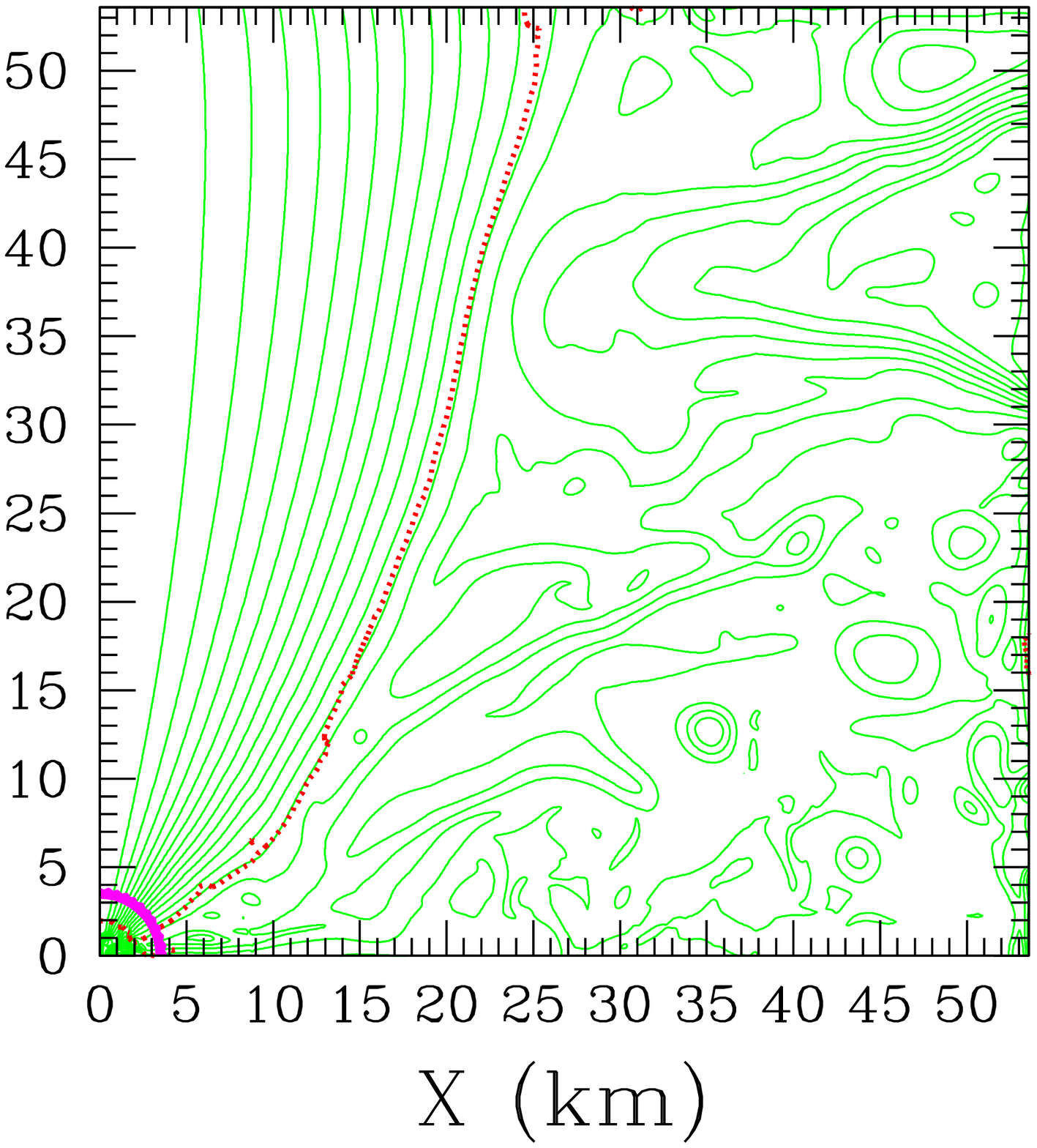}
\epsfxsize=2.2in
\leavevmode
\hspace{-0.5cm}\epsffile{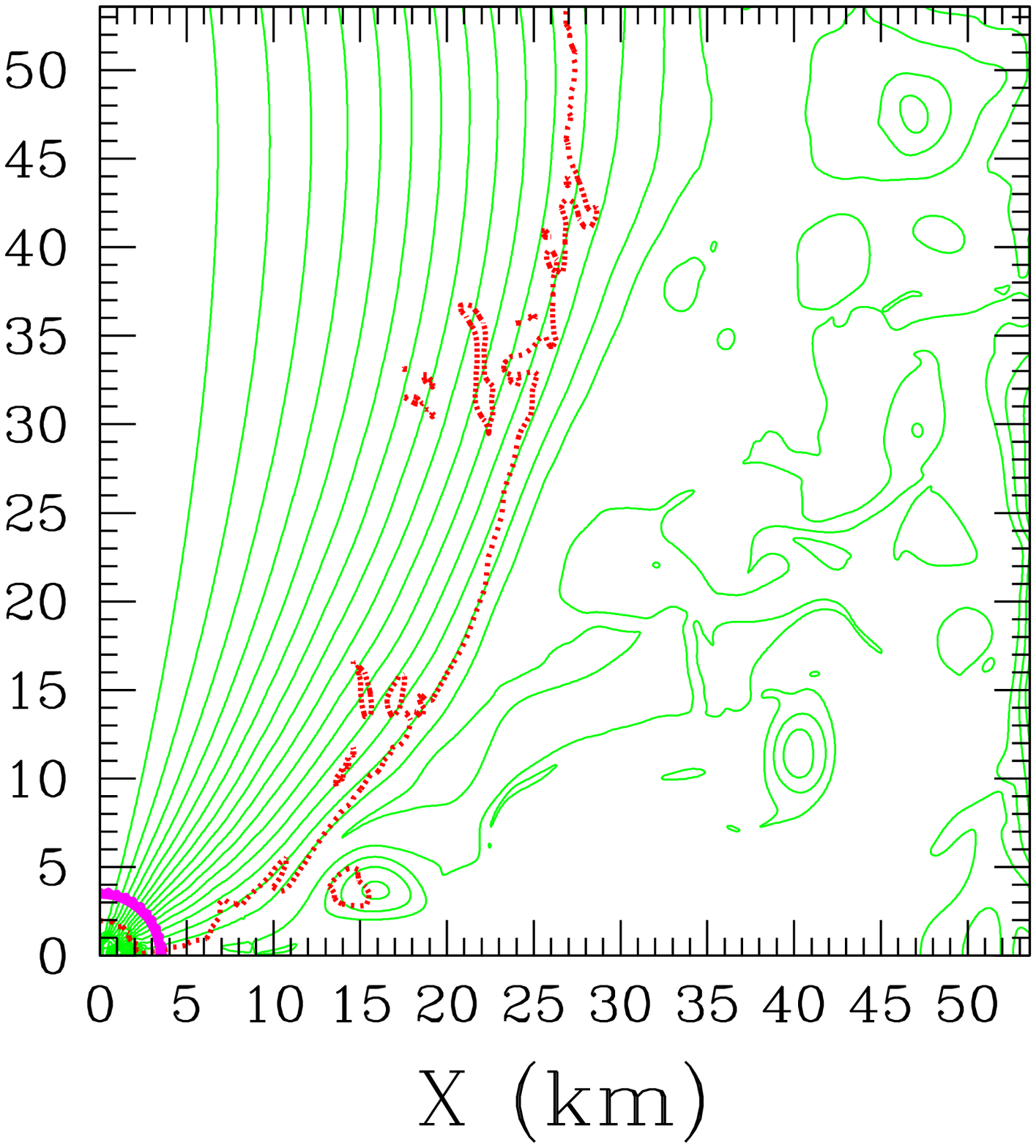}
\caption{Post-excision and Cowling evolution phases of case~C1, shown
at selected times.  The meanings of the lines are the same as in 
Fig~\ref{star_A_contours}, except that the $u_t = -1$ contours
have been left out for the sake of clarity.  
The first and last sets of panels represent the beginning of the post-excision 
phase and the end of the simulation, respectively. The panel
at $t-t_{\rm ex} = 280 M$ marks the beginning of the Cowling phase. 
\label{c1_contours}}
\end{center}
\end{figure*}

\begin{figure*}
\begin{center}
\epsfxsize=2.2in
\leavevmode
\hspace{-1.5cm}\epsffile{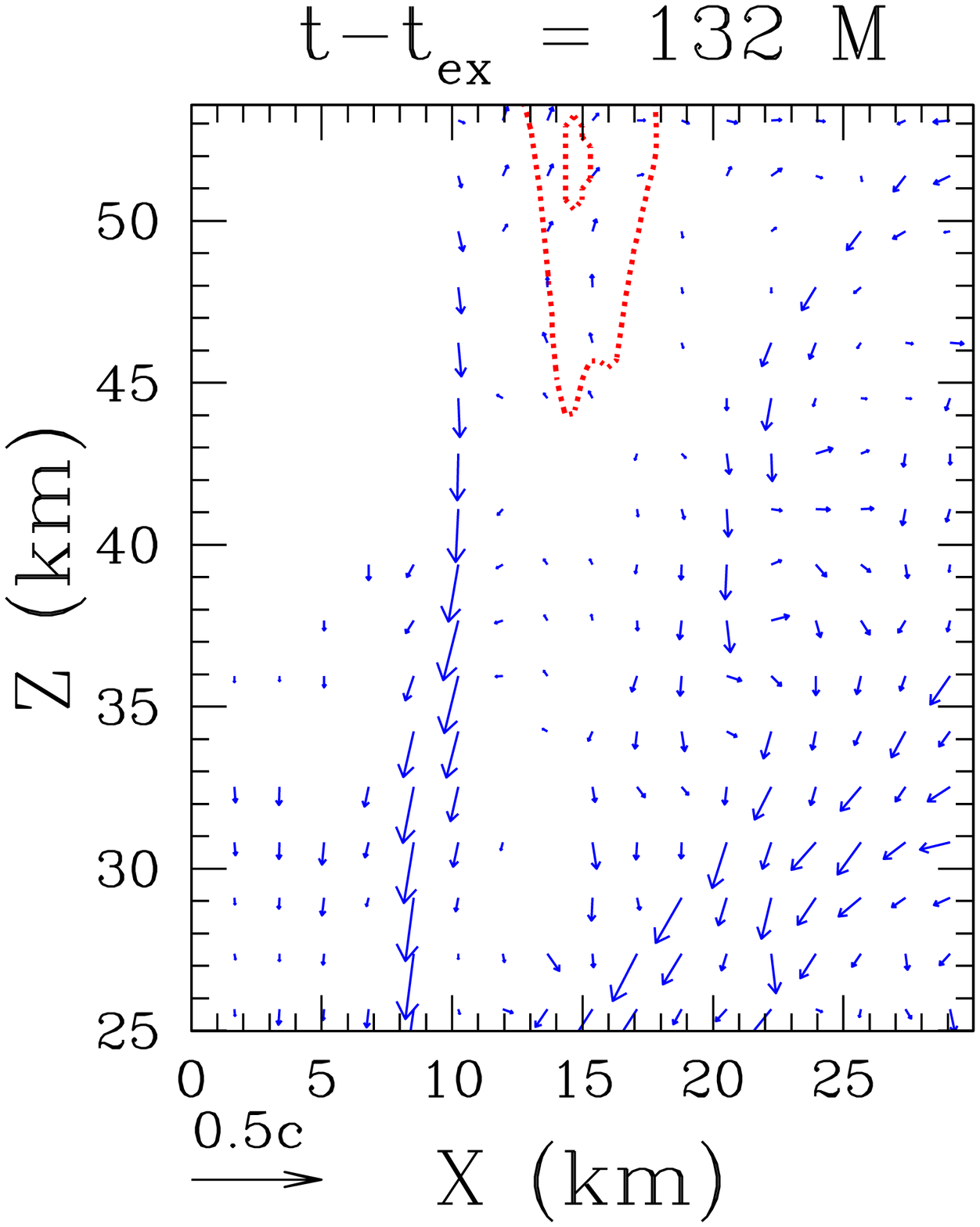}
\epsfxsize=2.2in
\leavevmode
\hspace{-1.5cm}\epsffile{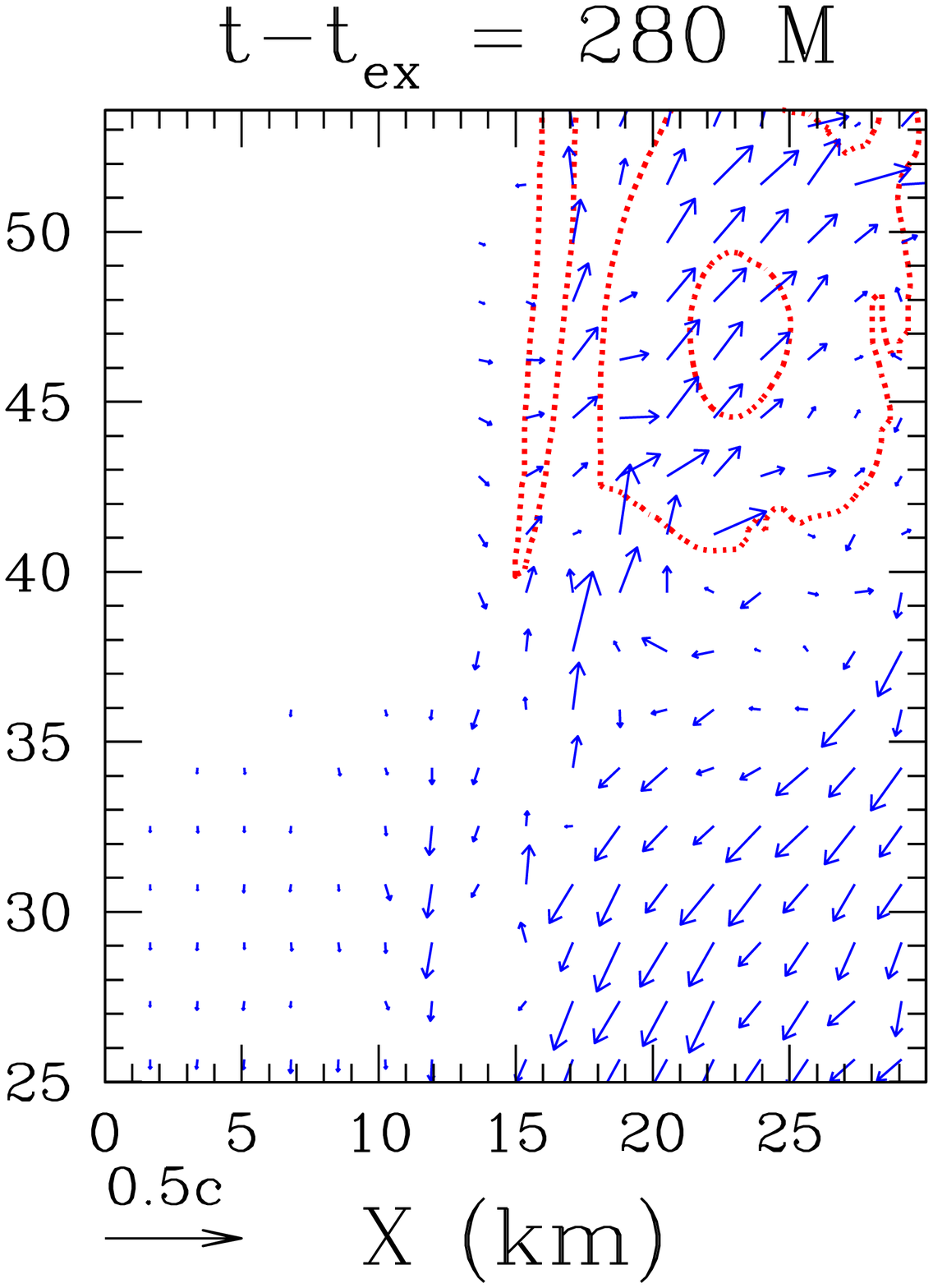}
\epsfxsize=2.2in
\leavevmode
\hspace{-1.5cm}\epsffile{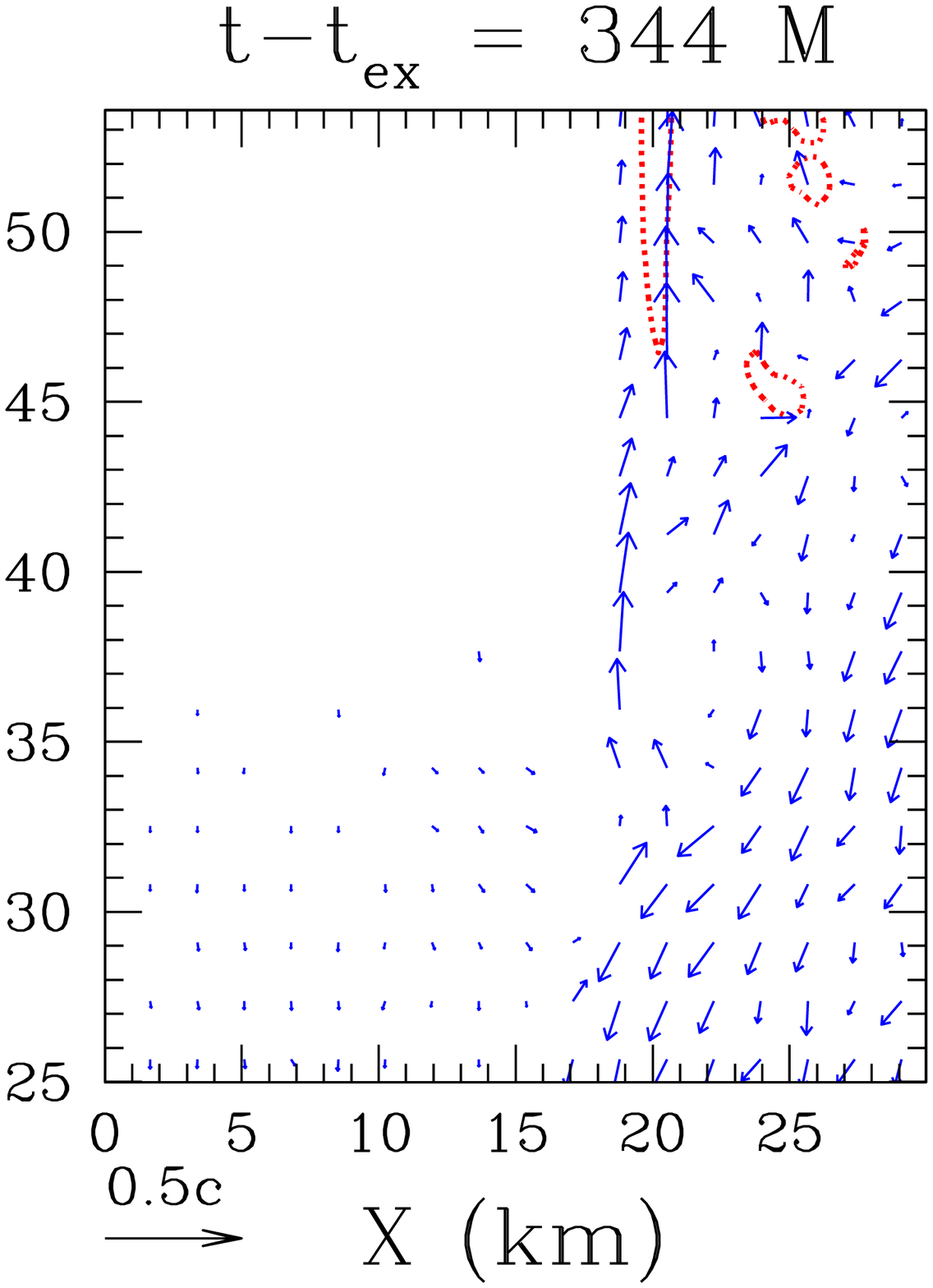}
\epsfxsize=2.2in
\leavevmode
\hspace{-1.5cm}\epsffile{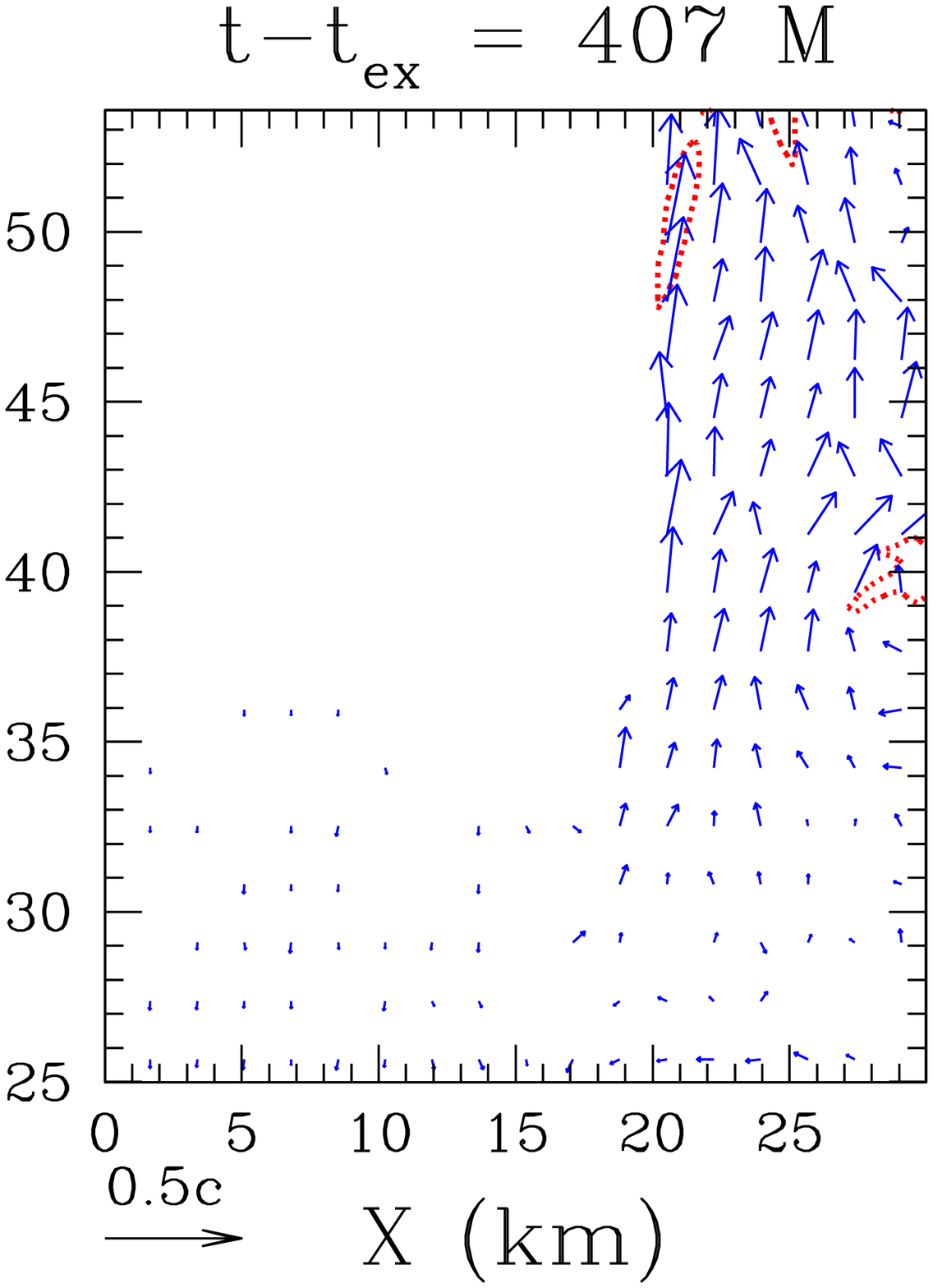}
\caption{Dotted contours demarcating regions with $-u_t \geq 1$,
and velocity vectors.  Note that the domain in these plots corresponds
to the upper left-hand corner of the domain shown in Fig.~\ref{c1_contours}.  
These snapshots demonstrate the
funnel wall outflow, and the first two times correspond to the
second and third sets of panels in Fig.~\ref{c1_contours}
\label{c1_contours_alt}}
\end{center}
\end{figure*}

We now describe the evolution of the hybrid EOS models C1 and C2.  
The behavior of both models up to the collapse and BH formation is 
represented in Fig.~\ref{preex}.  Secular MHD effects result in initially 
linear growth of $|B^y|_{\rm max}$ due to magnetic winding and sudden 
spurts of exponential growth of $|B^x|_{\rm max}$ due to the MRI 
(for a detailed discussion, see~\cite{DLSSS2}).
These MHD effects lead to collapse much later in the case of C2 than
in the case of C1.  This is due to the different magnetic 
field distributions in the two models.  Since $A_{\varphi}$ is proportional
to $\rho^{3/2}$ for~C2, the magnetic field is stronger in the outer
regions and weaker in the interior as compared with~C1 (see
Fig.~\ref{starC_init}).  However, collapse is triggered by 
transporting angular momentum from the interior to the exterior.
Since the interior magnetic field is weaker for case~C2, this process
takes longer and the collapse is delayed.  Both models eventually 
form BHs surrounded by magnetized accretion disks.  Constraint
violations remain less than $2\%$ during the pre-excision evolution for
both of these models.

\begin{figure}
\vspace{-4mm}
\begin{center}
\epsfxsize=3.in
\leavevmode
\epsffile{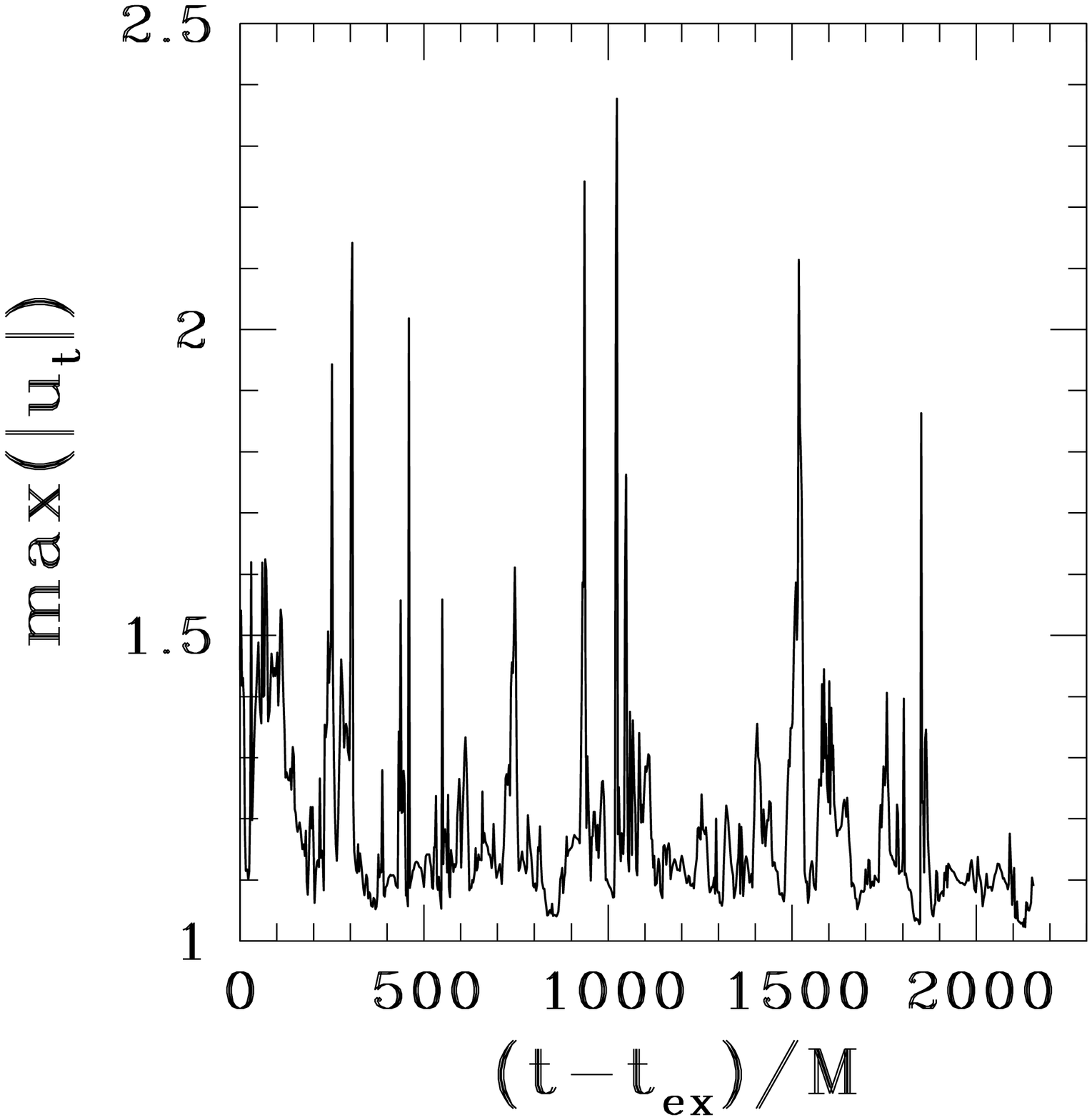}
\caption{The maximum value of $|u_t|$ on the grid as a function of time
since excision for C1.  This maximum is restricted to regions with outwardly 
directed velocity and with density greater than 10 times the atmosphere floor level.  
This quantity gives an approximation to the maximum asymptotic Lorentz factor
associated with the outflow.  Hence, the intermittent outflow is mildly relativistic.
\label{maxg}}
\end{center}
\end{figure}

For case C1, collapse occurs at $t \sim 939 M = 61 P_c$.
We excise the singularity at $t=t_{\rm ex}=941 M$  and freeze the spacetime
metric at $t=1221 M$.  Thus, the live BSSN evolution with 
excision lasts for $280 M$, and the maximum constraint violations 
during the post-excision phase are 
$(\mathcal{H},\mathcal{M}^x,\mathcal{M}^y,\mathcal{M}^z) = 
(0.5\%,4\%,8\%,3\%)$.  
 We evolve the system in the Cowling approximation for a further
$1876 M$ after the post-excision evolution ends. 
At the beginning of the Cowling phase, the 
rotation period midway through the disk is 
$\sim 148 M$, and thus the post-excision and Cowling phases 
together cover roughly 16 periods of the disk.   Following the 
collapse, the phase of extremely rapid accretion transitions to a 
slower rate at a time $t-t_{\rm ex} \sim 46 M$.  
Estimating the BH mass and angular momentum at this time 
[following Eqs.~(\ref{jhole}) and (\ref{mhole})] gives 
$M_h = 0.93 M$ and $J_h/M_h^2 = 0.71$. 
We note that the collapse time found here differs from the collapse 
time of $511 M = 33 P_c$ given for the corresponding model in~\cite{DLSSS2}.  
This is likely due to the sensitivity
of the system as it becomes marginally stable.

\begin{figure*}
\begin{center}
\epsfxsize=2.2in
\leavevmode
\hspace{-0.7cm}\epsffile{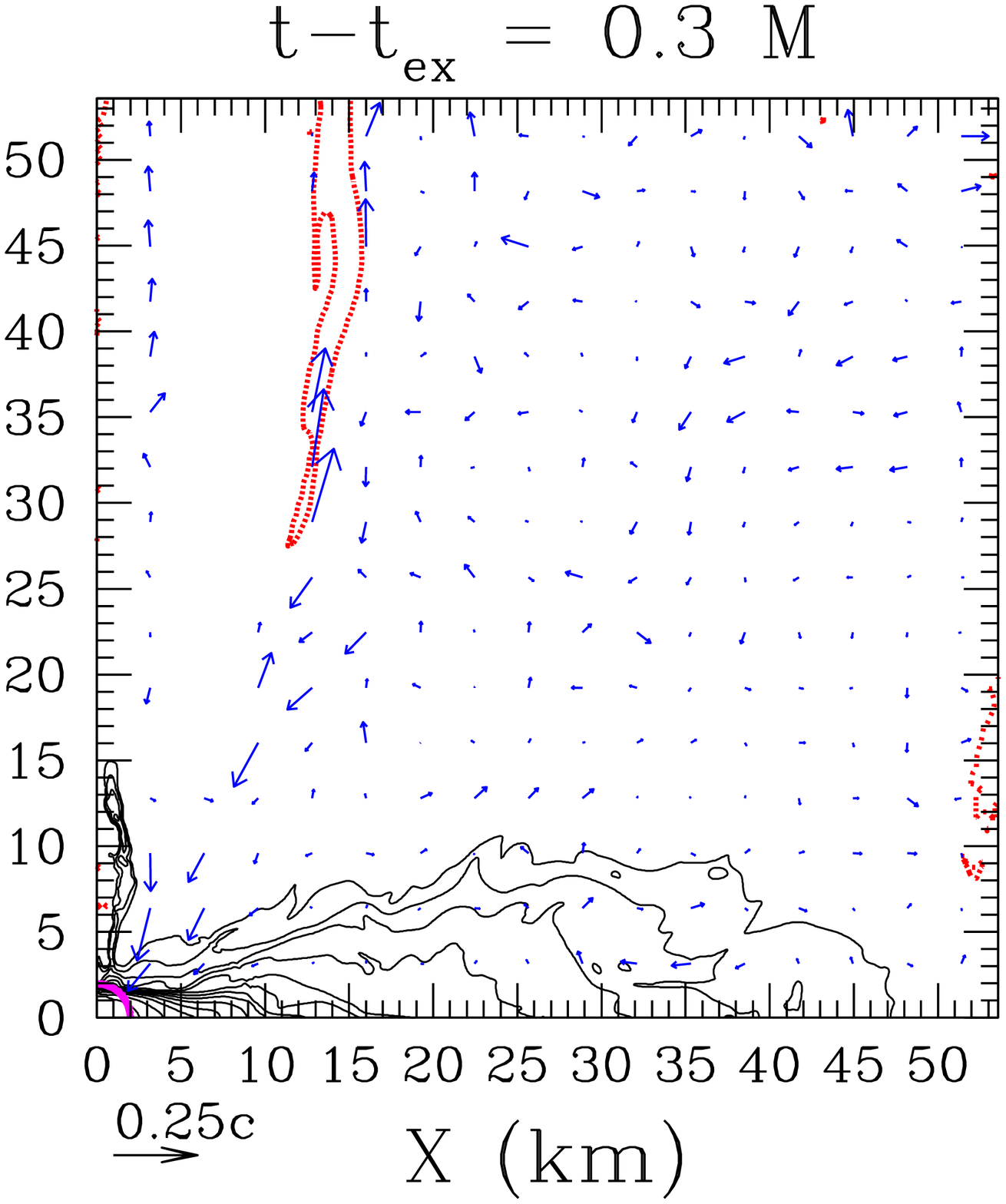}
\epsfxsize=2.2in
\leavevmode
\hspace{-0.5cm}\epsffile{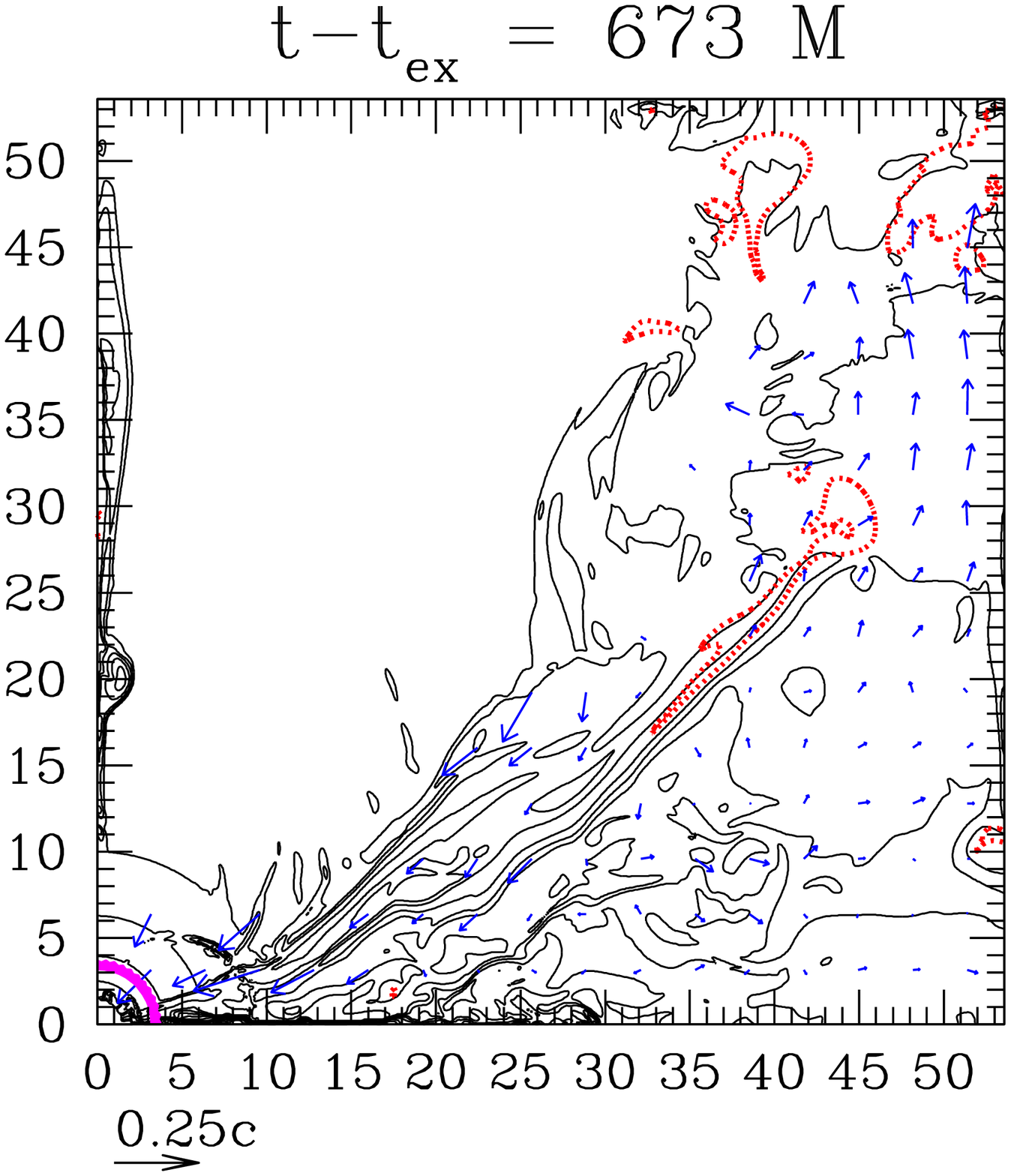}
\epsfxsize=2.2in
\leavevmode
\hspace{-0.5cm}\epsffile{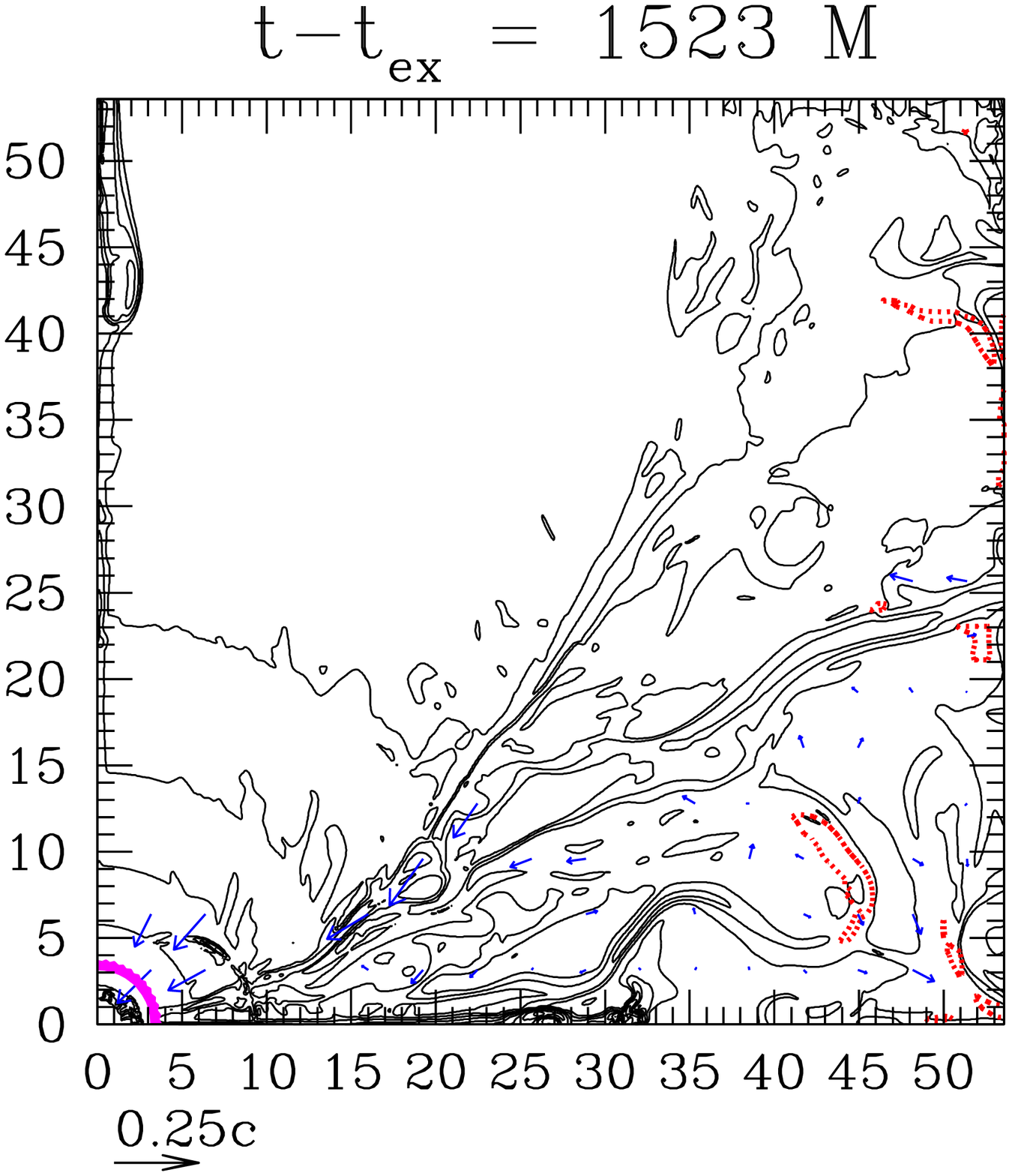} \\
\vspace{-0.5cm}
\epsfxsize=2.2in
\leavevmode
\hspace{-0.7cm}\epsffile{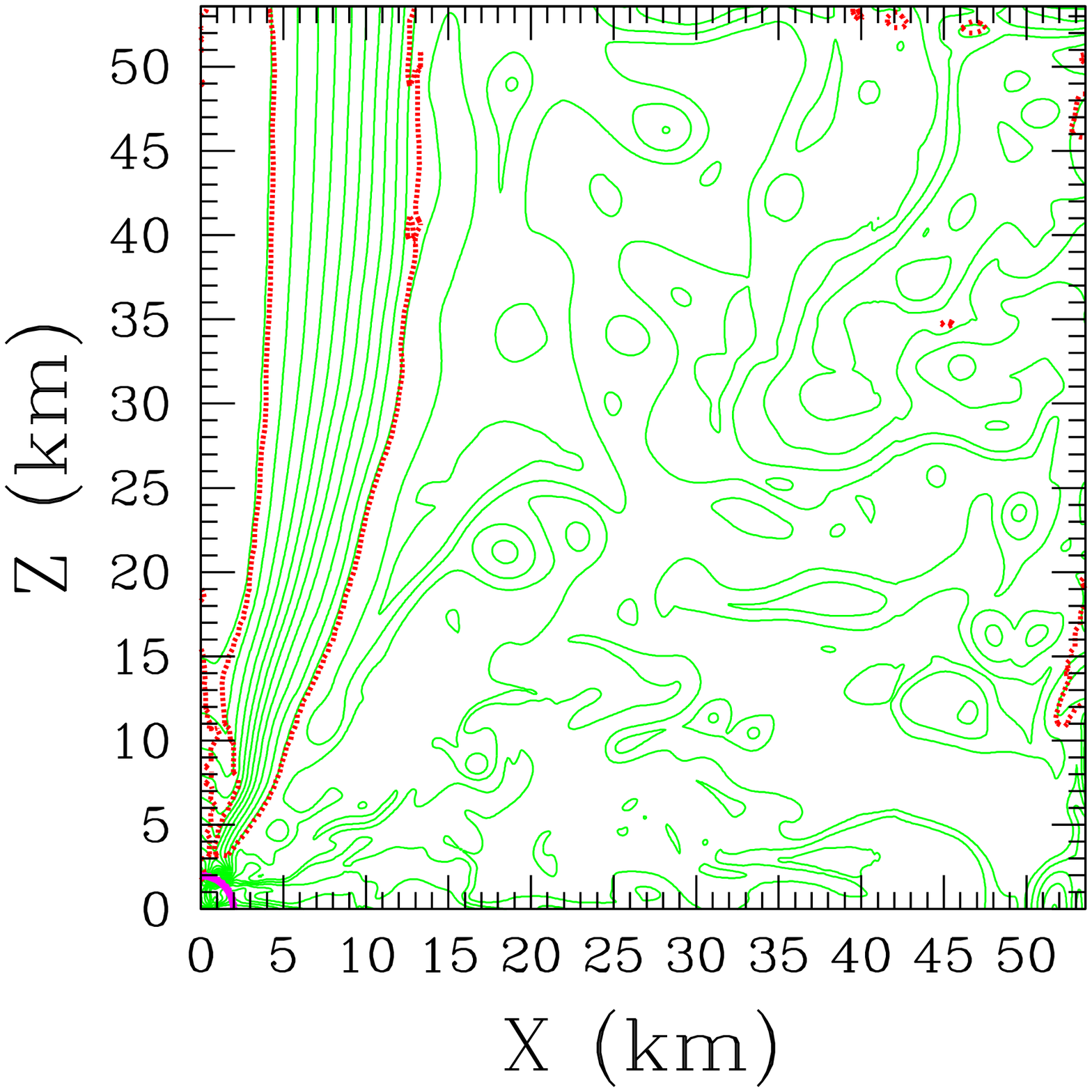}
\epsfxsize=2.2in
\leavevmode
\hspace{-0.5cm}\epsffile{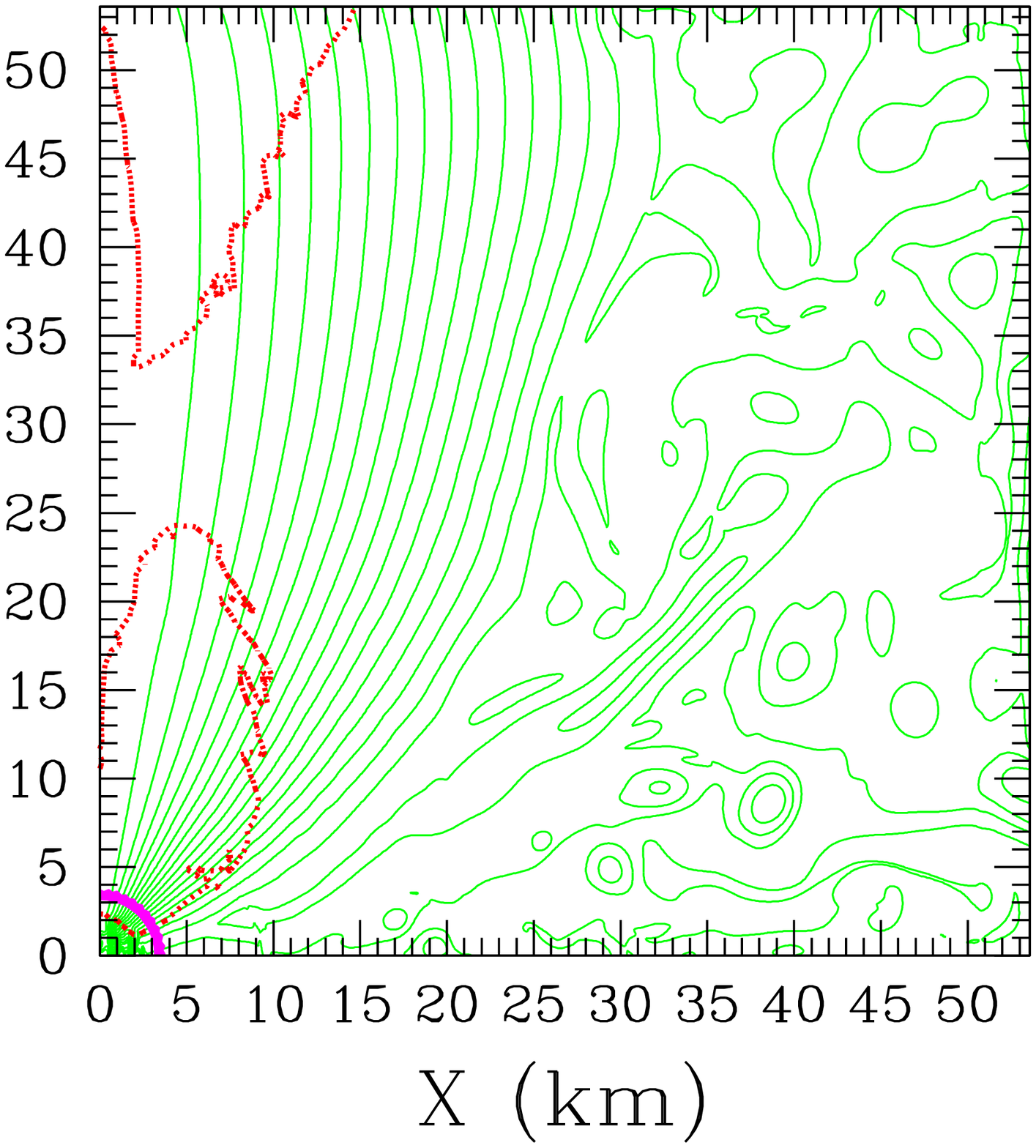}
\epsfxsize=2.2in
\leavevmode
\hspace{-0.5cm}\epsffile{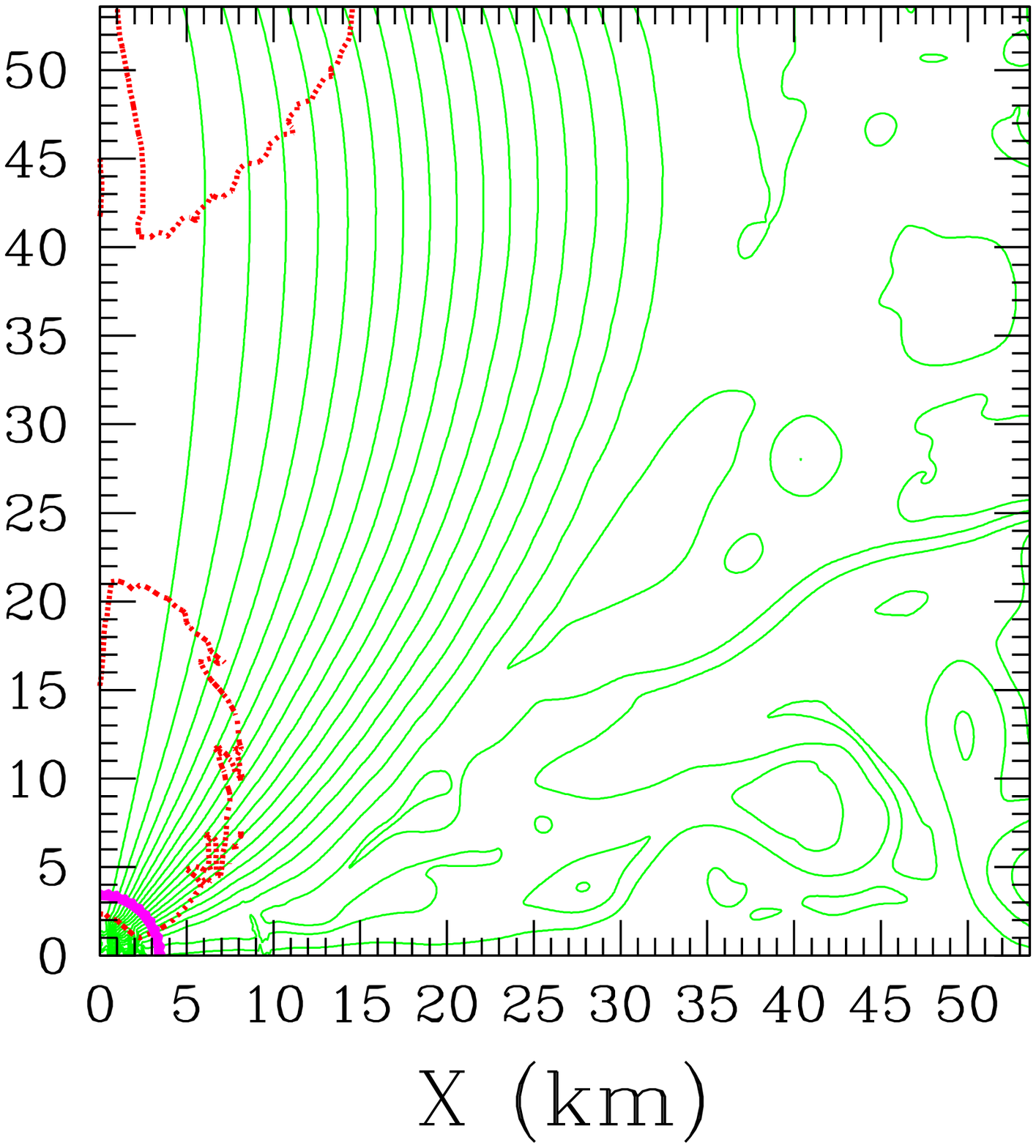}
\caption{Post-excision and Cowling evolution phases of case C2, shown
at selected times. The meanings of the lines are the same as in 
Fig~\ref{star_A_contours}.  The first and last sets of panels
mark the beginning of the post-excision phase and the end of the simulation,
respectively.  The middle panels show an intermediate time 
during the Cowling phase.  We note that the density feature near the 
rotation axis is not outbound or unbound, and thus does not 
constitute a jet.
\label{c2_contours}}
\end{center}
\end{figure*}

\begin{figure}
\vspace{-4mm}
\begin{center}
\epsfxsize=3.in
\leavevmode
\epsffile{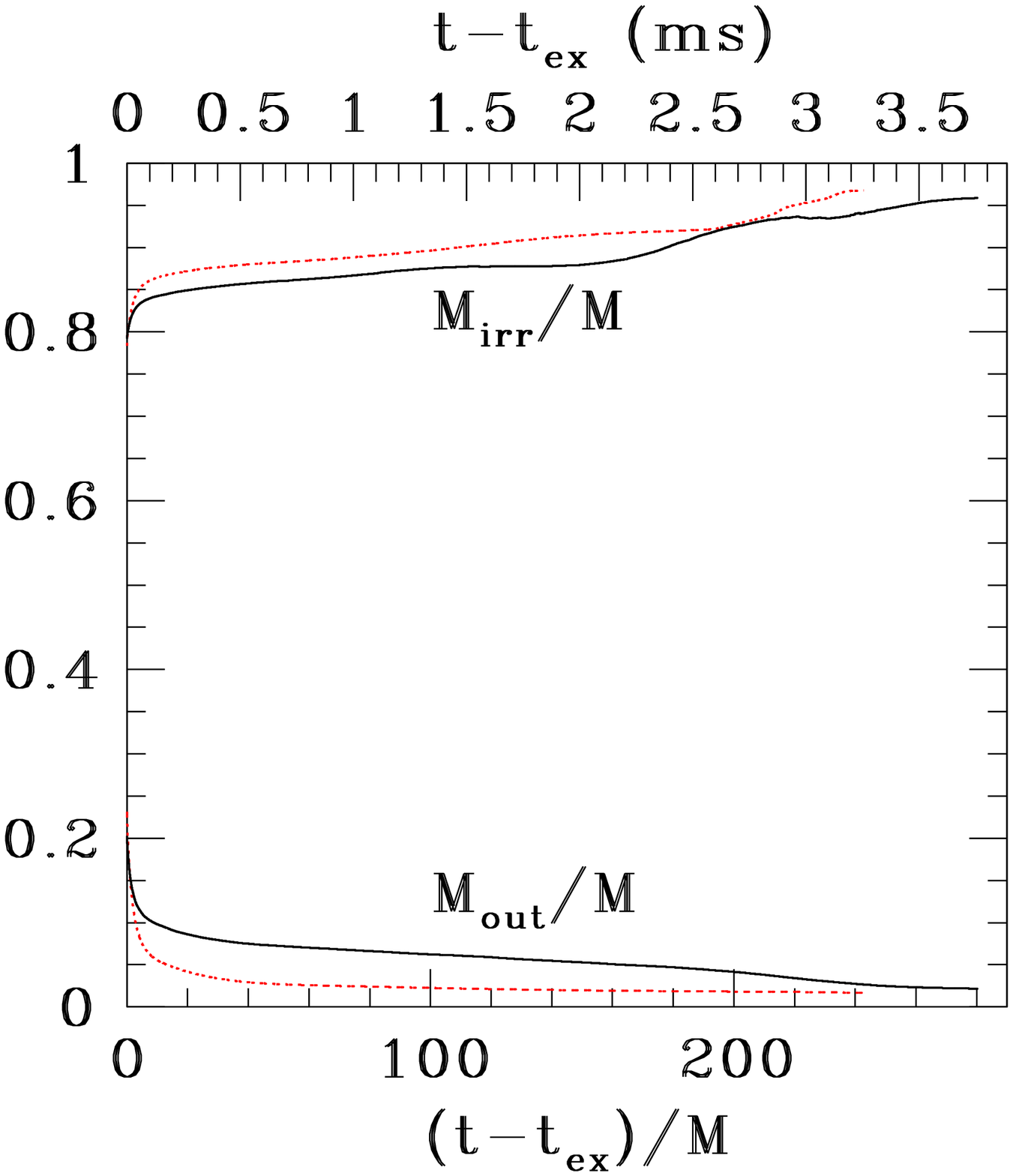}
\caption{Irreducible mass and rest mass outside the apparent horizon 
(normalized by the ADM mass at $t=0$) for cases~C1 (dashed red line)
and C2 (solid black line).  Notice 
that the mass remaining outside the apparent horizon decreases much 
more quickly for case~C2 than for~C1.
\label{excisionC}}
\end{center}
\end{figure}


Snapshots of the evolution during the post-excision and Cowling phases 
are shown in Fig.~\ref{c1_contours}.  (Contours showing $-u_t = 1$
are left out of this figure for the sake of clarity, but see 
Fig.~\ref{c1_contours_alt}.)  Several features of the 
evolution are immediately apparent.  As before, a low-density funnel 
region containing a collimated magnetic field has formed 
along the axis and persists through the evolution.  In the first 
three snapshots (post-excision and early Cowling phases), there is a 
considerable amount of material and magnetic field in a low density 
corona above the disk.  As material in this corona falls toward the 
central BH, the attached magnetic field lines make contact with 
the collimated magnetic field and reconnect.  Material attached
to the reconnected field line is then driven outward along the 
corona-funnel boundary by magnetic buoyancy.  (This process
is easiest to see in an animation, but can also be seen, for example,
in the $t-t_{\rm ex} = 280 M$ snapshot of Figure~\ref{c1_contours_alt} 
at $x\sim 20~{\rm km}$ and
$z \sim 47~{\rm km}$.  There, an unbound blob of material attached
to reconnected field lines moves toward the edge of the grid.)
Though we have not included a physical model for resistivity in
our code, this reconnection is ubiquitous in HRSC codes for
ideal MHD and physical reconnection is expected to operate in
systems with MHD turbulence such as this (see, e.g.~\cite{reconnection}).

This process leads to an intermittent outflow along the corona-funnel
wall boundary.  The outflow is mildly relativistic, with typical
Lorentz factors ranging between 1.2 and 1.5.  In Fig.~\ref{maxg}, we plot
the approximate maximum asymptotic Lorentz factors associated with the 
outflow for both the excision and Cowling phases.  The time-averaged value 
for the maximum is 1.2. We find that this outflow 
dies down as the Cowling evolution proceeds (see Fig.~\ref{outflux} below).  
As may be seen in the last three panels of Fig.~\ref{c1_contours}, the 
corona gradually empties of matter and magnetic fields, leaving a 
quasi-stationary disk surrounding the BH.  Without the turbulent driving 
of reconnection across the funnel-corona boundary, the outflow largely 
ceases.  In addition, the draining of pressure in the corona allows
the magnetic field in the funnel to expand outward, as
may be clearly seen in the last panel of Fig.~\ref{c1_contours}.  The
corona thus plays a role in confining the magnetically dominated 
funnel region, as found by McKinney~\cite{mcnum}.  It is possible 
that the corona would be sustained in a 3D evolution, for which 
the turbulent magnetic field growth is not limited by the 
anti-dynamo theorem~\cite{antidynamo}.  With our current computational
resources, it would not be feasible to perform this run in 3D
with high enough resolution to capture the MRI.  

\begin{figure}
\vspace{-4mm}
\begin{center}
\epsfxsize=3.in
\leavevmode
\epsffile{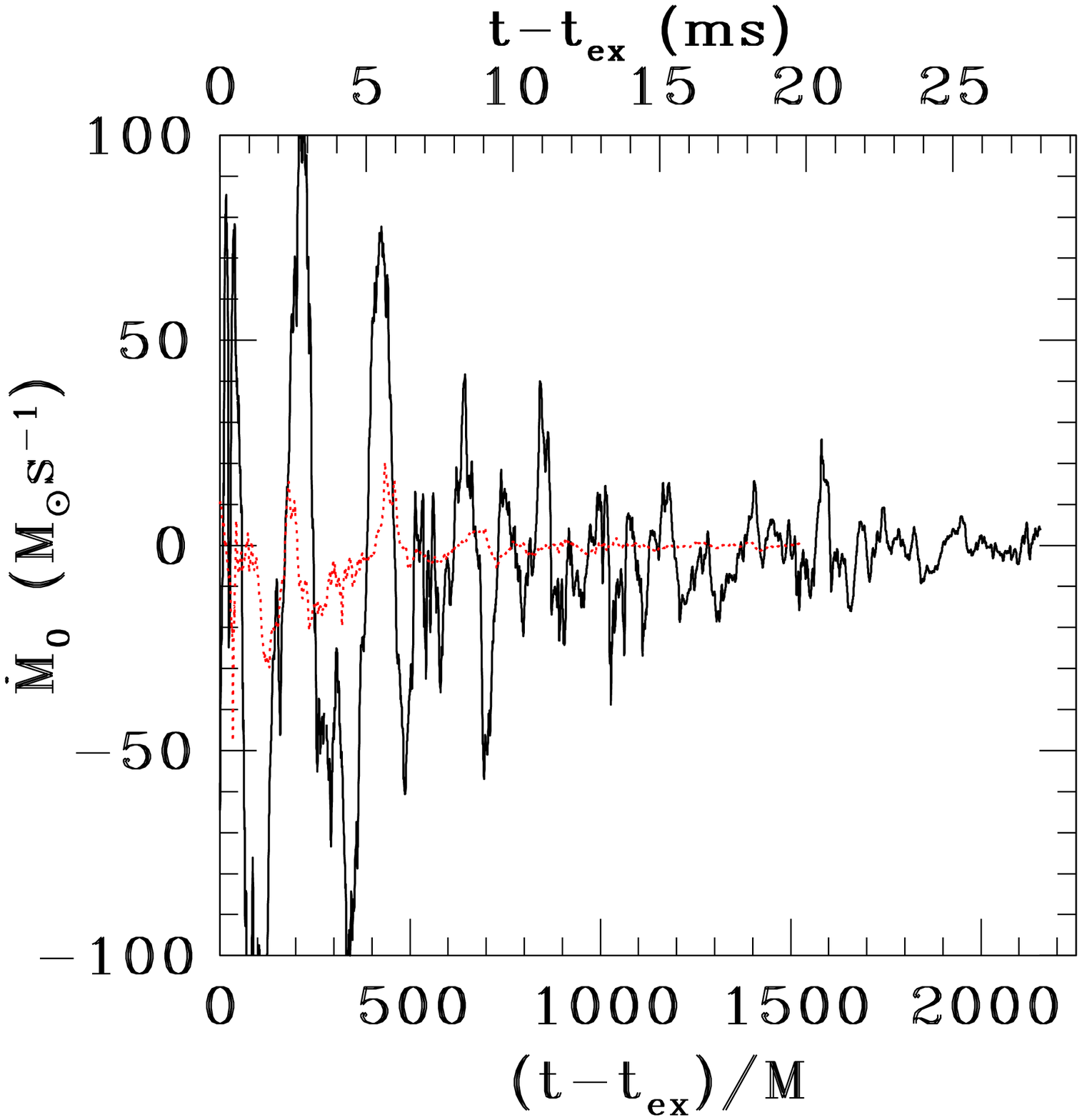}
\caption{Rest mass flux through a spherical shell located at 
$r = 12.2 M = 47.7~{\rm km}$ for cases~C1 (solid black line) and C2 (dashed 
red line).  This plot covers both the post-excision and Cowling phases.  
\label{outflux}}
\end{center}
\end{figure}

Case C2 undergoes collapse at $t \sim 3003 M = 194 P_c$ , and 
we excise the singularity at $t = t_{\rm ex} = 3008 M$.  The Cowling phase begins 
at $t= 3251 M$, giving a live BSSN post-excision run of duration 
$243 M$.  During the post-excision phase, the maximum normalized Hamiltonian 
and momentum constraints are
$(\mathcal{H},\mathcal{M}^x,\mathcal{M}^y,\mathcal{M}^z) = 
(0.4\%,1,5\%,13\%,3\%)$.  
The Cowling phase continues for an additional $1282 M$ beyond
the end of the post-excision run.  (A longer Cowling 
evolution is not necessary since the matter reaches an essentially quasi-stationary 
state.)  The rotation period midway through the disk at 
the beginning of the Cowling phase is $\sim 67 M$, and thus the post-excision 
and Cowling phases together cover roughly 22 periods of the disk.  
(Note that the disk in this case is considerably more compact than 
that of C1.)  Following the collapse,
the transition from extremely rapid to slower accretion occurs at 
$t \sim 3055 M = 197 P_c$.  Estimating the BH mass 
and angular momentum at this time gives $M_h = 0.97 M$ and 
$J_h/M_h^2 = 0.77$.  

\begin{figure}
\vspace{-4mm}
\begin{center}
\epsfxsize=3.in
\leavevmode
\epsffile{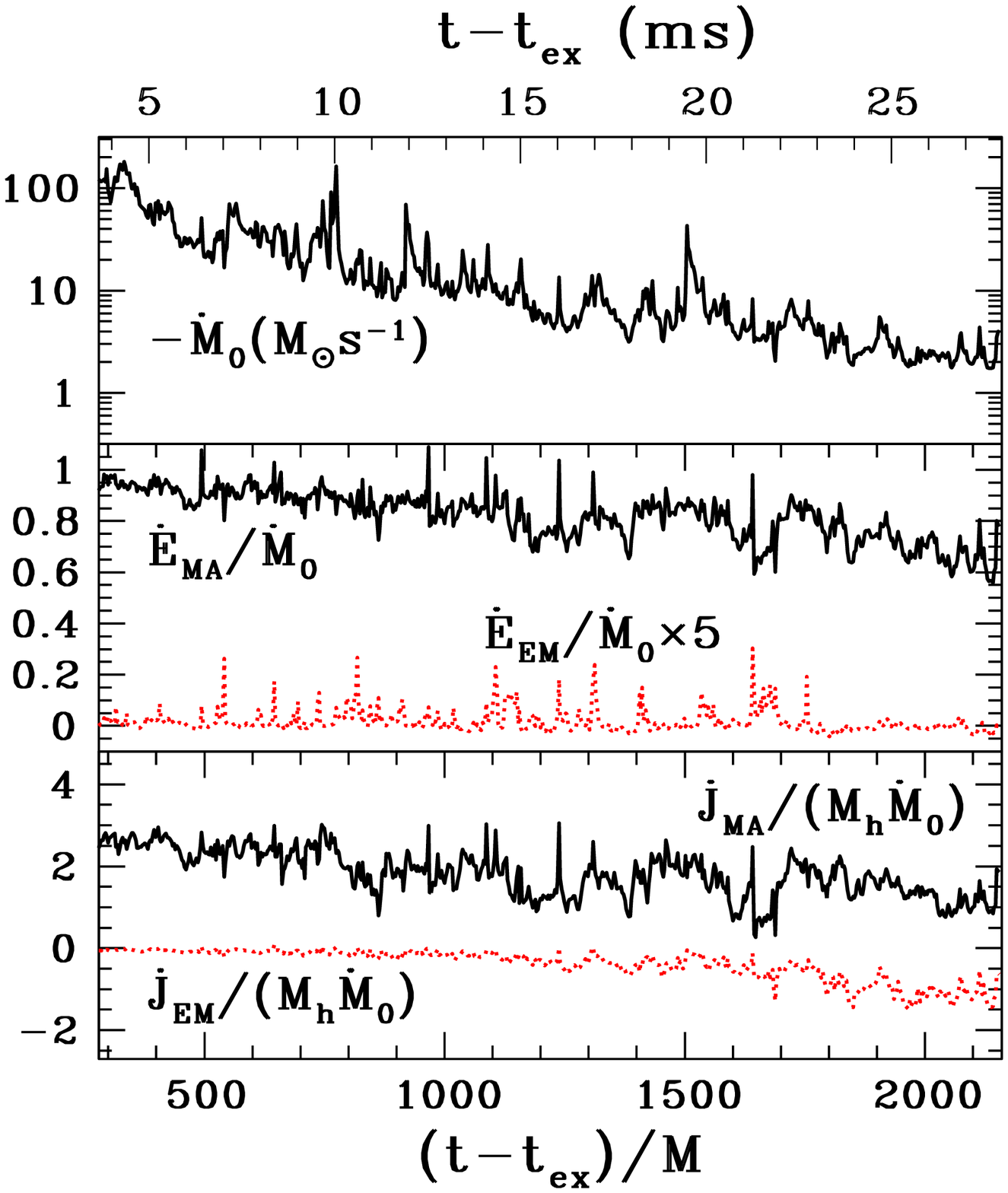}
\caption{Fluxes through the apparent horizon for case~C1 during 
the Cowling phase.  The
lines have the same meaning as in Fig.~\ref{fluxesA}.  Here, 
however, the electromagnetic contribution to the angular momentum 
flux in the lower panel is not multiplied by a factor in order to 
make it visible on the plot.
\label{fluxesC1}}
\end{center}
\end{figure}

\begin{figure}
\vspace{-4mm}
\begin{center}
\epsfxsize=3.in
\leavevmode
\epsffile{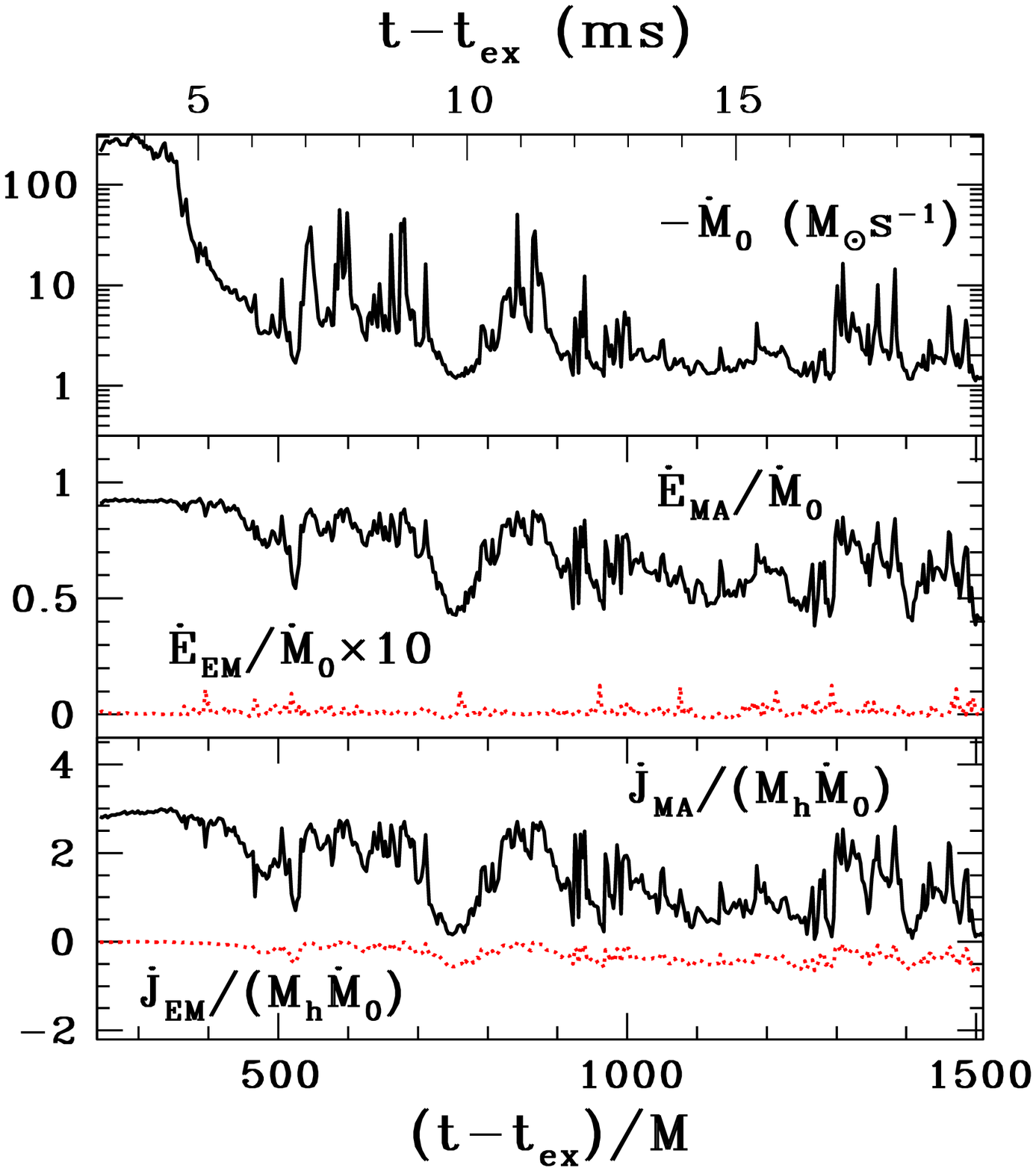}
\caption{Same as Fig.~\ref{fluxesC1} for case~C2.
\label{fluxesC2}}
\end{center}
\end{figure}

Selected snapshots from the post-excision and Cowling phases of 
evolution of case~C2 are shown in Fig.~\ref{c2_contours}.  
In contrast to case~C1, no significant outflows are seen for this 
case (except for a transient seen in the first panel).
The corona region contains
much less material and magnetic field at the beginning of the 
post-excision evolution than in case~C2.  This allows the funnel region
to expand, and it rapidly assumes the structure seen in the last 
two sets of panels.  The relative absence of material in the corona 
is likely due to the much more rapid accretion following collapse.
This can be seen in Fig.~\ref{excisionC}, which shows the evolution
of the irreducible mass and the mass remaining outside the horizon
in the two star~C cases.  For C2, $M_{\rm out}$ drops very
rapidly after excision (solid black
line), while the accretion takes place over a longer period of time
for case~C1. This may be understood from the difference in 
initial magnetic field configurations.  For C2, more of the star's mass
is initially threaded by magnetic field lines.  Since the magnetic field
remaining in the disk after collapse is thus stronger for case~C2, the
angular momentum transport is more efficient and the accretion is
more rapid.  

The comparative lack of outflows in C2 can also be seen 
by comparing the rest mass fluxes through an outer spherical shell for
cases~C1 and C2.  This is shown in Fig.~\ref{outflux}, where
the shell is placed at  $r = 12.2 M = 47.7~{\rm km}$.  This spherical 
shell cuts through the disk as well as the outflow region, and
thus there is a negative contribution from the inflow at lower
latitudes.  This overall inflow behavior is punctuated  by 
intermittent outflows in the C1 case, while the flux through this 
shell is much smaller in the C2 case.  These 
intermittent outflows eventually die down as the turbulence driving 
the outflow decays.  As in case~C2, the lack of outflows in case~A 
is likely caused by the lack of fluid and magnetic field in the 
corona region, which is in turn due to the EOS.  Star~A 
uses a $\Gamma = 2$ law for all density regimes, whereas the star~C 
EOS corresponds to $\Gamma = 1.3$ in the low density 
regions of interest in the disk and corona.  The softer EOS for 
star~C allows more efficient shock heating, which aids the ejection 
of material into the corona.  This interpretation is corroborated by 
the results of Mignone and McKinney~\cite{mm07}, who found that 
using $\Gamma = 4/3$ in FM torus simulations leads to more vigorous 
turbulence and a thicker corona than the same case with 
$\Gamma = 5/3$.  

Finally we plot the various flux quantities for the two star~C cases in 
Figures~\ref{fluxesC1} and \ref{fluxesC2}.  The upper panels show
the rest mass flux through the horizon, which again demonstrates
that, for case~C2, the accretion is very rapid at first but then 
decays rapidly.  The decrease in accretion rate for C1, on the 
other hand, is more gradual.  Averaging over the duration of
the Cowling evolution, we find that $\dot{E}/\dot{M}_0 = 0.90$ 
and $\dot{J}/(M_h\dot{M}_0) = 2.2$ for case C1.  For case C2, these values
are instead 0.90 and 2.7.  Cases C1 and C2 produce 
BHs with $J_h/M_h^2 = 0.72$ and 0.78, respectively.  For magnetized FM
tori surrounding Kerr BHs with spin parameters in the range 0.7-0.8, 
Table 2 of~\cite{mg04} suggests $\dot{E}/\dot{M}_0 \sim 0.9$ 
and $\dot{J}/(M_h\dot{M}_0) \gtrsim 2.0$.  Thus, for accretion
onto the central BH following collapse of star~C, the values
of $\dot{E}/\dot{M}_0$ and $\dot{J}/(M_h\dot{M}_0)$ are in 
rough agreement with the results of~\cite{mg04}.

\section{Conclusions}
\label{sec:conclusions}
We have used a code which solves the GRMHD equations in dynamical
spacetimes to simulate self-consistent disk formation and evolution following 
magnetized HMNS collapse.  Our simulations extend the results of~\cite{DLSSS2}
by following the disk evolution for a much longer time, as well as by
considering an alternative magnetic field geometry [see Eq.~(\ref{Aphi2})].
We evolve two HMNS models:  star~A with the simple $\Gamma-$law EOS of the
form $P = (\Gamma -1) \rho\varepsilon$ with $\Gamma = 2$, and star~C with 
the more realistic hybrid EOS given by Eqs.~(\ref{hybrid1}) and~(\ref{hybrid2}).  
Following collapse, star~A
quickly reaches a quasi-stationary accretion state, with very little matter
ejected from the system or churned up into the corona.  No significant 
outflow is observed from this system aside from a brief transient.  This is
likely due to the stiff EOS, which suppresses the formation of an extended
corona.  Without significant fluid and magnetic field above the disk, the
funnel wall outflow mechanism does not operate.  

We considered two cases for the star~C hypermassive model.
Case~C1 develops significant outflows along the boundary between the 
corona and the magnetically dominated funnel region near the axis.  These
outflows are triggered by reconnection across this boundary and the 
buoyant rising of the released field lines and matter.  These outflows
die down as the corona is gradually accreted.  In contrast, case~C2 
develops no significant outflows.  This model has a more extended initial 
magnetic field than C1, and thus the newly formed BH in the C2 case 
rapidly accretes most of the material remaining outside the apparent horizon, 
leaving little material left to the disk and corona.  Since the outflow
mechanism depends on interactions between the funnel and corona, outflows
are suppressed in this case.  We thus find that the presence of a funnel
wall outflow is sensitive both to the EOS and to the initial magnetic
field configuration.

As described in detail in~\cite{grbletter}, the remnant disk 
from the collapse of star~C may produce enough energy to power
a short GRB through neutrino-antineutrino annihilation alone.  However, 
we have also shown that the collapse results in the formation of a magnetically
dominated funnel, and the subsequent disk evolution (in the case of C1)
leads to a funnel-wall outflow.  The similarity of this morphology to 
previous studies of magnetized accretion disks~\cite{mg04,dvhk03} suggests 
that a Poynting dominated outflow in the funnel region may also be expected.  
That we do not see this feature may be due to the numerical difficulty of 
handling magnetically dominated regions, especially in a GRMHD code
for dynamical spacetimes (as discussed in~\cite{DLSS}),
and warrants further study.  In any case, a fully realistic evolution in this 
region requires more sophisticated treatment of microphysical processes 
(such as pair creation) and careful consideration of regions where the 
ideal MHD approximation may break down~\cite{mcnumalt,mcnum}.  
If the Blandford-Znajek process~\cite{bz77} does drive a Poynting-dominated 
outflow in the funnel region, the expected luminosity 
is~\cite{dimatteo,grbletter}
\beq 
L_{\rm BZ} \sim {\rm few} \times 10^{53} a^2 (B /10^{16}~{\rm G})^2
(M / 2.6 M_{\odot})^2~{\rm erg/s} \ .
\eeq
This luminosity would easily satisfy the typical energy
budget for a short GRB and would likely dominate the $\nu\bar{\nu}$ pair 
annihilation luminosity, which should be $L_{\nu\bar\nu}\sim 10^{50}~{\rm
ergs/s}$~\cite{dimatteo,grbletter}.  The mildly relativistic funnel wall 
outflow could play a role in collimating the inner fast jet~\cite{mcnum}, 
or even in stabilizing the jet against non-axisymmetric 
instabilities~\cite{mizuno3d}.

We also find that the magnetically dominated funnel region expands
at late times in the simulation as the corona density and pressure drop.
This could affect the opening angle at the base of a Poynting-dominated
outflow.  However, this emptying of the corona region may be due to 
the assumption of axisymmetry, since the corona depends on turbulent
churning of the disk and the disk turbulence must
decay by the anti-dynamo theorem~\cite{antidynamo}.  Ultimately, 
self-consistent 3D simulations encompassing the collapse and the 
subsequent disk and jet evolution will be required.  However, our
2D results should be qualitatively correct before the disk turbulence
begins to die down.

\acknowledgments
It is a pleasure to thank C.\ F.~Gammie, J.\ C.~McKinney, and S.\ 
C.~Noble for useful suggestions and discussions.  Numerical computations 
were performed at the National Center for Supercomputing Applications at 
the University of Illinois at Urbana-Champaign (UIUC).  This work was in 
part supported by NSF Grants PHY-0205155 and PHY-0345151, NASA Grants 
NNG04GK54G and NNG046N90H at UIUC.

\end{document}